\documentclass[
  BCOR=8mm,
  bibliography=totoc,
  DIV=12,
  fontsize=12pt,
  headings=small,
  headsepline=on,
  index=totoc,
  numbers=noenddot,
  paper=a4,
  parskip=false,
  twoside=on,
  ]{scrbook}

\pdfoutput=1
\DeclareOldFontCommand{\bf}{\normalfont\bfseries}{\mathbf}
\DeclareOldFontCommand{\tt}{\normalfont\ttfamily}{\mathtt}
\DeclareOldFontCommand{\it}{\normalfont\itshape}{\mathit}

\immediate\pdfobj stream attr{/N 3}  file{eciRGB_v2.icc}
\pdfcatalog{%
/OutputIntents [ <<
/Type /OutputIntent
/S/GTS_PDFA1
/DestOutputProfile \the\pdflastobj\space 0 R
/OutputConditionIdentifier (eciRGB v2)
/Info(eciRGB v2)
>> ]
}

\input glyphtounicode.tex
\input glyphtounicode-cmr.tex
\usepackage[ngerman, english]{babel}
\usepackage[T1]{fontenc}
\usepackage{graphicx}
\usepackage{epigraph}
\usepackage{setspace}
\usepackage{enumitem}
\usepackage{soul}
\usepackage{blindtext}
\usepackage[utf8]{inputenc}
\usepackage{makeidx}
\usepackage{hyperxmp}
\usepackage{float}
\usepackage{amsfonts}
\usepackage{amsthm}
\usepackage[]{algorithm2e}
\usepackage{mathtools}
\usepackage{axodraw4j}
\usepackage{color}
\usepackage[numbers,sort&compress]{natbib}
\usepackage[bookmarks=false]{hyperref}
\usepackage[]{hyperref}
\usepackage{tikz}
\usetikzlibrary{arrows,positioning, calc}

\hypersetup{
  pdftitle                       = {Algorithmic transformation of multi-loop Feynman integrals to a canonical basis},  
  pdfauthor                   = {Christoph Meyer}, 
  pdfsubject                  = {dissertation},
  pdfkeywords               = {Feynman integrals, canonical basis, algorithm},
  pdflang                       = en,
  bookmarksnumbered =true,
  pdfdisplaydoctitle       = true,
  colorlinks                    = false,
  plainpages                  = false,
  hypertexnames           = false,
  }

\SetKwRepeat{Do}{do}{end}

\def\eq#1{Eq.~\eqref{#1}}
\def\eqs{Eqs.~}
\def\chap#1{Chapter~\ref{#1}}
\def\sec#1{Sec.~\ref{#1}}
\def\fig#1{Fig.~\ref{#1}}
\def\app#1{App.~\ref{#1}}

\DeclareMathOperator{\ID}{ID}
\newcommand*{\defeq}{=} 
\newcommand*{\varset}{\epsilon, \{x_j\}} 
\newcommand*{\invariants}{\{x_j\}} 
\newtheorem{defn}{Definition}
\newtheorem{thm}{Theorem}
\newtheorem{corollary}{Corollary}
\newtheorem{lemma}{Lemma}
\newtheorem{claim}{Claim}
\newcommand*\justify{%
  \fontdimen2\font=0.4em
  \fontdimen3\font=0.2em
  \fontdimen4\font=0.1em
  \fontdimen7\font=0.1em
  \hyphenchar\font=`\-
}

\begin{document}

    \frontmatter
       \input{titlepagePREPRINTNR}
       \selectlanguage{english}
\chapter*{\begin{normalsize}\centerline{Abstract}\end{normalsize}}
\begin{quotation}
\noindent The evaluation of multi-loop Feynman integrals is one of the main challenges in the computation of precise theoretical predictions for the cross sections measured at the LHC. In recent years, the method of differential equations has proven to be a powerful tool for the computation of Feynman integrals. It has been observed that the differential equation of Feynman integrals can in many instances be transformed into a so-called canonical form, which significantly simplifies its integration in terms of iterated integrals.

The main result of this thesis is an algorithm to compute rational transformations of differential equations of Feynman integrals into a canonical form. Apart from requiring the existence of such a rational transformation, the algorithm needs no further assumptions about the differential equation. In particular, it is applicable to problems depending on multiple kinematic variables and also allows for a rational dependence on the dimensional regulator. First, the transformation law is expanded in the dimensional regulator to derive differential equations for the coefficients of the transformation. Using an ansatz in terms of rational functions, these differential equations are then solved to determine the transformation.

This thesis also presents an implementation of the algorithm in the \texttt{\justify Mathematica} package \textit{CANONICA}, which is the first publicly available program to compute transformations to a canonical form for differential equations depending on multiple variables. The main functionality and its usage are illustrated with some simple examples. Furthermore, the package is applied to state-of-the-art integral topologies appearing in recent multi-loop calculations. These topologies depend on up to three variables and include previously unknown topologies contributing to higher-order corrections to the cross section of single top-quark production at the LHC.


\end{quotation}
\begingroup
\selectlanguage{ngerman}
\chapter*{\begin{normalsize}\centerline{Zusammenfassung}\end{normalsize}}
\begin{quotation}
\noindent Die Auswertung von Mehrschleifen-Feynman-Integralen ist eine der gr\"o{\ss}ten Herausforderungen bei der Berechnung pr\"aziser theoretischer Vorhersagen f\"ur die am LHC gemessenen Wirkungsquerschnitte. In den vergangenen Jahren hat sich die Nutzung von Differentialgleichungen bei der Berechnung von Feynman-Integralen als sehr erfolgreich erwiesen. Es wurde dabei beobachtet, dass die von den Feynman-Integralen erf\"ullte Differentialgleichung oftmals in eine sogenannte kanonische Form transformiert werden kann, welche die Integration der Differentialgleichung mittels iterierter Integrale wesentlich vereinfacht.

Das zentrale Ergebnis der vorliegenden Arbeit ist ein Algorithmus zur Berechnung rationaler Transformationen von Differentialgleichungen von Feynman-Integralen in eine kanonische Form. Neben der Existenz einer solchen rationalen Transformation stellt der Algorithmus keinerlei weitere Bedingungen an die Differentialgleichung. Insbesondere ist der Algorithmus auf Mehrskalenprobleme anwendbar und erlaubt eine rationale Abh\"angigkeit der Differentialgleichung vom dimensionalen Regulator. Bei der Anwendung des Algorithmus wird zun\"achst das Transformationsgesetz im dimensionalen Regulator entwickelt, um Differentialgleichungen f\"ur die Koeffizienten in der Entwicklung der Transformation herzuleiten. Diese Differentialgleichungen werden dann mit einem rationalen Ansatz f\"ur die gesuchte Transformation gel\"ost.

Es wird zudem eine Implementation des Algorithmus in dem \texttt{\justify Mathematica} Paket \textit{CANONICA} vorgestellt, welches das erste ver\"offentlichte Programm dieser Art ist, das auf Mehrskalenprobleme anwendbar ist. Die wesentlichen Funktionen des Pakets werden zun\"achst mit einfachen Beispielen illustriert. \textit{CANONICAs} Potential f\"ur moderne Mehrschleifenrechnungen wird anhand mehrerer nicht trivialer Mehrschleifen-Integral\-topologien demonstriert. Die gezeigten Topologien h\"angen von bis zu drei Variablen ab und umfassen auch vormals ungel\"oste Topologien, die zu Korrekturen h\"oherer Ordnung zum Wirkungsquerschnitt der Produktion einzelner Top-Quarks am LHC beitragen.


\end{quotation}
\endgroup
\cleardoublepage

       \selectlanguage{english}
\chapter*{\begin{normalsize}\centerline{List of publications}\end{normalsize}}
This thesis is based on the following publications.
\begin{itemize}
\item C.~Meyer, \emph{{Evaluating multi-loop Feynman integrals using differential                                                                                                                                                                                                    
  equations: automatizing the transformation to a canonical basis}},
  {\emph{PoS} {\bf LL2016} (2016) 028}.
\item C.~Meyer, \emph{{Transforming differential equations of multi-loop Feynman                                                                                                                                                                                                     
  integrals into canonical form}},
  \href{http://dx.doi.org/10.1007/JHEP04(2017)006}{\emph{JHEP} {\bf 04} (2017)
  006}, [\href{https://arxiv.org/abs/1611.01087}{{\tt 1611.01087}}].
\item C.~Meyer, \emph{{Algorithmic transformation of multi-loop master integrals to a                   
  canonical basis with CANONICA}},
  \href{http://dx.doi.org/10.1016/j.cpc.2017.09.014}{\emph{Comput. Phys.                          
  Commun.} {\bf 222} (2018) 295--312},
  [\href{https://arxiv.org/abs/1705.06252}{{\tt 1705.06252}}].
  
\end{itemize}
\cleardoublepage

       \pagestyle{headings}
\pagenumbering{Roman}
\tableofcontents

    \mainmatter
       \chapter{Introduction}

The current knowledge of the fundamental constituents of matter and their interactions is largely based on scattering experiments. The pioneering gold foil experiment~\cite{Geiger495}, which led Rutherford to hypothesize the nuclear structure of the atom~\cite{Rutherford:1911zz}, was the first in a long series of scattering experiments conducted to improve the understanding of subatomic phenomena. With ever more sophisticated instruments, researchers were able to increase both the energy and the intensity of the involved particle beams by several orders of magnitude since the early experiments~\cite{Shiltsev:2014zga}. Over the course of the last century, this led to the discovery of a plethora of new particles~\cite{Olive:2016xmw}, which prompted the conception of the Standard Model of particle physics \cite{GLASHOW1961579, PhysRevLett.19.1264, Salam:1968rm, Fritzsch:1973pi, Gross:1973ju, Gross:1973id, Politzer:1973fx} to describe their interactions. In 2012, this development culminated in the observation \cite{Aad:2012tfa, Chatrchyan:2012xdj} of the Higgs boson \cite{Guralnik:1964eu, Englert:1964et, Higgs:1964ia, Higgs:1966ev} at CERN's Large Hadron Collider (LHC). With the Higgs boson being the last constituent of the Standard Model to be discovered, it is now considered to be complete in the sense that it is self-consistent up to energy scales far beyond current experimental reach. 

The Standard Model successfully describes almost all observations made at past and present collider experiments \cite{Altarelli:1997et, Liu:2273226}, often with remarkable precision. Although the conception of the Standard Model represents a great success of particle physics, it provides no explanation for some observed properties of the universe. Numerous astronomical observations strongly suggest the existence of dark matter in the universe \cite{Begeman:1991iy, Clowe:2006eq, Ade:2013zuv}. In the most commonly accepted scenario of cold dark matter \cite{Peebles:1982ff, Blumenthal:1984bp}, the dark matter is comprised of weakly interacting non-relativistic particles. However, the Standard Model does not offer any suitable candidates for these particles~\cite{Quigg:2008ab} and thus needs to be extended. A second open problem is posed by the observed asymmetry of matter and anti-matter in the universe \cite{Cohen:1997ac, Alcaraz:2000ss}, since it is unknown which dynamical mechanism, if any, has created it. The observed value of the Higgs boson mass and the amount of $CP$\nobreakdash-violation in the Standard Model render it very unlikely that the Standard Model can accommodate such a mechanism~\cite{Canetti:2012zc}. A further shortcoming of the Standard Model is that it does not account for the non-vanishing neutrino masses \cite{Pontecorvo:1957cp, Maki:1962mu, Schechter:1980gr}, which have been experimentally observed \cite{Fukuda:1998mi, Ahmad:2001an, An:2012eh}. Lastly, the Standard Model does not incorporate the gravitational interactions, which would have to be included in any fundamental theory of physics.

In order to address the aforementioned open problems, many extensions of the Standard Model have been proposed. Among the most popular are supersymmetric extensions \cite{Nilles:1983ge, Haber:1984rc, Barbieri:1987xf}, models with additional or composite Higgs bosons \cite{Lee:1973iz, Branco:2011iw, Bellazzini:2014yua, Panico:2015jxa, Csaki:2015hcd}, models with extra dimensions \cite{Appelquist:1987nr, ArkaniHamed:1998rs, Randall:1999ee} and those with heavy partners of the gauge bosons \cite{Senjanovic:1975rk, Langacker:2008yv}. However, the experimental data from the LHC does currently not show any significant deviations from the Standard Model. Since the LHC is operating almost at its design center of mass collision energy of 14 $\textup{TeV}$ and the construction of new colliders is likely to take decades~\cite{Zimmermann:2014qxa}, it will not be possible to directly probe the Standard Model at much higher energy scales in the near future. The only possibility left is to look for deviations from the Standard Model by increasing the precision of the comparisons between theory and experiment.
%
%
The principal observables used in this comparison are cross sections of the particle reactions taking place at the LHC. For some processes, the experimental uncertainties \cite{Aaboud:2016pbd, Aaboud:2016btc, Aaboud:2016zpd} have reached the level of the theoretical uncertainties and are predicted to drop further with the LHC accumulating more data~\cite{ATL-PHYS-PUB-2014-016}. Thus, more precise theoretical predictions for the background and the signal processes are necessary to harness the LHC's full potential.

The calculation of cross section predictions in quantum field theory is very challenging and mostly only accessible by perturbative methods. In the perturbative approach, the cross section is expanded as an asymptotic power series~\cite{PhysRev.85.631} for small values of the coupling constants and truncated at some finite order. This is only a good approximation if the respective coupling strength is small enough at the energy scale the process is considered at. For the typical energy scales of the hard particle reactions at the LHC, the couplings in the Standard Model are small enough for the perturbative approach to be feasible~\cite{Olive:2016xmw}. 


The accuracy of the theoretical predictions is limited by the accuracy of the experimentally determined input parameters and the order at which the perturbation series is truncated. Therefore, great effort is dedicated to the precise measurement of the input parameters and to the calculation of higher-order corrections in the perturbative expansion. These calculations are highly non-trivial and often take several years until completion. One of the main challenges is posed by the evaluation of integrals over unconstrained momenta, called \emph{Feynman integrals} or \emph{loop integrals}. Each order in the perturbative expansion introduces a further unconstrained momentum and thereby increases the difficulty of the respective Feynman integrals. 
%
%
%
%
%
%
%
%

Given the enormous complexity of higher-order corrections, computers have become an indispensable tool for their calculation. Since many calculational techniques apply to a wide range of scattering processes, it is worthwhile to automatize them as much as possible. Over the past decades, great progress has been made in the automation of next-to-leading order (NLO) corrections. There are numerous tools~\cite{Ossola:2015xga} publicly available allowing the automated calculation of NLO corrections for most processes of interest at the LHC. Among other insights \cite{Bern:1994zx, Bern:1994cg, Britto:2004nc, Forde:2007mi, Ellis:2007br, Giele:2008ve}, it was the explicit knowledge of the Feynman integrals occurring in NLO computations that made these developments possible.


In recent years, several advances \cite{GehrmannDeRidder:2005cm, Catani:2007vq, Somogyi:2008fc, Czakon:2010td, Henn:2013pwa, Boughezal:2015eha} allowed to also calculate the next-to-next-to-leading order (NNLO) corrections for many processes \cite{Ferrera:2011bk, Czakon:2013goa, Brucherseifer:2014ama, Cascioli:2014yka, Gehrmann:2014fva, Grazzini:2015nwa, Boughezal:2015dva, Boughezal:2015dra, Caola:2015wna, Czakon:2015owf, Boughezal:2015ded, Campbell:2016yrh, Grazzini:2016swo, Ridder:2015dxa, Ridder:2016nkl, Catani:2011qz, Abelof:2016pby, Chen:2016zka, Gehrmann-DeRidder:2016jns, Campbell:2016lzl}. Even higher-order corrections are known for the Higgs boson production cross section in the gluon fusion channel~\cite{Anastasiou:2015ema}. The recent progress has in part been enabled by new developments in the field of Feynman integrals. Most calculations of higher-order corrections are organized such that a huge number of Feynman integrals appear at intermediate stages of the calculation. These integrals are related by an enormous number of linear relations, called integration-by-parts (IBP) relations \cite{Tkachov:1981wb, Chetyrkin:1981qh}. By virtue of these relations, all integrals can be expressed in terms of a finite basis of independent integrals, the so-called \emph{master integrals}. After this reduction process, only the relatively small number of master integrals need to be evaluated.

A vast array of techniques has been developed for the evaluation of Feynman integrals. In practice, however, these techniques remain limited in their scope, and there is currently no general solution available for the problem of evaluating Feynman integrals. A rather general technique is to derive a differential equation \cite{Kotikov:1990kg, Remiddi:1997ny, Gehrmann:1999as} for the master integrals by differentiating them with respect to the kinematic invariants and masses they depend on. However, solving this differential equation in terms of known functions can, in general, be prohibitively difficult. In 2013 it was discovered by Henn~\cite{Henn:2013pwa} that the solution can often be simplified dramatically by using a particular basis of master integrals coined \emph{canonical basis}. The differential equation of a canonical basis of master integrals attains a simple so-called \emph{canonical form} that renders its integration in terms of iterated integrals a merely combinatorial task. With this remarkable observation, the evaluation of Feynman integrals is essentially reduced to the problem of constructing a canonical basis of master integrals, given it exists. This new technique has been successfully applied to the calculation of numerous previously unknown Feynman integrals \cite{Henn:2013pwa, Henn:2013fah, Henn:2013woa, Henn:2013nsa, Argeri:2014qva, Henn:2014lfa, Caron-Huot:2014lda, Gehrmann:2014bfa, Caola:2014lpa, Li:2014bfa, Hoschele:2014qsa, DiVita:2014pza, vonManteuffel:2014mva, Grozin:2014hna, Bell:2014zya, Huber:2015bva, Gehrmann:2015ora, Gehrmann:2015dua, Bonciani:2015eua, Anzai:2015wma, Grozin:2015kna, Gehrmann:2015bfy, Gituliar:2015iyq, Lee:2016htz, Henn:2016men, Bonciani:2016ypc, Eden:2016dir, Lee:2016lvq, Bonciani:2016qxi, Bonetti:2016brm, Henn:2016kjz, Lee:2016ixa, DiVita:2017xlr, Boels:2017skl, Lee:2017mip} and thereby contributed to the aforementioned proliferation of NNLO calculations.

Despite the recent advances, the automation of NNLO calculations has not yet reached the same level as NLO calculations have. In contrast to the fully automated frameworks available for NLO cross sections, there are only computer codes available to perform certain steps of the calculation. Concerning the evaluation of the Feynman integrals, the systematic application of the IBP relations for the reduction to master integrals is widely considered as a conceptually solved problem, and there is a number of programs available \cite{Anastasiou:2004vj, Studerus:2009ye, vonManteuffel:2012np, Lee:2012cn, Smirnov:2013dia, Smirnov:2014hma, Larsen:2015ped, Georgoudis:2016wff, Maierhoefer:2017hyi} to perform this computation. After the reduction to some basis of master integrals, the derivation of their differential equation is straightforward and has been implemented in \cite{vonManteuffel:2012np, Lee:2012cn}.

This leaves the process of constructing a canonical basis as the next step to be automated. In this thesis, an algorithm will be described to compute a rational transformation to a canonical basis from a given basis of master integrals, provided such a transformation exists. Prior to the publication of this algorithm~\cite{Meyer:2016slj}, some methods to attain a canonical basis had already been proposed \cite{Henn:2013pwa, Argeri:2014qva, Caron-Huot:2014lda, Gehrmann:2014bfa, Hoschele:2014qsa, Lee:2014ioa, Henn:2014qga, Eden:2016dir}. In particular, an algorithm to compute a transformation to a canonical form for differential equations depending on only one variable has been described in detail by Lee~\cite{ Lee:2014ioa}. Most of the other methods do not rise to the same level in terms of their algorithmic description, but rather represent recipes for specific cases. This is also reflected by the fact that Lee's algorithm is the only one with publicly available implementations \cite{Gituliar:2016vfa, Prausa:2017ltv, Gituliar:2017vzm}. The main drawback of Lee's algorithm is that it is only applicable to differential equations depending on one variable, which severely restricts the range of processes it can be applied to. For instance, most $2\rightarrow 2$ scattering processes depending on one or more mass scales are not accessible with this method. The motivation for the development of the algorithm described in this thesis is to overcome this restriction. To this end, the algorithm is devised such that it is applicable to differential equations depending on an arbitrary number of variables. In order to facilitate the application of this algorithm, it has been implemented and made publicly available~\cite{Meyer:2017joq} in a \texttt{\justify Mathematica} package called \textit{CANONICA}.

The outline of this thesis is as follows. After introducing some basic concepts related to Feynman integrals, \chap{chap:preliminaries} reviews the IBP reduction to master integrals and the derivation of the corresponding differential equations. The solution of this differential equation is then shown to simplify considerably by using a canonical basis of master integrals.

\chap{chap:algorithm} is dedicated to the problem of transforming a given differential equation of Feynman integrals into canonical form. After examining some general properties of such transformations, it is shown that they can be computed by solving a finite number of differential equations with a rational ansatz. Moreover, it is argued that this computation can be split into a series of smaller computations by exploiting certain structural properties of the differential equation.  Altogether, \chap{chap:algorithm} lays out an algorithm to compute rational transformations to canonical bases, which is applicable to differential equations depending on an arbitrary number of variables. 

The implementation of the aforementioned algorithm in the \texttt{\justify Mathematica} package \textit{CANONICA} is presented in \chap{chap:package}. The usage of its main features is explained with a number of simple examples along with a discussion of its limitations.

The power of \textit{CANONICA} and the underlying algorithm are demonstrated in \chap{chap:applications}, which presents the application of \textit{CANONICA} to a variety of non-trivial multi-loop Feynman integrals. In particular, this includes differential equations depending on up to three variables and previously unknown integrals. 

The conclusions are drawn in the final \chap{chap:conclusion}.

       \chapter{Aspects of multi-loop calculations}
\label{chap:preliminaries}

The main part of this thesis is devoted to the presentation of an algorithm related to the evaluation of Feynman integrals. While Feynman integrals are an interesting topic in their own right, the main motivation for the techniques developed in this thesis is the calculation of higher-order corrections to cross section predictions for the LHC. After showing how Feynman integrals arise in these calculations, this chapter discusses the techniques for treating Feynman integrals used in modern calculations of higher-order corrections. Particular emphasis will be on those technical aspects related to the method of differential equations, which is a powerful technique for the evaluation of Feynman integrals. The development of an algorithm to transform these differential equations into a canonical form, which is the main result of this thesis, is then motivated by illustrating the tremendous benefits such a form provides for the integration of the differential equation.

Most of the material presented in this chapter is well established and can be found in much more detail in the references given below. The exposition here aims to provide the practical context for the following more abstract chapters by introducing the relevant concepts and illustrating them with a simple example.

\section{From cross sections to Feynman integrals}

The most frequently used observables in collider experiments are cross sections of the various scattering processes. This section reviews the relation of cross sections to scattering amplitudes and shows how Feynman integrals arise in their perturbative calculation.






\subsection{Cross sections and Feynman diagrams}

%

Scattering processes are modeled in quantum field theory by the transition of an initial state $|i\rangle$ to a final state $\langle f|$, which are considered as Heisenberg picture states in the infinite past and infinite future, respectively. The time evolution operator $\mathcal{S}=U(\infty,-\infty)$ encodes the interaction and depends on the specific quantum field theory used, which can, for example, be the Standard Model. The scattering amplitudes $\langle f|\mathcal{S}|i\rangle$ are conveniently separated into a trivial and an interacting part by defining the transition operator $\mathcal{T}$ by
\begin{equation}
\mathcal{S}=\mathbb{I}+\textup{i}(2\pi)^4\delta^{(4)}\left(\Sigma p_f - \Sigma p_i\right)\mathcal{T},
\end{equation}
where the $\delta$\nobreakdash-function enforces momentum conservation. The non-trivial part of the interaction is then contained in the \emph{matrix elements}
\begin{equation}
\mathcal{M}_{fi}=\langle f|\mathcal{T}|i\rangle.
\end{equation}
The matrix elements are directly related to the total cross section via
\begin{equation}
\sigma \sim \int |\mathcal{M}_{fi}|^2\textup{d}\Pi,
\end{equation}
where the integration is over the phase space of the final state and the constant of proportionality depends on the kinematics of the specific process. Predictions for hadron colliders require additional integrations over the parton momenta in order to relate the partonic cross section to the hadronic cross section.

\begin{figure}[ht]
\centering
\begin{minipage}{.5\textwidth}
  \centering
\includegraphics[scale=0.95]{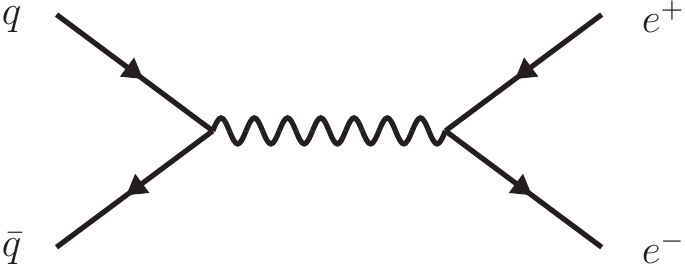}
\end{minipage}%
\begin{minipage}{.5\textwidth}
  \centering
\includegraphics[scale=0.95]{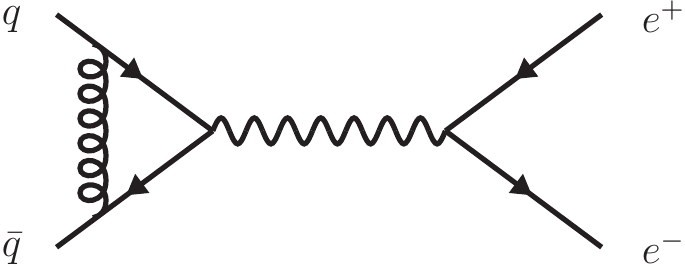}
\end{minipage}
\caption{Drell--Yan tree level and one-loop Feynman diagrams.}
  \label{fig:DrellYanMiniExample}
\end{figure}

In perturbation theory, matrix elements are expanded as a power series in the respective coupling strength, for instance, the coupling strength of the strong interactions~$\alpha_s$
\begin{equation}
\mathcal{M}_{fi}=\alpha_s^n\left(\mathcal{M}_{fi}^{(0)}+\alpha_s\mathcal{M}_{fi}^{(1)}+\mathcal{O}(\alpha_s^2)\right).
\end{equation}
The coefficients of this expansion are the building blocks for the calculation of higher-order corrections to the cross section \cite{Kinoshita:1962ur, Lee:1964is}. The individual terms contributing to the calculation of $\mathcal{M}_{fi}^{(l)}$ have a diagrammatic representation in terms of so-called \emph{Feynman diagrams}. These are graphs comprised of a specific set of edges and vertices connecting the initial and final state external legs, which is illustrated in \fig{fig:DrellYanMiniExample} by the Drell--Yan process $q\bar{q}\rightarrow e^{+}e^{-}$. 
The allowed vertices and edges are determined by the specific quantum field theory via a set of rules known as \emph{Feynman rules}, which translate Feynman diagrams into their corresponding analytic expressions. Generally, each additional loop in a Feynman diagram raises the power of the coupling strength in the corresponding analytic expression by one. As a consequence, only Feynman diagrams of a fixed loop order contribute to a given order in the perturbative expansion of a matrix element. In practice, Feynman diagrams provide a convenient way to generate the analytic representation of a given matrix element $\mathcal{M}_{fi}^{(l)}$. First, all Feynman diagrams with the given loop order and the right initial and final state are generated. These diagrams are then converted into analytic expressions by virtue of the Feynman rules.

%
%
%
%
%
%

\subsection{Dimensionally regulated Feynman integrals}

\begin{figure}[ht]
\begin{center}
\includegraphics[scale=1]{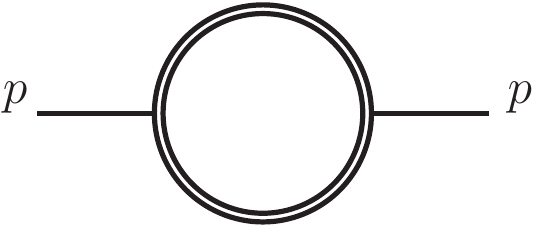}
\end{center}
\caption{One-loop massive bubble integral.}
\label{fig:OneLoopBubble}
\end{figure}

The independent loops in a Feynman diagram are each associated with the integration over an unconstrained so-called \emph{loop momentum}. In general, these \emph{Feynman integrals} are of the form
\begin{equation}
\label{defTensorFI}
\int\prod_{k=1}^L\frac{\textup{d}^dl_k}{\textup{i}\pi^{d/2}}\frac{l_1^{\mu_1}\cdots l_1^{\mu_{r_1}}\cdots l_L^{\kappa_1}\cdots l_L^{\kappa_{r_L}}}{P_1\cdots P_t},
\end{equation}
where the \emph{inverse propagators} $P_i$ are given by\footnote{An additional term of $+\textup{i}\delta$ in the inverse propagator ensuring the correct time-ordering of the propagator by shifting its poles away from the real axis is omitted here and in the following.}
\begin{equation}
\label{invPropdefn}
P_i=q_i^2-m_i^2,
\end{equation}
with $m_i$ denoting the mass of the propagator and $q_i$ being a linear combination of the loop momenta and the momenta of the initial and final state particles, which are referred to as \emph{external momenta}. Since momentum conservation always allows to eliminate one of the external momenta, the term external momenta is in the following understood to refer to the remaining $N_\textup{ex}$ momenta after momentum conservation has been enforced. A typical example of a one-loop Feynman integral is given by
\begin{equation}
I^\mu(p^2)=\int\frac{\textup{d}^dl}{\textup{i}\pi^{d/2}}\frac{l^\mu}{[l^2-m^2][(l-p)^2-m^2]},
\end{equation}
which corresponds to the Feynman diagram in \fig{fig:OneLoopBubble}.
The naive evaluation of Feynman integrals in four space-time dimensions leads, in general, to divergent results. Therefore, it is necessary to regularize these integrals. While there are several different regularization schemes available, modern calculations almost exclusively employ variants of \emph{dimensional regularization}~\cite{tHooft:1972tcz}. In dimensional regularization, the four-dimensional loop integrations are rendered convergent by promoting them to integrations in
\begin{equation}
\label{reldeps}
d=4-2\epsilon
\end{equation}
dimensions. The divergencies in four dimensions are then reflected by poles in the regulator $\epsilon$. The integration over non-integer dimensional vector spaces is, of course, not to be understood literally. Instead, $d$\nobreakdash-dimensional integration can be defined as a functional of the integrand satisfying the following axioms~\cite{Wilson:1972cf}:
\begin{description}
\item[Linearity:]{\begin{equation}
\int\textup{d}^dl[af(l)+bg(l)]=a\int\textup{d}^dlf(l)+b\int\textup{d}^dlg(l),\quad a,b\in\mathbb{C},
\end{equation}}
\item[Scaling:]{\begin{equation}
\int\textup{d}^dlf(sl)=|s^{-d}|\int\textup{d}^dlf(l),\quad s\in\mathbb{C},
\end{equation}}
\item[Translation invariance:]{\begin{equation}
\int\textup{d}^dlf(l+q)=\int\textup{d}^dlf(l),\quad q=\mathrm{const.},
\end{equation}}
\end{description}
which resemble properties of ordinary integration if $d$ is a positive integer. The axioms above can be shown \cite{Wilson:1972cf, Collins:105730} to uniquely fix the values of all dimensionally regulated integrals up to a universal normalization, and therefore all explicit constructions of such a functional must yield equivalent results up to normalization. The universal normalization is often fixed by defining the value of the integral over the $(d-1)$\nobreakdash-dimensional unit sphere
\begin{equation}
\int\textup{d}\Omega_{d-1}\mathrel{\vcentcolon=}\frac{2\pi^{d/2}}{\Gamma\left(\frac{d}{2}\right)},
\end{equation}
which also holds for ordinary integration in positive integer dimensions. In practice, it is rarely necessary to resort to an explicit construction of the integration functional\footnote{Explicit constructions may be obtained by a parametric representation of Feynman integrals~\cite{Smirnov:2006ry} or by a prescription using integrations over finite dimensional subspaces of an infinite dimensional vector space~\cite{Collins:105730}.}. Instead, it is often sufficient to use the axioms above and some additional properties derived from them. In particular, the integration-by-parts \cite{Tkachov:1981wb, Chetyrkin:1981qh} property given by
\begin{description}
\item[Integration-by-parts:]{\begin{equation}
\label{IBPpropertydimreg}
\int\textup{d}^dl\frac{\partial}{\partial l^\mu}f(l)=0,
\end{equation}}
\end{description}
is widely used in practice, as will be explained in \sec{sec:Reduction}. The properties outlined above are sufficient for the purposes of this thesis; for a more extensive account of the properties of Feynman integrals the reader is referred to \cite{Collins:105730, Smirnov:2006ry}.

\subsection{The projection method}
In the calculation of matrix elements, the tensor structures in the numerator of the Feynman integrals in \eq{defTensorFI} are either contracted with loop momenta, external momenta, polarization vectors or with Dirac gamma matrices. This section presents a technique frequently employed in multi-loop calculations to separate the spin degrees of freedom from the loop integrals. As a result, all loop momenta in the numerator of the Feynman integrals are only contracted with either loop momenta or external momenta. These scalar products in the numerators can then be reduced to a minimal set of \emph{irreducible scalar products}. 

The first step is to identify an independent set of spin structures $D_j$ sufficient to decompose the matrix element (cf. e.g., \cite{Anastasiou:1999bn, Glover:2004si})
\begin{equation}
\label{AmpDecomp}
\mathcal{M}_{fi}^{(l)}=\sum_j m_j D_j.
\end{equation}
An efficient way to organize the computation of the scalar coefficients $m_j$ is to define projection operators $\mathcal{P}_j$ to project the Feynman diagrammatic representation of the matrix element onto the spin structures. The projectors can be decomposed with respect to the basis of spin structures:
\begin{equation}
\mathcal{P}_j=\sum_kc_{jk}D_k^\dagger.
\end{equation}
The coefficients $c_{jk}$ are determined by the condition
\begin{equation}
\label{ApplyProjector}
\sum_{\textup{spins}}\mathcal{P}_j\mathcal{M}_{fi}^{(l)}=\sum_{k,r}c_{jk}\sum_{\textup{spins}}D_{k}^\dagger D_r m_r=m_j.
\end{equation}
It is convenient to define the matrix
\begin{equation}
\label{spinsumProjector}
\mathcal{D}_{ij}=\sum_{\textup{spins}}D_i^\dagger D_j,
\end{equation}
which allows to express the coefficients of the projectors through its inverse
\begin{equation}
c_{ij}=\left(\mathcal{D}^{-1}\right)_{ij}.
\end{equation}
Using the representation of $\mathcal{M}_{fi}^{(l)}$ in terms of Feynman diagrams, the contribution of each Feynman diagram to the coefficients $m_j$ can be extracted by applying the respective projectors as in \eq{ApplyProjector}. The resulting scalar coefficients $m_j$ are linear combinations of Feynman integrals of the form
\begin{equation}
\label{sIntsproj}
\int\prod_{k=1}^L\frac{\textup{d}^dl_k}{\textup{i}\pi^{d/2}}\frac{Q_{t+1}^{-\nu_{t+1}}\cdots Q_{t+N_s}^{-\nu_{t+N_s}}}{P_1\cdots P_t},
\end{equation}
with non-positive integer powers $\nu_{t+1},\dots,\nu_{t+N_s}$. The numerator factors $Q_i$ are given by the 
\begin{equation}
N_s=\frac{L(L+1)}{2}+N_\textup{ex}L
\end{equation}
different scalar products of loop momenta and external momenta, involving at least one loop momentum. 
%
The inverse propagators $P_1,\dots,P_t$ are independent linear combinations of the scalar products $Q_i$ and terms independent of the loop momenta. Therefore, there are $N_s-t$ linear combinations of the scalar products $Q_i$ and terms independent of the loop momenta which are linearly independent of the inverse propagators. Upon choosing such a set of $N_s-t$ so-called \emph{irreducible scalar products} $P_{t+1},\dots,P_{N_s}$, the integrals in \eq{sIntsproj} can be uniquely written as linear combinations of the integrals 
%
\begin{equation}
\label{sIntdef}
I(\nu_1,\dots,\nu_{N_s})=\int\prod_{k=1}^L\frac{\textup{d}^dl_k}{\textup{i}\pi^{d/2}}\frac{P_{t+1}^{-\nu_{t+1}}\cdots P_{N_s}^{-\nu_{N_s}}}{P_1^{\nu_1}\cdots P_t^{\nu_t}},
\end{equation}
where the powers $\nu_i$ of the inverse propagators are now allowed to assume any integer value\footnote{For one-loop integrals, there are no irreducible scalar products since $N_s=t$. This fact is exploited in the Passarino--Veltman~\cite{Passarino:1978jh} reduction procedure, which is widely used in one-loop calculations to relate tensor integrals to scalar integrals.}. 
In state-of-the-art computations, the number of integrals of the form in \eq{sIntdef} necessary to express the whole matrix element is often of the order of several thousands or more. Thus, it is clearly desirable to treat these with an automatized procedure rather than attempting a case by case analysis.


\section{Reduction to master integrals}
\label{sec:Reduction}

The integrals of the form in \eq{sIntdef} are related by a class of linear relations known as \emph{integration-by-parts} (IBP) identities, which allow to express all such integrals as linear combinations of a relatively small number of so-called \emph{master integrals}. As a result of this reduction, it is sufficient to evaluate only the master integrals. This section reviews the basic concepts related to the IBP reduction as well as some aspects of the practical organization of such calculations. 

\subsection{Topologies and sectors}

It is beneficial to group the integrals occurring in a particular multi-loop calculation into sets of integrals that can be expressed by the same set of propagators and irreducible scalar products. Allowing for arbitrary integer powers of the inverse propagators and non-positive integer powers of the irreducible scalar products, each of these sets contains an infinite number of integrals of the form in \eq{sIntdef}. These infinite sets of integrals are called \emph{topologies}\footnote{Some authors also use the term \emph{integral family}.}. The integrals within a given topology can be further divided into different \emph{sectors}, where a sector is a set of integrals which share the same set of propagators with positive exponent. Therefore, there are $2^t$ sectors in a topology with $t$ propagators. A sector is said to be a \emph{subsector} of another sector if its set of propagators is a subset of the other sector's set of propagators. Since two sectors may have disjoint sets of propagators, the sector-subsector relation defines only a partial ordering on the set of sectors. This partial ordering can be turned into a total ordering by defining an integer-valued function on the integrals of the topology:
\begin{align}
\ID[I]&=\sum_{k=1}^t2^{k-1}\Theta(\nu_k),\\
\Theta(x)&=\begin{cases} 
      1 & x > 0, \\
      0 & x\leq 0.
   \end{cases}
\end{align}
This function is constant on sectors, and therefore it can be understood as assigning an integer to each sector, called \emph{sector-id}. Moreover, the sector-id is compatible with the partial ordering induced by the sector-subsector relation, because the sector-id of a sector is always greater than the sector-ids of all of its subsectors. Note that the definition of the sector-id depends on the ordering of the propagators. 

\subsection{Integration by parts identities}
While a topology contains an infinite number of integrals, there also exists an infinite number of linear relations among them. These integration-by-parts (IBP) relations arise from the property in \eq{IBPpropertydimreg} of dimensionally regularized Feynman integrals \cite{Tkachov:1981wb, Chetyrkin:1981qh}, which can be cast in the form
\begin{equation}
\label{IBP}
\int\prod_{k=1}^L\frac{\textup{d}^dl_k}{\textup{i}\pi^{d/2}}\frac{\partial}{\partial l_j^\mu}\left(q^\mu\frac{P_{t+1}^{-\nu_{t+1}}\cdots P_{N_s}^{-\nu_{N_s}}}{P_1^{\nu_1}\cdots P_t^{\nu_t}}\right)=0, \quad j=1,\dots,L.
\end{equation}
This relation holds for any value of the propagator powers and for $q$ being any loop or external momentum. The derivative in the integrand can be carried out explicitly to generate a relation between integrals with different powers of the inverse propagators and irreducible scalar products. Using the fact that all scalar products which may occur due to the contraction with $q$ can be written as a linear combination of the inverse propagators and irreducible scalar products, \eq{IBP} can be expressed as a linear combination of integrals of the same topology with their propagator powers possibly lowered or raised by one:
\begin{equation}
\sum_j c_jI_j(\nu_1+\Delta^j_1,\dots,\nu_{N_s}+\Delta^j_{N_s})=0,\quad \Delta^j_i\in\{-1,0,1\}.
\end{equation}
The coefficients of these relations are linear functions of the scalar products of external momenta, the internal masses and the space-time dimension $d$. Since there is one IBP relation for each allowed value of the powers $\nu_i$ and each choice of $q$ and the derivative, there exists an infinite number of such relations between the infinite number of integrals in each topology.

It has been shown that by virtue of the IBP relations all integrals in a topology can be expressed as a linear combination of a \emph{finite} number \cite{Smirnov:2010hn, Lee:2013hzt} of integrals with the coefficients being rational functions of the external momenta, the masses and $d$. The choice of this finite basis of so-called \emph{master integrals} is not unique. In fact, the master integrals can be chosen to be any set of linear combinations of integrals of the form in \eq{sIntdef} that is both independent with respect to the IBP relations and suffices to express all other integrals.

The following one-loop integral illustrates the use of IBP relations and will be used in later sections of this chapter as well. The integral topology is defined by its two propagators
\begin{equation}
\label{extopology}
I(\nu_1,\nu_2)=\int\frac{\textup{d}^dl}{\textup{i}\pi^{d/2}}\frac{1}{[l^2-m^2]^{\nu_1}[(l-p)^2-m^2]^{\nu_2}},
\end{equation}
which corresponds to the Feynman diagram in \fig{fig:OneLoopBubble}.
There is no need for additional irreducible scalar products in this situation because, for $L=1$ and $N_\textup{ex}=1$, the number $N_s=2$ of possible scalar products is equal to the number of propagators. Both scalar products involving the loop momentum $l$ can thus be expressed as linear combinations of the inverse propagators:
\begin{align}
\label{SPasLCofProp1}
l^2&=P_1+m^2,\\
\label{SPasLCofProp2}
l\cdot p&=\frac{1}{2}(P_1-P_2+p^2).
\end{align}
In order to generate IBP relations, consider \eq{IBP} for $q=l$ 
\begin{align}
\label{oneloopbubbleint}
0&=\int\frac{\textup{d}^dl}{\textup{i}\pi^{d/2}}\frac{\partial}{\partial l^\mu}\left(l^\mu\frac{1}{P_1^{\nu_1}P_2^{\nu_2}}\right)\\
&=d\cdot I(\nu_1,\nu_2)+\int\frac{\textup{d}^dl}{\textup{i}\pi^{d/2}}l^\mu\frac{\partial}{\partial l^\mu}\frac{1}{P_1^{\nu_1}P_2^{\nu_2}}\\
&=d\cdot I(\nu_1,\nu_2)-\nu_1\int\frac{\textup{d}^dl}{\textup{i}\pi^{d/2}}\frac{2l^2}{P_1^{\nu_1+1}P_2^{\nu_2}}-\nu_2\int\frac{\textup{d}^dl}{\textup{i}\pi^{d/2}}\frac{2(l^2-l\cdot p)}{P_1^{\nu_1}P_2^{\nu_2+1}}.
\end{align}
The scalar products in the numerators can be rewritten in terms of the propagators by virtue of \eqs \eqref{SPasLCofProp1} and \eqref{SPasLCofProp2}
\begin{align}
\label{genIBP1}
0=\,\,&(d-2\nu_1-\nu_2)I(\nu_1,\nu_2)-2\nu_1m^2I(\nu_1+1,\mu_2)\\
&-\nu_2I(\nu_1-1,\nu_2+1)-\nu_2(2m^2-p^2)I(\nu_1,\nu_2+1)\nonumber.
\end{align}
In addition to these IBP relations, the integral topology also enjoys the symmetry 
\begin{equation}
I(\nu_1, \nu_2)=I(\nu_2, \nu_1),
\end{equation}
which corresponds to the change 
\begin{equation}
l\rightarrow -l-p
\end{equation}
of the loop momentum integration variable in \eq{extopology}. This symmetry and the IBP relations are sufficient to relate all integrals of the topology to two master integrals, which may be chosen to be 
\begin{equation}
g_1=I(1,0),\quad g_2=I(1,1).
\end{equation}
For instance, by setting $\nu_1=1$ and $\nu_2=0$ in \eq{genIBP1}, the integral $I(2,0)$ can be reduced to the master integral $I(1,0)$:
\begin{equation}
\label{IBPred1}
I(2,0)=\frac{(d-2)}{2m^2}I(1,0).
\end{equation}
Using this relation and the IBP relation obtained from \eq{genIBP1} for $\nu_1=\nu_2=1$, the integral $I(2,1)$ is reduced to the master integrals as follows:
\begin{equation}
\label{IBPred2}
I(2,1)=\frac{(d-3)}{(4m^2-p^2)}I(1,1)-\frac{(d-2)}{2m^2(4m^2-p^2)}I(1,0).
\end{equation}
The reduction relations \eq{IBPred1} and \eq{IBPred2} are sufficient to calculate the differential equation of this topology, which is demonstrated in \sec{sec:DiffEqnsFI}. In practice, the IBP reduction is used to generate such relations for all integrals occurring in the matrix element of interest.

\subsection{Lorentz invariance identities}
The Lorentz invariance of the integrals in \eq{sIntdef} implies a further set of relations~\cite{Gehrmann:1999as}, which is widely used in practice for the reduction to master integrals. In addition to that, these relations are useful for the differentiation of Feynman integrals with respect to kinematic invariants and are therefore reviewed in the following. Consider the action of an infinitesimal Lorentz transformation 
\begin{equation}
\Lambda_{~\nu}^\mu=\delta_{~\nu}^\mu+\omega_{~\nu}^\mu, \quad \omega^{\mu}_{~\nu}=-\omega^{\nu}_{~\mu}
\end{equation}
on one of the $N_\textup{ex}$ external momenta
\begin{equation}
p_j^{\mu\prime}=\Lambda_{~\nu}^\mu p_j^\nu=p_j^\mu+\omega^{\mu}_{~\nu} p_{j}^\nu.
\end{equation}
Then, Lorentz invariance implies for any scalar integral $I$ that
\begin{align}
I(p_j)&=I(p_j^\prime)\\
&=I(p_j)+\omega^{\mu}_{~\nu}\sum_{j=1}^{N_\textup{ex}}p_{j}^\nu\frac{\partial}{\partial p_j^\mu}I(p_j)
\end{align}
holds for all infinitesimal $\omega^{\mu}_{~\nu}$. Using the antisymmetry of $\omega^{\mu}_{~\nu}$, the above equation implies
\begin{equation}
\sum_{j=1}^{N_\textup{ex}}\left(p_{j\nu}\frac{\partial}{\partial p_j^\mu}-p_{j\mu}\frac{\partial}{\partial p_j^\nu}\right)I(p_j)=0,
\end{equation}
which can be turned into scalar relations by contracting with antisymmetric combinations of external momenta. For instance, for $N_\textup{ex}=2$, there is only the identity
\begin{equation}
\label{LIInvRelEx}
(p^\nu_1p_2^\mu-p_1^\mu p_2^\nu)\sum_{j=1}^2\left(p_{j\nu}\frac{\partial}{\partial p_j^\mu}-p_{j\mu}\frac{\partial}{\partial p_j^\nu}\right)I(p_j)=0.
\end{equation}
Since at most $d$ of the $N_\textup{ex}$ external momenta can be linearly independent, the number of linearly independent external momenta after using momentum conservation is given by $N_\textup{ind}=\textup{min}(d,N_\textup{ex})$. Thus, there are $N_\textup{ind}(N_\textup{ind}-1)/2$ independent Lorentz invariance relations of the above form, because this is the number of antisymmetric combinations of the independent external momenta.

It has been shown~\cite{Lee:2008tj} that the Lorentz invariance identities are not linearly independent of the IBP relations and thus not strictly necessary for the reduction to master integrals. In practice, however, they can speed up the reduction process and are therefore widely used.
\subsection{Systematic reduction strategies}

As mentioned before, the number of Feynman integrals contributing to a particular matrix element can be relatively large. It is thus desirable to automate their reduction to master integrals. One strategy to attempt an automatized reduction is to combine IBP identities with symbolic propagator powers, such as \eq{genIBP1}, into symbolic reduction rules \cite{Lee:2008tj, Lee:2012cn, Smirnov:2013dia, Lee:2013mka, Ruijl:2017cxj}, which may be interpreted as ladder operators acting on the propagator powers. Applied recursively, the reduction rules relate all integrals of a given topology to master integrals. Once the reduction rules have been found, the reduction itself is very efficient. However, a systematic way of constructing symbolic reduction rules has not yet been found, and thus implementations of this strategy have to resort to heuristic methods.

Laporta proposed~\cite{Laporta:2001dd} the completely systematic but rather brute-force strategy of considering the IBP relations for a finite range of integer values of the propagator powers. The integrals within this range are called \emph{seed integrals}. For each seed integral, \eq{IBP} generates an IBP relation for all of the $L(N_\textup{ex}+L)$ choices of $q$ and the derivative, which usually results in a large system of equations for the seed integrals. The next step is to essentially perform a Gaussian elimination to triangularize the system of equations. By defining a so-called \emph{Laporta ordering} on the integrals that reflects their complexity, the elimination can be performed such that more complex integrals are eliminated in favor of less complicated ones. Usually, this ordering is chosen to be compatible with the ordering of the sectors induced by the sector-id. If this is the case, every integral is reduced to master integrals from the same or lower sectors.

Variations of the Laporta strategy have been implemented in numerous publicly available programs \cite{Anastasiou:2004vj, Studerus:2009ye, vonManteuffel:2012np, Smirnov:2013dia, Smirnov:2014hma, Maierhoefer:2017hyi}. In practice, these calculations suffer from the fact that most of the IBP relations generated from the seed integrals are linearly dependent. In fact, in the limit of large ranges of seed integrals only one relation per $L(N_\textup{ex}+L)$ IBP relations can be linearly independent, since the number of master integrals remains fixed. 
The unnecessary computations due to linearly dependent IBP relations can be avoided by eliminating such relations with finite field techniques \cite{Kant:2013vta, vonManteuffel:2014ixa, Maierhoefer:2017hyi} prior to the Gaussian elimination step. By mapping the time-consuming computations over the field of rational functions to the modular arithmetic of finite fields, the linearly dependent relations can be eliminated very efficiently. 

Altogether the IBP reduction to master integrals can be considered as a conceptionally solved problem, but in practice, the computations are often limited by the computational resources at disposal.

\section{Differential equations of Feynman integrals}
\label{sec:DiffEqnsFI}

After the reduction to master integrals, it remains to evaluate the relatively small number of master integrals as functions of the kinematic invariants. This problem has been approached in numerous ways, but despite many advances, a general solution has not yet been found and still appears to be far out of reach. However, in recent years the method of differential equations \cite{Kotikov:1990kg, Remiddi:1997ny, Gehrmann:1999as} has been successfully applied to a large class of Feynman integrals. This development has been enabled by the observation~\cite{Henn:2013pwa} that a particular choice of the basis of master integrals drastically simplifies the solution of the corresponding differential equation. The key ideas of the differential equations approach are reviewed in this section.

\subsection{Differentiation of Feynman integrals}
\label{sec:DiffofFIs}

The master integrals are functions of $N_\textup{ex}$ external momenta and a number of internal mass scales. Lorentz invariance implies that Feynman integrals can only depend on the external momenta via kinematic invariants $X_1,\dots,X_E$, which are independent Lorentz invariant functions of the external momenta. In addition to these invariants, the integrals may also depend on internal masses $X_{E+1},\dots,X_Q$. The goal is to evaluate the master integrals as functions of all kinematic invariants $X_1,\dots,X_Q$.

The basic idea of the method of differential equations is to derive a system of differential equations for the master integrals by calculating the derivatives of all master integrals with respect to the kinematic invariants $X_1,\dots,X_Q$ and then solve it in terms of known functions. The derivative of an integral with respect to one of the internal masses $X_{E+1},\dots,X_Q$ is straightforward to perform in the representation \eq{sIntdef} since the mass dependence is explicit. By interchanging the derivative with the loop momentum integration~\cite{collins1984renormalization} and using the product rule on the propagators, the derivative results in a linear combination of integrals from the same sector with possibly raised propagator powers. 

The derivatives with respect to the kinematic invariants $X_1,\dots,X_E$ are related to the derivatives with respect to the external momenta by the chain rule
\begin{equation}
\frac{\partial}{\partial p_i^\mu}=\sum_{j=1}^E\frac{\partial X_j}{\partial p_i^\mu}\frac{\partial}{\partial X_j},\quad i=1,\dots,N_\textup{ex}.
\end{equation}
Each of these relations can be contracted with any of the $N_\textup{ind}$ independent external momenta to give the scalar relations 
\begin{equation}
\label{contractedDerivatives}
p^\mu_k\frac{\partial}{\partial p_i^\mu}=\sum_{j=1}^Ep^\mu_k\frac{\partial X_j}{\partial p_i^\mu}\frac{\partial}{\partial X_j},\quad i=1,\dots,N_\textup{ex},\quad k=1,\dots,N_\textup{ind}.
\end{equation}
The contracted derivatives on the left-hand side are related by Lorentz invariance relations of the form in \eq{LIInvRelEx}, if they are applied to Lorentz invariant scalars. Therefore, only
\begin{equation}
N_\textup{ex}N_\textup{ind}-\frac{N_\textup{ind}(N_\textup{ind}-1)}{2}
\end{equation}
of the relations in \eq{contractedDerivatives} are independent, which precisely corresponds to the number $E$ of independent Lorentz invariants that can be formed from the $N_\textup{ex}$ external momenta.

Thus, after choosing a set of $E$ independent relations from \eq{contractedDerivatives}, these can be solved for the $E$ derivatives with respect to the kinematic invariants $X_1,\dots,X_E$, which are then expressed as linear combinations of the contracted derivatives with respect to the external momenta. This allows to evaluate the derivatives with respect to the kinematic invariants by acting on the integrals with the derivatives with respect to the external momenta, which can be interchanged with the integration and evaluated directly on the integrand.

The derivative of the integrand with respect to the external momenta and the subsequent contraction with an external momentum can be written as a linear combination of inverse propagators and irreducible scalar products, which results in a linear combination of integrals with shifted propagator powers. Note that this operation can never generate new propagators with positive powers. Thus, the derivative of a scalar integral with respect to an external invariant can always be written as a linear combination of integrals from the same or lower sectors. 

As an example, consider the derivative of $g_2=I(1,1)$ with respect to the kinematic invariant $s=p^2$. There is only one relation of the form in \eq{contractedDerivatives}
\begin{equation}
\label{excontrderiv}
p^{\mu}\frac{\partial}{\partial p^{\mu}}=p^\mu\frac{\partial s}{\partial p^\mu}\frac{\partial}{\partial s}=2s\frac{\partial}{\partial s}.
\end{equation}
The derivative with respect to $p^\mu$ raises the power of $P_2$ and generates a linear combination of scalar products in the numerator upon contraction with $p^\mu$. By virtue of \eq{SPasLCofProp2}, the scalar product $l\cdot p$ is rewritten in terms of the inverse propagators, which allows to express the derivative as a linear combination of scalar integrals with shifted propagator powers
\begin{equation}
\label{derivPTOscalar}
p^\mu\frac{\partial I(1,1)}{\partial p^\mu}=\int\frac{\textup{d}^dl}{\textup{i}\pi^{d/2}}\frac{2l\cdot p-2p^2}{P_1P_2^2}=I(0,2)-I(1,1)-p^2I(1,2).
\end{equation}
Solving \eq{excontrderiv} for the derivative with respect to $s$ leads to
\begin{equation}
\label{derivSTOscalar}
\frac{\partial I(1,1)}{\partial s}=\frac{1}{2s}\bigl(I(0,2)-I(1,1)\bigr)-I(1,2).
\end{equation}
Since the master integral $g_1=I(1,0)$ does not depend on the external momentum, its derivative with respect to $s$ vanishes due to \eq{excontrderiv}
\begin{equation}
\frac{\partial I(1,0)}{\partial s}=0.
\end{equation}
The strategy described here to calculate derivatives of Feynman integrals with respect to their invariants in terms of integrals of the same topology is completely algorithmic and has for example been implemented in \cite{Studerus:2009ye, vonManteuffel:2012np}.

\subsection{Differential equations and canonical bases}
\label{sec:DEofFIs}
In the previous section, it has been shown that derivatives of Feynman integrals with respect to the kinematic invariants can be expressed as linear combinations of scalar integrals from the same or lower sectors. By applying the IBP reduction to those integrals, derivatives of scalar integrals can always be written as linear combinations of master integrals. If the differentiated integrals are master integrals, this results in a coupled first-order linear system of differential equations for the master integrals. Solving these differential equations in terms of known functions, and imposing appropriate boundary conditions then achieves the goal of evaluating the master integrals as functions of the kinematic invariants.

In the case of the previously considered example, the scalar integrals on the right-hand side of \eq{derivSTOscalar} are reduced to master integrals with \eq{IBPred1} and \eq{IBPred2}
\begin{equation}
\frac{\partial g_2}{\partial s}=\frac{(d-2)}{s(4m^2-s)}g_1-\frac{1}{2}\left(\frac{1}{s}+\frac{(d-3)}{(4m^2-s)}\right)g_2,
\end{equation}
\begin{equation}
\frac{\partial g_1}{\partial s}=0.
\end{equation}
A differential equation of this form can be derived for all of the kinematic invariants. However, the dependence of the master integrals on one of the kinematic invariants can always be reconstructed from the mass dimension $\textup{dim}(I)$ of the integral, which is easily determined by counting the powers of its propagators and irreducible scalar products. Then, by choosing any of the kinematic invariants, for instance, $X_Q$, the basis of master integrals can be normalized such that all integrals have mass dimension zero
\begin{equation}
f_i=(X_Q)^{-\textup{dim}(g_i)/\textup{dim}(X_Q)}g_i.
\end{equation}
Due to their trivial mass dimension, the integrals $f_i$ must be functions of $M=Q-1$ dimensionless functions of the kinematic invariants. These can, for example, be chosen to be the dimensionless ratios
\begin{equation}
x_i=\frac{X_i}{X_Q^{\textup{dim}(X_i)/\textup{dim}(X_Q)}},\quad i=1,\dots, M.
\end{equation}
Using dimensionless master integrals is straightforward in practice and reduces the number of variables by one. 

The integrals in the above example depend on the two dimensionful invariants $s$ and $m$. By choosing $m$ as the variable to be factored out, a dimensionless basis of master integrals is obtained by
\begin{equation}
\label{masterchangefg}
f_1=(m)^{-(d-2)}e^{\epsilon\gamma_E}\epsilon g_1,\quad f_2=(m)^{-(d-4)}e^{\epsilon\gamma_E}\epsilon g_2,
\end{equation}
where $\epsilon$ denotes the dimensional regulator as introduced in \eq{reldeps}. The integral vector has been multiplied with the overall factor $e^{\epsilon\gamma_E}$ because it conveniently removes terms involving the Euler--Mascheroni constant $\gamma_E$ from the expansion of $\vec{f}$. The additional factor of $\epsilon$ ensures that the $\epsilon$\nobreakdash-expansion of the $f_i$ starts at a non-negative order. Due to their trivial mass dimension, the integrals $\vec{f}$ must be functions of the dimensionless ratio
\begin{equation}
x=\frac{s}{m^2}.
\end{equation}
Using the chain rule and \eq{derivSTOscalar} yields for the derivatives with respect to $x$
\begin{align}
\frac{\partial f_2}{\partial x}=m^{6-d}\frac{\partial g_2}{\partial s}&=m^{6-d}\left(\frac{(d-2)}{s(4m^2-s)}g_1-\frac{1}{2}\left(\frac{1}{s}+\frac{(d-3)}{(4m^2-s)}\right)g_2\right)\\
&=\frac{(d-2)}{x(4-x)}f_1-\frac{1}{2}\left(\frac{1}{x}+\frac{(d-3)}{(4-x)}\right)f_2
\end{align}
and 
\begin{equation}
\frac{\partial f_1}{\partial x}=m^{4-d}\frac{\partial g_1}{\partial s}=0.
\end{equation}
Upon replacing the dimension with $d=4-2\epsilon$ and considering the full vector of master integrals, the derivative can be written as
\begin{equation}
\label{xcoordinateBubbleDEQ}
\frac{\partial \vec{f}}{\partial x}=
\left(\begin{array}{cc}
0 & 0 \\
\frac{(2-2\epsilon)}{x(4-x)} & -\frac{1}{2}\left(\frac{1}{x}+\frac{(1-2\epsilon)}{(4-x)}\right)
\end{array}\right)\vec{f}.
\end{equation}
For a general topology with $m$ master integrals depending on $M$ dimensionless invariants, the derivative can be taken with respect to all dimensionless invariants, which results in a coupled system of linear differential equations for the master integrals
\begin{equation}
\label{DEQNonDifferentialForm}
\partial_i\vec{f}(\varset)=a_{i}(\varset)\vec{f}(\varset),\quad i=1,\dots, M,
\end{equation}
with the $a_i(\varset)$ being $m\times m$ matrices of rational functions in the kinematic invariants $\invariants$ and $\epsilon$. The fact that the matrices $a_i(\varset)$ are rational functions of the kinematic invariants and $\epsilon$ is evident from the structure of the integration-by-parts relations in \eq{IBP}. In \sec{sec:DiffofFIs} it was argued that the derivative of a Feynman integral can be represented as a linear combination of integrals from the same or lower sectors. If the Laporta ordering is compatible with the sector structure, each of these integrals has a representation in terms of master integrals from the same or lower sectors. Altogether this leads to a lower-left block-triangular form of the $a_i(\varset)$ matrices if the vector of master integrals $\vec{f}$ is ordered according to the sector-id. It is convenient to use the more compact differential notation for the system of differential equations in \eq{DEQNonDifferentialForm}\footnote{The differential equation \eq{DEQDifferentialForm} is completely determined by the differential form $a(\varset)$. Therefore, $a(\varset)$ is also frequently referred to as the \emph{differential equation} in the following.}
\begin{equation}
\label{DEQDifferentialForm}
\textup{d}\vec{f}(\varset)=a(\varset)\vec{f}(\varset),
\end{equation}
with 
\begin{equation}
a(\varset)=\sum_{i=1}^Ma_i(\varset)\textup{d}x_i.
\end{equation}
The choice of the basis of master integrals is not unique. Changing the basis of master integrals to a new basis $\vec{f}^\prime$ that is related to the original basis by an invertible transformation~$T$
\begin{equation}
\vec{f}=T(\varset)\vec{f}^\prime,
\end{equation}
as suggested in~\cite{Henn:2013pwa}, leads to a transformation law for $a(\varset)$:
\begin{equation}
\label{DEQTrafo}
a^\prime = T^{-1}aT-T^{-1}\textup{d}T.
\end{equation}
In the following, some notation and terminology related to particular forms of $a(\varset)$ is introduced. The differential equation is said to be in \emph{dlog-form} if the differential form $a(\varset)$ can be written as
\begin{equation}
\label{defdlog1}
a(\varset)=\textup{d}A(\varset),
\end{equation}
with
\begin{equation}
\label{defdlog2}
A(\varset)=\sum_{l=1}^NA_l(\epsilon)\log(L_l(\invariants)).
\end{equation}
Here $L_l(\invariants)$ denotes functions of the invariants, and the $A_l$ are $m\times m$ matrices, which solely depend on $\epsilon$. The set of functions
\begin{equation}
\mathcal{A}=\{L_1(\invariants),\dots,L_N(\invariants)\}
\end{equation}
is commonly referred to as the \textit{alphabet} of the differential equation. The individual $L_l(\invariants)$ are called \textit{letters} of the differential equation. In~\cite{Henn:2013pwa} it was observed that with a suitable change of the basis of master integrals it is often possible to arrive at a dlog-form in which the dependence on $\epsilon$ factorizes
\begin{equation}
\label{EpsForm}
a(\varset)=\epsilon\,\textup{d}\tilde{A}(\invariants)=\epsilon\sum_{l=1}^N\tilde{A}_l\textup{d}\log(L_l(\invariants)),
\end{equation}
with $\tilde{A}_l$ being constant $m\times m$ matrices. In this form, which is called \textit{canonical form} or \textit{$\epsilon$\nobreakdash-form}, the integration of the differential equation simplifies significantly as will be shown in \sec{sec:IntofCanForm}. A basis of master integrals for which the differential equation assumes a canonical form is called a \emph{canonical basis}.

Note that in the derivation of the differential equation \eq{DEQDifferentialForm} described in the previous sections, the master integrals and the invariants $\invariants$ can always be chosen such that the resulting differential form $a(\varset)$ is rational in the invariants and the regulator. For rational transformations to the canonical form, it follows that the resulting canonical form is rational as well and thus the letters are polynomials in the invariants. However, there are differential equations for which the transformation to a canonical form necessarily contains roots of polynomials in the invariants, which may lead to letters containing these roots.

The differential equation \eq{xcoordinateBubbleDEQ} of the example considered above can also be transformed into canonical form. To this end, it is advantageous to first change the coordinates to
\begin{equation}
\label{varchangexy}
x=-\frac{(1-y)^2}{y},
\end{equation}
because this allows for a rational transformation of the differential equation to a canonical form to exist. With respect to the new coordinate $y$, the differential equation \eq{xcoordinateBubbleDEQ} reads in differential form
\begin{equation}
\label{ycoordinateBubbleDEQ}
\textup{d}\vec{f}=
\left(\begin{array}{cc}
0 & 0 \\
\frac{1-\epsilon}{-1+y}+\frac{-1+\epsilon}{1+y} & \frac{-1}{-1+y}+\frac{\epsilon}{y}+\frac{1-2\epsilon}{1+y}
\end{array}\right)\textup{d}y\vec{f}.
\end{equation}
The transformation given by
\begin{equation}
\label{canonicalTrafoex}
T=\left(\begin{array}{cc}
\frac{1}{1-\epsilon} & 0 \\
\frac{1}{1-2\epsilon} & \frac{1+y}{2(1-2\epsilon)(-1+y)}
\end{array}\right)
\end{equation}
transforms the differential equation into the canonical form
\begin{align}
\textup{d}\vec{f}^\prime&=
\epsilon\left(\begin{array}{cc}
0 & 0 \\
-\frac{2}{y} & \frac{1}{y}-\frac{2}{1+y}
\end{array}\right)\textup{d}y\vec{f}^\prime\\
&=
\epsilon\left[\left(\begin{array}{cc}
0 & 0 \\
-2 & 1
\end{array}\right)\textup{d}\log(y)+\left(\begin{array}{cc}
0 & 0 \\
0 & -2
\end{array}\right)\textup{d}\log(1+y)\right]\vec{f}^\prime.\label{examplesDEQ}
\end{align}
The main part of this thesis will be devoted to the problem of finding such a transformation for a given differential equation. 

\section{Solving differential equations in canonical form}
\label{sec:IntofCanForm}
The main part of this thesis is concerned with the problem of finding a transformation to a canonical basis for a given differential equation of Feynman integrals. In this section, it will be shown that once a canonical form is found, the integration of the corresponding differential equation is essentially reduced to a simple combinatorial procedure (c.f. e.g., \cite{Henn:2013woa, Henn:2014lfa, Caron-Huot:2014lda, Gehrmann:2014bfa, Caola:2014lpa}). 

\subsection{Integrating differential equations in canonical form}
Consider a differential equation of Feynman integrals in canonical form
\begin{equation}
\label{canonicalDEQcopy}
\textup{d}\vec{f}^\prime(\varset)=\epsilon\,\textup{d}\tilde{A}(\invariants)\vec{f}^\prime(\varset),
\end{equation}
with
\begin{equation}
\textup{d}\tilde{A}(\invariants)=\sum_{l=1}^N\tilde{A}_l\textup{d}\log(L_l(\invariants)).
\end{equation}
The differential form $\textup{d}\tilde{A}$ is singular on the zero-sets of the letters
\begin{equation}
V_l\defeq \left\{\invariants\in\mathcal{M}\,\,\, | \,\,\, L_l(\invariants)=0\right\},
\end{equation}
where $\mathcal{M}=\mathbb{C}^M$ denotes the manifold of the $M$ invariants. Therefore, $\textup{d}\tilde{A}$ is well-defined on 
\begin{equation}
\overline{\mathcal{M}}\defeq \mathcal{M}\setminus\left(\bigcup_{l=1}^NV_l\right),
\end{equation}
which is the natural domain for the following considerations about the solutions of \eq{canonicalDEQcopy}. Upon normalizing the vector of master integrals with appropriate powers of $\epsilon$, it can be assumed to have an expansion starting at the constant order
\begin{equation}
\label{vecfexpansion}
\vec{f}^\prime(\varset)=\sum_{n=0}^\infty \epsilon^n \vec{f}^{\prime(n)}(\invariants).
\end{equation}
Due to the factorization of the $\epsilon$ dependence on the right-hand side of the differential equation \eq{canonicalDEQcopy}, its expansion is of the simple form
\begin{equation}
\label{expandedfDEQ}
\textup{d}\vec{f}^{\prime(n)}(\invariants)=\textup{d}\tilde{A}(\invariants)\vec{f}^{\prime(n-1)}(\invariants).
\end{equation}
A recursive relation for the coefficients $\vec{f}^{\prime(n)}$ is obtained by integrating \eq{expandedfDEQ} 
\begin{equation}
\label{recursiveIntSoln}
\vec{f}^{\prime(n)}(\invariants)=\int_\gamma\textup{d}\tilde{A}(\invariants)\vec{f}^{\prime(n-1)}(\invariants)+\vec{f}^{\prime(n)}(\{x_{j0}\}),
\end{equation}
where the integration path is a smooth path $\gamma:[0,1]\rightarrow \overline{\mathcal{M}}$ with $\gamma(0)=\{x_{j0}\}$ and $\gamma(1)=\invariants$. The integration does only depend on the homotopy class of $\gamma$ and is otherwise path independent if the differential form $\textup{d}\vec{f}^\prime$ is closed~\cite{lee2003introduction}. For a generic basis of master integrals $\vec{f}$, demanding $\textup{d}\vec{f}=a\vec{f}=a\wedge\vec{f}$ to be closed implies 
\begin{align}
0=\textup{d}\textup{d}\vec{f}&=\textup{d}a\wedge\vec{f}-a\wedge\textup{d}\vec{f}\\
&=\textup{d}a\wedge\vec{f}-a\wedge a\wedge\vec{f}=\left(\textup{d}a-a\wedge a\right)\wedge\vec{f}.
\end{align}
Since the master integrals are assumed to be linearly independent over the field of rational functions, the above equation implies the integrability condition
\begin{equation}
\label{diffIntegralityCondition}
\textup{d}a-a\wedge a=0.
\end{equation}
In practice, this condition may be used as consistency check of the differential equation. Employing \eq{recursiveIntSoln}, the solution can be constructed by iterated integration of the integration kernel $\textup{d}\tilde{A}$. According to \eqs \eqref{vecfexpansion} and \eqref{recursiveIntSoln}, the iteration starts at order $n=0$ with constant coefficients
\begin{equation}
\vec{f}^{\prime(0)}(\{x_{j}\})=\vec{f}^{\prime(0)}(\{x_{j0}\}).
\end{equation}
For all practical purposes, it is sufficient to stop this iteration at a finite order in the $\epsilon$\nobreakdash-expansion.

An alternative but equivalent way to represent the solution of \eq{canonicalDEQcopy} to all orders is given by the formal expression
\begin{equation}
\label{formalDEsoln}
\vec{f}^\prime(\varset)=\mathbb{P}\exp\left(\epsilon\int_\gamma\textup{d}\tilde{A}\right)\vec{f}^\prime(\epsilon,\{x_{j0}\}),
\end{equation}
where the operator $\mathbb{P}$ is a path-ordering operator, which is defined below. First, consider the pull-back of $\textup{d}\tilde{A}$ by $\gamma$ to a differential form $\gamma^*(\textup{d}\tilde{A})$ on $[0,1]$. After choosing a coordinate $s$ on $[0,1]$, the pull-back may be written as
\begin{equation}
\gamma^*(\textup{d}\tilde{A})=\tilde{\alpha}(s)\textup{d}s
\end{equation}
and thus
\begin{equation}
\int_\gamma\textup{d}\tilde{A}=\int_{[0,1]}\tilde{\alpha}(s)\textup{d}s.
\end{equation}
With this notation, the exponential and the path ordering in \eq{formalDEsoln} can be defined by
\begin{align}
\vec{f}^\prime(\epsilon,\{x\})&=\sum_{k=0}^\infty\frac{\epsilon^k}{k!}\int_{[0,1]^k}\mathbb{P}\left[\tilde{\alpha}(s_k)\cdots\tilde{\alpha}(s_1)\right]\textup{d}s_1\cdots\textup{d}s_k\vec{f}^\prime(\epsilon,\{x_{j0}\})\\
&=\sum_{k=0}^\infty\epsilon^k\int_{0\leq s_1\leq\dots\leq s_k\leq 1}\tilde{\alpha}(s_k)\cdots\tilde{\alpha}(s_1)\textup{d}s_1\cdots\textup{d}s_k\vec{f}^\prime(\epsilon,\{x_{j0}\}),\label{IteratedDEQSoln}
\end{align}
where the integral for $k=0$ is understood to be the $m\times m$ unit matrix. Integrals of this type are called \emph{Chen iterated integrals} and have been studied in~\cite{Chen:1977oja}. See also~\cite{brown_2013} for a pedagogical review.
The recursive integration formula \eq{recursiveIntSoln} can be recovered by first inserting the expansion \eq{vecfexpansion} in \eq{IteratedDEQSoln}
\begin{align}
\vec{f}^{\prime(n)}(\invariants)=\,&\sum_{k=0}^n \int\tilde{\alpha}(s_k)\cdots\tilde{\alpha}(s_1)\textup{d}s_1\cdots\textup{d}s_k \vec{f}^{\prime(n-k)}(\{x_{j0}\})\\
=\,&\sum_{k=0}^{n-1}\int\tilde{\alpha}(s_{k+1})\cdots\tilde{\alpha}(s_1)\textup{d}s_1\cdots\textup{d}s_{k+1} \vec{f}^{\prime(n-1-k)}(\{x_{j0}\})\\\,&+\vec{f}^{\prime(n)}(\{x_{j0}\}),\nonumber
\end{align}
where the integration domains have been omitted for brevity. Separating the outer integration then leads to \eq{recursiveIntSoln}
\begin{align}
\vec{f}^{\prime(n)}(\invariants)&=\int_{[0,1]}\tilde{\alpha}(s_{k+1})\vec{f}^{\prime(n-1)}(s_{k+1})\textup{d}s_{k+1}+\vec{f}^{\prime(n)}(\{x_{j0}\})\\
&=\int_\gamma\textup{d}\tilde{A}\vec{f}^{\prime(n-1)}+\vec{f}^{\prime(n)}(\{x_{j0}\}),
\end{align}
which proves that \eq{formalDEsoln} is equivalent to \eq{recursiveIntSoln} and therefore satisfies the differential equation as well.

\subsection{Multiple polylogarithms}
\label{sec:mpls}
The solution of differential equations in canonical form has been shown to be constructible by the iterated integrations in  \eq{IteratedDEQSoln} in the previous section. This representation is convenient from a theoretical point of view, as it makes the homotopy invariance of the result manifest. However, for applications in phenomenology, it is necessary to have a representation of the result that is suitable for fast and stable numerical evaluation. Such a representation in terms of classes of functions for which numerical routines exist can usually be obtained by choosing particular integration paths. The rather general class of \emph{multiple polylogarithms} \cite{Goncharov:1998kja, borwein2001special} is of particular interest here, since many Feynman integrals can be represented in terms of these functions. Moreover, there is a library available allowing for the numerical evaluation of general multiple polylogarithms with arbitrary precision~\cite{Vollinga:2004sn}. For various special cases, there are other routines available as well \cite{Kolbig1970, Gehrmann:2001pz, Gehrmann:2001jv, Gehrmann:2002zr}. The goal of this section is to introduce multiple polylogarithms and elucidate the relation between their series and integral representation, which is used in~\cite{Vollinga:2004sn} for the numerical evaluation of multiple polylogarithms.

Multiple polylogarithms may be defined by the following representation in terms of nested sums\footnote{Several conventions regarding the ordering of the arguments and indices are used in the literature. The exposition here adopts the conventions in~\cite{Vollinga:2004sn}.}
\begin{equation}
\label{LIseriesrep}
\textup{Li}_{m_1,\dots,m_k}(x_1,\dots,x_k)=\sum_{i_1>i_2>\cdots> i_k>0}\frac{x_1^{i_1}\cdots x_k^{i_k}}{i_1^{m_1}\cdots i_k^{m_k}}
\end{equation}
which converges for $|x_i|<1$. Special cases \cite{Nielsen, lewin1981polylogarithms, borwein2001special} of multiple polylogarithms include
\begin{description}
\item[Classical polylogarithms:]
\begin{equation}
\textup{Li}_n(x)=\sum_{i=1}^\infty\frac{x^i}{i^n},
\end{equation}
\item[Multiple zeta values:] 
\begin{equation}
\zeta_{m_1,\dots,m_k}=\sum_{i_1>i_2>\cdots>i_k>0}\frac{1}{i_1^{m_1}\cdots i_k^{m_k}},
\end{equation}
\item[Nielsen's polylogarithms:] 
\begin{equation}
S_{n,p}(x)=\textup{Li}_{n+1,1,\dots,1}(x,\underbrace{1,\dots,1}_{p-1}).
\end{equation}
\end{description}
The representation of multiple polylogarithms in terms of nested sums in \eq{LIseriesrep} is very useful to evaluate Feynman integrals by applying symbolic summation techniques \cite{Moch:2001zr, Weinzierl:2002hv, Moch:2005uc, Blumlein:2010zv, Ablinger:2015tua}. However, for the integration of differential equations of the form in \eq{canonicalDEQcopy} the integral representation of multiple polylogarithms is more convenient. The integral representation of multiple polylogarithms, introduced by Goncharov in~\cite{Goncharov:1998kja}, is recursively defined by
\begin{equation}
\label{GdefItInt}
G(z_1, \dots, z_k; y)=\int_0^y\frac{\textup{d}t}{t-z_1}G(z_2, \dots, z_k;t)
\end{equation}
for $(z_1,\dots,z_k)\neq \vec{0}_k$. The variables $z_1,\dots,z_k$ are called \emph{indices} and $y$ the \emph{argument} of the multiple polylogarithms. The empty index set, $G(;y)$ is defined as 
\begin{equation}
G(;y)=1.
\end{equation}
For the all-zero index set, the definition reads
\begin{equation}
\label{Gdefallzero}
G(\underbrace{0,\dots, 0}_k ;y)=\frac{1}{k!}\log(y)^k.
\end{equation}
In order to make contact with the series representation in \eq{LIseriesrep}, it is convenient to introduce the following shorthand notation
\begin{equation}
\label{GdefTrailingZeros}
G_{m_1, \dots, m_k}(z_1, \dots, z_k; y)=G(\underbrace{0, \dots, 0}_{m_1-1}, z_1, \dots, z_{k-1}, \underbrace{0, \dots, 0}_{m_k-1}, z_k; y).
\end{equation}
The integral representation is related to the series representation inside its radius of convergence by
\begin{equation}
\label{relationLiTOG}
\textup{Li}_{m_1,\dots,m_k}(x_1,\dots,x_k)=(-1)^k G_{m_1, \dots, m_k}\left(\frac{1}{x_1},\frac{1}{x_1 x_2},\dots,\frac{1}{x_1\cdots x_k}; 1\right).
\end{equation}
This relation can be proven by rewriting the integrands of the integral representation in terms of a geometric series. For instance, in the case of the dilogarithm, the following relation has to be proven
\begin{equation}
\label{dilogRel}
\textup{Li}_2(x)=-G_2\left(\frac{1}{x};1\right).
\end{equation}
According to the definitions \eq{GdefItInt} and \eq{GdefTrailingZeros}, the right-hand side is given by the iterated integral
\begin{align}
-G_2\left(\frac{1}{x};1\right)&=-\int_0^1\frac{\textup{d}t}{t}\int_0^t\frac{\textup{d}t^\prime}{t^\prime-\frac{1}{x}}\\
&=\int_0^1\frac{\textup{d}t}{t}\int_0^{xt}\frac{\textup{d}\tilde{t}}{1-\tilde{t}},\quad\quad \tilde{t}=t^\prime x.
\end{align}
The integrand of the inner integration may be rewritten as a geometric series
\begin{equation}
\frac{1}{1-\tilde{t}}=\sum_{n\geq 0}\tilde{t}^n,
\end{equation}
which converges for all $|\tilde{t}|=|t|\cdot|x|<1$. On the domain of the outer integration $|t|<1$ holds and therefore the series converges for $|x|<1$. Interchanging the summation with the integrations then proves \eq{dilogRel}
\begin{align}
-G_2\left(\frac{1}{x};1\right)&=\int_0^1\frac{\textup{d}t}{t}\int_0^{xt}\textup{d}\tilde{t}\sum_{n\geq 0}\tilde{t}^n=\sum_{n>0}\frac{x^n}{n}\int_0^1\textup{d}t\,\, t^{n-1}\\
&=\sum_{n>0}\frac{x^n}{n^2}=\textup{Li}_2(x).
\end{align}
The proof of the general statement \eq{relationLiTOG} proceeds along the same lines and establishes the equality of the integral representation and the series representation within its radius of convergence. Both the integral and the series representation lead to large classes of functional relations among multiple polylogarithms. The study of these functional relations is a rich subject on its own, but since it is not directly relevant for the later chapters of this thesis, the interested reader is referred to the vast literature on the subject \cite{lewin1981polylogarithms, Vollinga:2004sn, Brown:2009qja, brown_2013, Duhr:2014woa}.

\subsection{Solution in terms of multiple polylogarithms}
\label{sec:solitopoly}
The iterated integrals occurring in the general solution \eq{recursiveIntSoln} of differential equations in canonical form can often be cast in the form of the integral representation of multiple polylogarithms in \eq{GdefItInt} upon choosing an appropriate integration path. Due to the homotopy invariance of \eq{recursiveIntSoln}, different but homotopy equivalent paths will generally produce different but equivalent representations of the result. Typically, the choice of piecewise linear integration paths recovers the definition of multiple polylogarithms in \eq{GdefItInt}, but this obviously depends on the form of the letters present in $\textup{d}\tilde{A}$.

In the following, the integration procedure is illustrated for the one-loop bubble integral considered in previous sections. The corresponding differential equation in canonical form in \eq{examplesDEQ} is a linear combination of the differential forms
\begin{equation}
\label{exOneforms}
\omega_0=\frac{\textup{d}y}{y},\quad \omega_{-1}=\frac{\textup{d}y}{1+y}.
\end{equation}
Let the integration path $\gamma$ be given by the linear path along the real axis from some $y_0>0$ to $y>y_0$. Since the resulting integrations will appear repeatedly, the computation can be streamlined by first examining all possible cases. Integrations of $\omega_0$ against $G(0,\dots,0;y)$ lead to
\begin{equation}
\int_\gamma\frac{\textup{d}y}{y}G(\underbrace{0,\dots, 0}_k ;y)=\frac{\log^{k+1}(y)}{(k+1)!}-\frac{\log^{k+1}(y_0)}{(k+1)!}=G(\underbrace{0,\dots, 0}_{k+1}; y)+\mathrm{const.},
\end{equation}
for all $k\geq 0$ by virtue of the definition \eq{Gdefallzero} and partial integration. In all other cases, it is useful to split the integration path into a path $\gamma_1$ from $y_0$ to $0$ and a path $\gamma_2$ from $0$ to $y$, where it is demanded that the concatenation of the segments $\gamma_1\star\gamma_2$ is homotopy equivalent to the original path $\gamma$. This allows to split the integration as follows
\begin{align}
\int_\gamma\frac{\textup{d}y}{y}G(z_1,\dots,z_k ;y)&=\int_0^y\frac{\textup{d}t}{t}G(z_1,\dots, z_k ;t)+\mathrm{const.}\\
&=G(0,z_1,\dots,z_k;y)+\mathrm{const.},
\end{align} 
where the constant corresponds to the integration along $\gamma_1$, and it is assumed that at least one index is non-zero:
\begin{equation}
(z_1,\dots,z_k)\neq\vec{0}_k.
\end{equation}
Similarly, the integration of $\omega_{-1}$ yields 
\begin{align}
\int_\gamma\frac{\textup{d}y}{1+y}G(z_1,\dots,z_k ;y)&=\int_0^y\frac{\textup{d}t}{t-(-1)}G(z_1,\dots, z_k ;t)+\mathrm{const.}\\
&=G(-1,z_1,\dots,z_k;y)+\mathrm{const.},
\end{align} 
but in this case, the restriction $(z_1,\dots,z_k)\neq\vec{0}_k$ is not necessary since the integration of $\omega_{-1}$ leads to a non-zero index. Altogether, this shows that the integration of any polylogarithm against one of the forms in \eq{exOneforms} amounts to prepending the corresponding index to the list of indices of the polylogarithm and adding an unknown integration constant. Applying this strategy to the integration on the right-hand side of \eq{recursiveIntSoln}, generates constants, which are combined with $\vec{f}^{\prime(n)}(y_0)$ to an unknown constant denoted by $\vec{c}^{(n)}$. By a slight abuse of notation, the vector $\vec{c}^{(n)}$ will also be used to denote the constants before performing the integration:
\begin{equation}
\label{iteratesolConst}
\vec{f}^{\prime(n)}(\{y\})=\int_\gamma\textup{d}\tilde{A}(\{y\})\vec{f}^{\prime(n-1)}(y)+\vec{c}^{(n)}.
\end{equation}
With this preparation, the computation of the one-loop bubble integral becomes entirely combinatoric in nature. Starting at order $n=0$ of \eq{iteratesolConst} and using that the master integrals $\vec{f}^\prime$ are normalized such that their expansion starts at the order $\epsilon^0$, leads to
\begin{equation}
\vec{f}^{\prime(0)}(y)=\vec{c}^{(0)}.
\end{equation}
The constants $\vec{c}^{(0)}$ have to be determined by boundary conditions. Inserting this result in the next order $n=1$ yields
\begin{align}
\vec{f}^{\prime(1)}(y)&=\int_\gamma\left(\begin{array}{cc}
0 & 0 \\
-2 & 1
\end{array}\right)\vec{c}^{(0)}\frac{\textup{d}y}{y}+\int_\gamma\left(\begin{array}{cc}
0 & 0 \\
0 & -2
\end{array}\right)\vec{c}^{(0)}\frac{\textup{d}y}{1+y}+\vec{c}^{(1)}\\
&=\left(\begin{array}{c}
c^{(1)}_1 \\
(-2c^{(0)}_1+c^{(0)}_2)G(0;y)-2c^{(0)}_2G(-1;y)+c^{(1)}_2
\end{array}\right).
\end{align}
It is beneficial to determine the integration constants before proceeding with the next order to keep the expressions compact. The integral $f^\prime_1$ is given by the constants~$c_1$
\begin{equation}
f_1^{\prime(n)}(y)=c^{(n)}_1,\quad n\geq 0,
\end{equation}
because its derivative with respect to $y$ is zero, which is reflected in the vanishing first row of $\textup{d}\tilde{A}$. In this case, the integral is entirely determined by the boundary conditions, which amounts to solving the integral with other methods. Typically, this is only necessary for a small number of relatively simple integrals. The integral $g_1$ is calculated in \app{App:Tadpole} by other means and can be related to $f^\prime_1$ by \eq{masterchangefg} and the transformation in \eq{canonicalTrafoex} 
\begin{equation}
\label{f1generalcondition}
f^\prime_1=-\epsilon(1-\epsilon) e^{\epsilon\gamma_E}\Gamma(\epsilon-1),
\end{equation}
which fixes all constants $c_1^{(n)}$. In particular, the lowest orders of $f^\prime_1$ evaluate to
\begin{equation}
\label{c1conditions}
c_1^{(0)}=1,\quad c_1^{(1)}=0, \quad c_1^{(2)}=\frac{1}{2}\zeta_2.
\end{equation}
The remaining integration constants can be fixed by exploiting the regularity of the integral $g_2=I(1,1)$ for $s=0$, which implies the regularity of $f_2$ at $y=1$ via \eq{masterchangefg} and \eq{varchangexy}. Using the inverse of the transformation $T$ from \eq{canonicalTrafoex}, $f_2^\prime$ can be expressed in terms of $f_1$ and $f_2$
\begin{equation}
f^\prime_2=\frac{2(-1+\epsilon)(-1+y)}{1+y}f_1+\frac{2(1-2\epsilon)(-1+y)}{1+y}f_2.
\end{equation}
Since $f_1$ is constant and $f_2$ is regular at $y=1$, this relation implies 
\begin{equation}
\label{c2generalcondition}
f^\prime_2(1)=0,
\end{equation}
which fixes all remaining boundary conditions. At order $n=0$ this condition implies $c_2^{(0)}=0$. Together with \eq{c1conditions} this leads to
\begin{equation}
\label{f2order1}
f_2^{\prime(1)}(y)=-2G(0; y)+c_2^{(1)}.
\end{equation}
The constant $c_2^{(1)}$ is fixed by evaluating \eq{f2order1} at $y=1$ and applying \eq{c2generalcondition}
\begin{equation}
0=f_2^{\prime(1)}(1)=c_2^{(1)},
\end{equation}
where $G(0;1)=\log(1)=0$ was used. Since all orders of $f^\prime_1$ are fixed by \eq{f1generalcondition}, only $f_2^\prime$ needs to be considered for the next order in the iterated integration
\begin{equation}
f_2^{\prime(2)}=\int_\gamma\frac{\textup{d}y}{y}(-2c_1^{(1)}+f_2^{\prime(1)})-2\int_\gamma\frac{\textup{d}y}{1+y}f_2^{\prime(1)}+c_2^{(2)}.
\end{equation}
Inserting $c_1^{(1)}=0$ and $f_2^{(1)}=-2G(0;y)$ yields
\begin{equation}
f_2^{\prime(2)}=-2\int_\gamma\frac{\textup{d}y}{y}G(0;y)+4\int_\gamma\frac{\textup{d}y}{1+y}G(0;y)+c_2^{(2)},
\end{equation}
which integrates to
\begin{equation}
f_2^{\prime(2)}=-2G(0,0;y)+4G(-1,0;y)+c_2^{(2)}.
\end{equation}
Again, the integration constant is fixed by \eq{c2generalcondition}, which in this case leads to the following evaluations of polylogarithms
\begin{equation}
G(0,0;1)=\frac{\log(1)^2}{2}=0
\end{equation}
and
\begin{equation}
G(-1,0;1)=-\frac{1}{2}\zeta_2,
\end{equation}
where the former directly follows from \eq{Gdefallzero} and the latter may be verified by using relations among polylogarithms \cite{lewin1981polylogarithms, Vollinga:2004sn, Brown:2009qja, brown_2013, Duhr:2014woa}. Thus, the boundary condition yields
\begin{equation}
c_2^{(2)}=2\zeta_2.
\end{equation}
In summary, the first orders of the master integral $f_2^\prime$ are given by
\begin{equation}
f_2^{\prime(0)}=0,\quad f_2^{\prime(1)}=-2G(0;y),\\
\end{equation}
\begin{equation}
f_2^{\prime(2)}=-2G(0,0;y)+4G(-1,0;y)+2\zeta_2.
\end{equation}
The recursive integration in this example illustrates the combinatorial nature of the integration --- essentially integrating amounts to adding a new index to the list of indices of the Goncharov polylogarithms of the previous order. This also shows that the class of functions is fixed to all orders in the $\epsilon$\nobreakdash-expansion, since the set of letters, or equivalently, the set of indices they correspond to is fixed. The constant matrices $\tilde{A}_l$ of the canonical form encode the prefactors of the resulting polylogarithms and thereby the distribution of the indices among the master integrals. The integration procedure illustrated in this section can be generalized to cases with several variables and is well suited for the implementation in a recursive routine.

\subsection{Determination of boundary conditions}

For differential equations of Feynman integrals in canonical form, the problem of solving them is essentially reduced to the determination of the boundary conditions, as was illustrated in \sec{sec:solitopoly}. The information needed to fix the boundary conditions corresponds to calculating the value of the vector of master integrals at one fixed kinematic point using, for instance, a Mellin--Barnes representation of the integrals \cite{Smirnov:1999gc, Tausk:1999vh}. While this is simpler than the original problem of calculating the integral vector with full dependence on the kinematics, it is not the way most calculations actually proceed. In practice, researchers often combine several approaches to determine all boundary conditions (cf. e.g., \cite{Henn:2013fah, Henn:2013woa, Henn:2013nsa, Argeri:2014qva, Henn:2014lfa, Caron-Huot:2014lda, Gehrmann:2014bfa, Caola:2014lpa}). Usually, the evaluation of some of the master integrals is elementary to obtain in terms of $\Gamma$\nobreakdash-functions, such as the integral $f_1$ in the example considered in the previous section. For the other integrals, the boundary conditions can often be inferred by imposing their regularity in certain kinematic points. In addition to these methods, the expansion of master integrals in certain kinematic limits with the expansion by regions approach \cite{Beneke:1997zp, Smirnov:1998vk, Smirnov:1999bza, Smirnov:2002pj, Pak:2010pt, Jantzen:2012mw} may be used to generate additional boundary conditions.
       \chapter{Algorithm}
\label{chap:algorithm}
In the previous chapter it has been illustrated that the evaluation of Feynman integrals can be drastically simplified by using a canonical basis. In recent years this approach has been successfully applied to many phenomenologically relevant integral topologies \cite{Henn:2013pwa, Henn:2013fah, Henn:2013woa, Henn:2013nsa, Argeri:2014qva, Henn:2014lfa, Caron-Huot:2014lda, Gehrmann:2014bfa, Caola:2014lpa, Li:2014bfa, Hoschele:2014qsa, DiVita:2014pza, vonManteuffel:2014mva, Grozin:2014hna, Bell:2014zya, Huber:2015bva, Gehrmann:2015ora, Gehrmann:2015dua, Bonciani:2015eua, Anzai:2015wma, Grozin:2015kna, Gehrmann:2015bfy, Gituliar:2015iyq, Lee:2016htz, Henn:2016men, Bonciani:2016ypc, Eden:2016dir, Lee:2016lvq, Bonciani:2016qxi, Bonetti:2016brm, Henn:2016kjz, Lee:2016ixa, DiVita:2017xlr, Boels:2017skl, Lee:2017mip}, demonstrating the broad scope of this approach.

The methods used to determine canonical bases can be broadly divided in those which solely operate on the differential equation and those using in addition some other representation of the master integrals. The latter methods usually proceed by identifying candidate integrals with certain properties that have been observed to lead to a canonical form of their differential equation. Most notably, it has been advocated \cite{Henn:2013pwa, Henn:2014qga} that integrals with constant leading singularities~\cite{Cachazo:2008vp} and those admitting a certain dlog-representation of their integrand lead to a canonical form. It has also been argued that suitable integrals can be identified by using a parametric representation of the master integrals \cite{Henn:2013pwa, Hoschele:2014qsa}. The identified candidate integrals may only lead to a form of the differential equation close to a canonical form making additional transformations necessary. 

Methods relying solely on the differential equation include an approach using Magnus and Dyson series~\cite{Magnus1954} applicable to differential equations with polynomial dependence on the regulator~\cite{Argeri:2014qva} and a technique exploiting the block-triangular structure of the differential equations that requires a linear dependence of the homogeneous part of the differential equation on the regulator~\cite{Gehrmann:2014bfa}. Another approach based on the differential equation uses factorization properties of the differential operator~\cite{Adams:2017tga}. For differential equations depending on one variable, an algorithm to compute a transformation to a canonical basis has been described in detail by Lee~\cite{Lee:2014ioa}. Most of the aforementioned approaches lack such a detailed algorithmic description, which is reflected by the fact that Lee's algorithm is the only one with publicly available implementations \cite{Gituliar:2016vfa, Prausa:2017ltv, Gituliar:2017vzm}. Since Lee's algorithm can only compute transformations depending on one variable, the range of processes it can be used for is severely restricted.

The goal of this chapter is to overcome this restriction by developing an algorithm applicable to differential equations depending on an arbitrary number of scales. Regarding the distinction made above, the presented algorithm falls into the second category of methods relying solely on the differential equation. It is clear that the differential equation must contain all information necessary to compute a transformation to a canonical form, if it exists at all. However, to the best of the author's knowledge there is no necessary and sufficient criterion for the existence of a canonical form known that is readily computable from the differential equation. In fact, it is not even known whether a canonical form can always be achieved for Feynman integrals evaluating to Chen iterated integrals, although there appears to be no counterexample either. Moreover, it is well known \cite{Caffo:1998du, Laporta:2004rb, Bloch:2013tra, Adams:2014vja, Bloch:2014qca, Adams:2015gva, Bloch:2016izu, Remiddi:2016gno, Adams:2016xah, Bonciani:2016qxi, Adams:2017ejb} that Feynman integrals exist that do not evaluate to Chen iterated integrals and therefore a canonical form as in \eq{EpsForm} cannot exist for these integrals. 

The strategy throughout this chapter is to \emph{assume} the existence of a transformation to a canonical form and develop the algorithm based on general properties of such transformations avoiding unnecessary assumptions as much as possible. These general properties are studied in \sec{sec:GenProps}. Then, it is shown in \sec{sec:ExpTrafo} that transformations to a canonical form are determined by a finite number of differential equations, which are obtained by expanding a reformulated version of the transformation law \eq{DEQTrafo}. In \sec{sec:OffDiagPart} this strategy is adapted to be applicable recursively by exploiting the block-triangular structure of the differential equations. A generalized partial fractions technique is used in \sec{sec:ansatz} to devise a rational ansatz to solve the aforementioned differential equations for a rational transformation. 

The limitations of the algorithm are discussed together with the limitations of its implementation in \chap{chap:package}.

The material presented in this chapter is based on the publications \cite{Meyer:2016zeb, Meyer:2016slj, Meyer:2017joq}.

\section{General properties of the transformation}
\label{sec:GenProps}
This section explores general properties of transformations from a given basis of master integrals to a canonical basis, which are later exploited for the construction of the algorithm. Assuming the existence of a transformation $T$ that transforms the differential equation $a(\varset)$ into the canonical form
\begin{equation}
\label{atildedlog2}
a^\prime(\varset)=\epsilon\,\textup{d}\tilde{A}(\invariants)=\epsilon\sum_{l=1}^N \tilde{A}_l\textup{d}\log(L_l(\invariants))
\end{equation}
is equivalent to demanding the transformation law 
\begin{equation}
\label{aprimeINeps}
\epsilon\,\textup{d}\tilde{A}=T^{-1}aT-T^{-1}\textup{d}T
\end{equation}
to be satisfied for some invertible $T$. The problem of computing such a transformation can thus be rephrased as finding a $T$ and a $\textup{d}\tilde{A}$ for a given $a(\varset)$ such that \eq{aprimeINeps} holds.

\subsection{Trace formula}
In the following, it will be shown that the determinant of $T$ and the trace of $\textup{d}\tilde{A}$ can be inferred from the trace of the given $a(\varset)$. Taking the trace on both sides of \eq{aprimeINeps} leads to
\begin{equation}
\epsilon\textup{Tr}[\textup{d}\tilde{A}]=\textup{Tr}[a]-\textup{Tr}[T^{-1}\textup{d}T].
\end{equation}
Applying Jacobi's formula for the differential of determinants
\begin{equation}
\textup{d}\det(T)=\det(T)\textup{Tr}[T^{-1}\textup{d}T],
\end{equation}
leads to
\begin{equation}
\label{TrOfTrafo}
\textup{d}\log(\det(T))=\textup{Tr}[a]-\epsilon\textup{Tr}[\textup{d}\tilde{A}].
\end{equation}
It follows immediately that the differential form $\textup{Tr}[a]$ necessarily has to be closed in order for a canonical form to exist:
\begin{equation}
\label{necessaryGeneralCanonicalF}
\textup{Tr}[a]=\textup{d}\left(\epsilon\textup{Tr}[\tilde{A}]+\log(\det(T))\right).
\end{equation}
In fact, with \eq{atildedlog2} it is evident that $\textup{Tr}[a]$ has to be in form
\begin{equation}
\label{Tradlog}
\textup{Tr}[a]=\epsilon\sum_{l=1}^N\textup{Tr}[\tilde{A}_l]\textup{d}\log(L_l(\invariants))+\textup{d}\log(\det(T)).
\end{equation}
This form is a slightly more general dlog-form than in the definition of the term in \eq{defdlog2}, since here the argument $\det(T)$  of the logarithm, in general, also depends on the regulator.

Assuming the transformation $T$ to be rational in the invariants and $\epsilon$, it follows that $\det(T)$ is rational as well. Then, the summands of $\det(T)$ can be put on a common denominator, and the resulting numerator and denominator polynomials can be factorized into irreducible polynomials in $K[\varset]$. Here, $K[\varset]$ denotes the ring of polynomials in the invariants and $\epsilon$ with coefficients in a field $K$. There is no need to further specify the field at this point, for the present application the real or complex numbers are most relevant. Denoting the irreducible factors depending only on the invariants by $p$ and those depending on both the invariants and the regulator by $q$, allows to write the factorization of $\det(T)$ as 
\begin{equation}
\label{detFactorization}
\det(T)=F(\epsilon)p_1(\invariants)^{e_1}\cdots p_K(\invariants)^{e_K}q_1(\varset)^{d_1}\cdots q_L(\varset)^{d_L},
\end{equation}
with $e_i,d_j\in\mathbb{Z}$. The product of all factors solely depending on $\epsilon$ is denoted by $F(\epsilon)$.  The factorization of $\det(T)$ then allows to rewrite \eq{Tradlog}:
\begin{equation}
\label{NecessaryCondition}
\textup{Tr}[a]=\epsilon X(\invariants)+Y(\varset),
\end{equation}
with
\begin{equation}
X(\invariants)=\sum_{l=1}^N\textup{Tr}[\tilde{A}_l]\textup{d}\log(L_l(\invariants)),
\end{equation}
and 
\begin{equation}
Y(\varset)=\sum_{i=1}^Ke_i\textup{d}\log(p_i(\invariants))+\sum_{j=1}^Ld_j\textup{d}\log(q_j(\varset)).
\end{equation}
This equation can be understood as a necessary condition on the form of $\textup{Tr}[a]$ for a \emph{rational} transformation $T$ to exist that transforms the differential equation into canonical form. Note that \eq{NecessaryCondition} is violated for non-integer $e_i$ or $d_i$, while the condition in \eq{necessaryGeneralCanonicalF} would still be satisfied. This leaves the possibility of rational differential equations to exist that require non-rational transformations to achieve a canonical form. While such differential equations do indeed exist and are briefly discussed in \sec{sec:TestsNLimits}, the differential equations considered here are assumed to admit a rational transformation to a canonical form and thus satisfy \eq{NecessaryCondition} with integer $e_i$ and $d_j$. From the right-hand side of \eq{NecessaryCondition} it is apparent that
\begin{equation}
\textup{Tr}[a^{(k)}]=0, \quad\forall\,k<0,
\end{equation}
where the $a^{(k)}$ denote the coefficients of the $\epsilon$\nobreakdash-expansion of $a(\varset)$. The coefficients on the right-hand side of \eq{NecessaryCondition} of the dlog-terms stemming from $\det(T)$ are independent of $\epsilon$, whereas the coefficients of the dlog-terms from $\textup{Tr}[\textup{d}\tilde{A}]$ are proportional to $\epsilon$. The determinant of $T$ can therefore be extracted from $\textup{Tr}[a]$ up to a rational function $F(\epsilon)$. Moreover, the traces of the $\tilde{A}_l$ in the resulting canonical form can be read off as well. In practice, it can be tested whether $\textup{Tr}[a]$ is of the form in \eq{NecessaryCondition}. If this is not the case, a rational transformation that transforms $a(\varset)$ into canonical form cannot exist. Otherwise, it is possible to extract
\begin{align}
\label{DetIsFixed}
\det(T)&=F(\epsilon)\exp\left(\int_\gamma Y(\varset)\right),\\
\label{TraceIsFixed}
\textup{Tr}[\textup{d}\tilde{A}]&=X(\invariants),
\end{align}
from the coefficients of the dlog-terms. As will be argued later, both equations provide useful information for the determination of $T$. Often, the factors $q_j$ are absent and therefore
$Y(\varset)=Y(\invariants)$. In this case, the above observations turn into
statements about the coefficients of the $\epsilon$-ex\-pan\-sion of
$a(\varset)$:
\begin{align}
\det(T)&=F(\epsilon)\exp\left(\int_\gamma \textup{Tr}[a^{(0)}]\right),\\
\textup{Tr}[\textup{d}\tilde{A}]&=\textup{Tr}[a^{(1)}].
\end{align}
Furthermore, \eq{NecessaryCondition} implies in this case
\begin{equation}
\textup{Tr}[a^{(k)}]=0, \quad\forall\,\, k\neq 0, 1.
\end{equation}
As an example, consider the differential equation \eq{ycoordinateBubbleDEQ} for which the trace may be written as
\begin{align}
\textup{Tr}[a]=\,\,&\epsilon\left(\textup{d}\log(y)-2\,\textup{d}\log(1+y)\right)\\
\,\,&+\left(-\textup{d}\log(-1+y)+\textup{d}\log(1+y)\right),
\end{align}
and thus
\begin{align}
X(\{y\})&=\textup{d}\log(y)-2\,\textup{d}\log(1+y),\\
Y(\epsilon,\{y\})&=-\textup{d}\log(-1+y)+\textup{d}\log(1+y).
\end{align}
Employing \eq{DetIsFixed} and \eq{TraceIsFixed} yields the following results:
\begin{equation}
\det(T)=F(\epsilon)(-1+y)^{-1}(1+y),
\end{equation}
\begin{equation}
\textup{Tr}[\tilde{A}_y]=1,\quad \textup{Tr}[\tilde{A}_{1+y}]=-2,
\end{equation}
which are easily verified by comparing them to the transformation in \eq{canonicalTrafoex} and the resulting canonical form in \eq{examplesDEQ}.

Note that for one-dimensional sectors \eq{DetIsFixed} already fixes the transformation up to a rational function in $\epsilon$, which is irrelevant for the resulting differential equation. Thus, upon fixing this function, the transformation is completely fixed. In \sec{subsec:AnsatzDB}, the determinant is used to extract valuable information for the computation of $T$ for higher-dimensional sectors as well.

\subsection{On the uniqueness of canonical bases}
\label{subsec:uniqueness}

The transformation of a differential equation in canonical form with a constant invertible transformation $C$, in general, leads to a different but still canonical form of the differential equation:
\begin{equation}
a^\prime=\epsilon\sum_{l=1}^N\left(C^{-1}\tilde{A}_lC\right)\textup{d}\log(L_l).
\end{equation}
This raises the question whether \emph{all} other canonical forms can be obtained in this way from a given one. The following claim shows that indeed every canonical form can be obtained by a constant transformation from any other canonical form. In this sense, the canonical form of a given differential equation is unique up to constant transformations.

\begin{claim}
\label{claim:uniqueC}
Let $a(\varset)$ be a differential equation of Feynman integrals and $T_1(\varset)$ and $T_2(\varset)$ be invertible rational transformations, which transform $a(\varset)$ into the canonical forms $\epsilon\,\textup{d}\tilde{A}_1(\{x_j\})$ and $\epsilon\,\textup{d}\tilde{A}_2(\{x_j\})$, respectively. Then there exists a constant invertible transformation $C$ that transforms $\epsilon\,\textup{d}\tilde{A}_1(\{x_j\})$ into $\epsilon\,\textup{d}\tilde{A}_2(\{x_j\})$.
\end{claim}
Consider the transformation $T=T^{-1}_1T_2$, which transforms $\epsilon\,\textup{d}\tilde{A}_1$ into $\epsilon\,\textup{d}\tilde{A}_2$. First, the transformation $T$ has to be shown to be independent of the invariants. The corresponding transformation law reads
\begin{equation}
\label{TrafoLaweps2eps}
\epsilon\,\textup{d}\tilde{A}_2=T^{-1}\epsilon\,\textup{d}\tilde{A}_1T-T^{-1}\textup{d}T.
\end{equation}
It is instructive to rewrite this equation:
\begin{align}
\textup{d}T&=\epsilon\left(\textup{d}\tilde{A}_1T-T\textup{d}\tilde{A}_2\right)\\
&=\epsilon\sum_{l=1}^N\left(\tilde{A}_{1l}T-T\tilde{A}_{2l}\right)\textup{d}\log(L_l).
\end{align}
The summation over the letters is meant to run over the union of the sets of letters of the two canonical forms since it is a priori not clear that they both have exactly the same set of letters. The letters are assumed to be irreducible polynomials, and the union is meant to remove all scalar multiples of letters as well. Since the transformation law is invariant under the multiplication of $T$ with any rational function $g(\epsilon)$, the $\epsilon$\nobreakdash-expansion of $T$ can be assumed to start at the order $\epsilon^0$. Then the first order in the expansion of the above equation reads
\begin{equation}
\textup{d}T^{(0)}=0
\end{equation}
and therefore $T^{(0)}$ has to be constant. At any order $n>0$ the expansion of the above equation is given by
\begin{equation}
\textup{d}T^{(n)}=\sum_{l=1}^N\left(\tilde{A}_{1l}T^{(n-1)}-T^{(n-1)}\tilde{A}_{2l}\right)\textup{d}\log(L_l).
\end{equation}
Assuming $T^{(n-1)}$ to be constant, this equation can easily be integrated
\begin{equation}
T^{(n)}=\sum_{l=1}^N\left(\tilde{A}_{1l}T^{(n-1)}-T^{(n-1)}\tilde{A}_{2l}\right)\log(L_l)+\mathrm{const.}
\end{equation}
Since $T_1$ and $T_2$ are assumed to be rational in $\epsilon$ and the invariants, the same holds for $T$, and therefore the coefficients of its $\epsilon$\nobreakdash-expansion have to be rational as well. This implies 
\begin{equation}
\tilde{A}_{1l}T^{(n-1)}-T^{(n-1)}\tilde{A}_{2l}=0, \quad \forall\, l,
\end{equation}
and consequently $T^{(n)}$ has to be constant. By induction, these arguments imply that all coefficients of the $\epsilon$\nobreakdash-expansion of $T$ are constant and therefore $T=T(\epsilon)$. As $T$ is independent of the invariants, the transformation law \eq{TrafoLaweps2eps} has the form
\begin{equation}
\textup{d}\tilde{A}_2=T(\epsilon)^{-1}\textup{d}\tilde{A}_1T(\epsilon).
\end{equation}
It can be concluded that $T(\epsilon)$ transforms $\textup{d}\tilde{A}_1$ into $\textup{d}\tilde{A}_2$ for all non-singular values of $\epsilon$, because the left-hand side does not depend on $\epsilon$. Upon choosing such a value $\epsilon_0$, a constant invertible transformation $C=T(\epsilon_0)$ is obtained, which concludes the proof of the claim. The same argument also holds for the more general case of an algebraic dependence of $T_1$ and $T_2$ on $\epsilon$ and the invariants. Altogether, canonical forms have been shown to be unique modulo $GL(m, K)$ transformations. 

In \sec{subsec:NLequations} this result is utilized for the construction of the algorithm. The uniqueness of canonical forms can also be used for the comparison of two different canonical forms of the same problem, provided they are expressed in the same set of invariants. In this situation, claim \ref{claim:uniqueC} asserts the existence of a constant transformation relating the two canonical forms. This can be tested by checking whether the following system of linear equations
\begin{equation}
C\tilde{A}_{2l}=\tilde{A}_{1l}C,\quad l=1,\dots,N,
\end{equation}
has a non-singular solution for the components of $C$.

\section{Algorithm for diagonal blocks}
\label{sec:ExpTrafo}

%


Every invertible transformation $T$ to a canonical form has to satisfy \eq{aprimeINeps} for some $\textup{d}\tilde{A}$, which has to be determined as well. For invertible $T$, \eq{aprimeINeps} can equivalently be written as
\begin{equation}
\label{DEQTrafoAlternativ}
\textup{d}T-aT+\epsilon T\textup{d}\tilde{A}=0.
\end{equation}
The general idea to find a solution of this equation is to expand \eq{DEQTrafoAlternativ} in $\epsilon$ and solve for the coefficients of the expansion of $T$ and $\textup{d}\tilde{A}$ order by order with a rational ansatz. The form of the transformation law in \eq{DEQTrafoAlternativ} has the advantage of not containing the inverse of $T$ and thus does not require the ansatz for $T$ to be inverted.

In general, the $\epsilon$\nobreakdash-expansion of $T$ may have infinitely many non-vanishing coefficients. This poses a problem for the algorithmic computation of these coefficients since in this case solving the expansion of \eq{DEQTrafoAlternativ} order by order would not stop after a finite number of orders. In the following, it will be shown how this problem can be circumvented by reformulating \eq{DEQTrafoAlternativ} in terms of quantities with finite $\epsilon$\nobreakdash-expansion.

\subsection{Reformulation in terms of quantities with finite expansion}

It is evident that for any solution $T$ of \eq{DEQTrafoAlternativ}, the multiplication with a rational function $g(\epsilon)$ leads to another solution. Thus, there is the freedom to fix the normalization of the solution with a function $g(\epsilon)$ without loss of generality. Any rational function $g(\epsilon)$ can be written as a product of some power of $\epsilon$ and a rational function $\eta(\epsilon)$ with non-vanishing constant coefficient
\begin{align}
g(\epsilon)&=\epsilon^\tau\eta(\epsilon),\quad \tau\in\mathbb{Z},\\
\eta(\epsilon)&=c_0+\epsilon c_1+\mathcal{O}(\epsilon), \quad c_0\neq 0.
\end{align}
The freedom to choose $\tau$ can be exploited by demanding the expansion of $T$ to start at order $\epsilon^0$
\begin{equation}
\label{TStartsAtZero}
T=\sum_{n=0}^\infty\epsilon^nT^{(n)}, \quad T^{(0)}\neq 0.
\end{equation}
The value of the rational function $\eta(\epsilon)$ is not affected by this condition and will be fixed later on. 

As $a(\varset)$ is required to be rational in both the invariants and $\epsilon$, a polynomial $h(\varset)$ exists such that $\hat{a}\defeq ah$ has a finite Taylor expansion in $\epsilon$
\begin{equation}
\label{aStartsAtZero}
\hat{a}=\sum_{k=0}^{k_\textup{max}}\epsilon^k\hat{a}^{(k)}.
\end{equation}
Likewise, there exists a polynomial $f(\varset)$ such that $\tilde{T}\defeq Tf$ has a finite expansion in $\epsilon$
\begin{equation}
\label{TtildeStartsAtZero}
\tilde{T}=\sum_{q=0}^{q_\textup{max}}\epsilon^q\tilde{T}^{(q)},\quad \tilde{T}^{(0)}\neq 0.
\end{equation}
Note that Eqs. \eqref{TStartsAtZero} and \eqref{TtildeStartsAtZero} imply that the expansion of $f$ starts at the constant term
\begin{equation}
\label{fStartsAtZero}
f(\varset)=f^{(0)}(\invariants)+\mathcal{O}(\epsilon), \quad f^{(0)}(\invariants)\neq 0,
\end{equation}
whereas the condition \eq{aStartsAtZero} may require the expansion of $h$ to start at some higher order $l_\textup{min}$
\begin{equation}
h(\varset)=\sum_{l=l_\textup{min}}^{l_\textup{max}}\epsilon^l h^{(l)}(\invariants),\quad l_\textup{min}\geq 0,
\end{equation}
because $a(\varset)$ can, in general, have negative powers of $\epsilon$ in its expansion, which in the case of $T$ have already been absorbed by the choice of $\tau$. In addition to the above conditions, $h$ and $f$ are required to be \textit{minimal} in the sense that they shall have the smallest possible number of irreducible factors for which $\hat{a}$ and $\tilde{T}$ have finite $\epsilon$\nobreakdash-expansions of the form in \eq{aStartsAtZero} and \eq{TtildeStartsAtZero}, respectively. This fixes $h$ and $f$ up to multiplicative constants, which are irrelevant here. Let the factorizations of $h$ and $f$ into irreducible factors in $K[\varset]$ be denoted by
\begin{equation}
f=\prod_{i=1}^{N_f}f_i,\quad h=\prod_{i=1}^{N_h}h_i.
\end{equation}

\subsection{Investigating the relation of \texorpdfstring{$f$}{f} and \texorpdfstring{$h$}{h}}
\label{sec:relationfh}
It is straightforward to compute $h$ for a given $a(\varset)$. However, $f$ could only be computed directly if $T$ was known. Since this is not the case, it is explored in the following which information about $f$ can be extracted from $h$. In terms of the quantities defined above, \eq{DEQTrafoAlternativ} may be rewritten as
\begin{equation}
\label{DEQTrafof}
\frac{h\tilde{T}\textup{d}f}{f}=\left(\textup{d}\tilde{T}+\epsilon\tilde{T}\textup{d}\tilde{A}\right)h-\hat{a}\tilde{T}.
\end{equation}
The right-hand side of \eq{DEQTrafof} only consists of sums and products of quantities with finite expansions. Therefore, both sides of the above equation have a finite expansion. For the left-hand side, this implies that
\begin{equation}
\label{FiniteSummands}
\frac{h\tilde{T}\textup{d}f}{f}=\sum_{i=1}^{N_f}\frac{h\tilde{T}\textup{d}f_i}{f_i}
\end{equation}
has a finite expansion. In fact, already each individual summand of the above sum can be shown to have a finite expansion. To this end, it is sufficient to show that there is no number $n$ of such terms with infinite expansion that can sum up to give a finite expansion. For $n=1$ this is obvious, and therefore it remains to be shown that if the assertion holds for $n$ terms, it also holds for $n+1$ terms. Consider $f_1, \dots, f_{n+1}$ and assume that the assertion is not true, i.e., each $h\tilde{T}\textup{d}f_i/f_i$ has an infinite expansion but the sum of all of these terms has a finite expansion. Defining $F_n=f_1\cdots f_n$ allows to write
\begin{equation}
\label{Ffsum}
\frac{h\tilde{T}\textup{d}F_n}{F_n}+\frac{h\tilde{T}\textup{d}f_{n+1}}{f_{n+1}}=\frac{h\tilde{T}\textup{d}(F_nf_{n+1})}{F_nf_{n+1}}.
\end{equation}
The second term on the left-hand side has an infinite expansion by assumption, and the first term has to have an infinite expansion because the assertion holds for $n$ terms. Since the right-hand side is assumed to have a finite expansion, both $F_n$ and $f_{n+1}$ have to be canceled by corresponding factors in the numerator. However, neither $h$ nor $\tilde{T}$ can be a product of one or both of these factors with a quantity with finite expansion, since this would render the expansions of the terms on the left-hand side finite. Thus, the only possibility left to investigate is 
\begin{equation}
\label{dfargument}
\textup{d}(F_nf_{n+1})=r(\varset)F_nf_{n+1},
\end{equation}
where $r(\varset)$ denotes a rational differential form with finite expansion. Upon integration, this relation leads to 
\begin{equation}
F_nf_{n+1}=\rho(\epsilon)\exp\left(\int_\gamma r(\varset)\right),
\end{equation}
with $\rho$ denoting a polynomial in $\epsilon$. Since $F_n\cdot f_{n+1}$ is polynomial in the invariants and $\epsilon$, the finiteness of the expansion of $r$ implies
\begin{align}
r^{(k)}(\invariants)&=0,\quad\forall\,\,k\neq 0,\\
r^{(0)}(\invariants)&=\textup{d}\log\left(p(\invariants)\right),
\end{align}
with $p$ being a polynomial in the invariants. However, since $f$ is required to be minimal, it cannot contain any irreducible factors independent of $\epsilon$. Therefore, $p$ has to be a constant which implies
\begin{equation}
\label{Fnfnp1epspoly}
F_nf_{n+1}=\rho(\epsilon).
\end{equation}
Both, $F_n$ and $f_{n+1}$ need to have non-vanishing differentials, because otherwise the terms on the left-hand side of \eq{Ffsum} would have a finite expansion. Consequently, both factors have a non-trivial dependence on the invariants. Since $F_n$ and $f_{n+1}$ are polynomials, their product has a non-trivial dependence on the invariants as well, which contradicts \eq{Fnfnp1epspoly}. Thus, the assertion has to be true for $n+1$ terms as well and therefore, by induction, hold for all $n>0$. Altogether, this shows that each summand in \eq{FiniteSummands} has to have a finite expansion.

The minimality of $f$ implies that $\tilde{T}$ cannot be of the form $\tilde{T}=rf_i$ for some rational $r(\varset)$ with finite expansion because otherwise the factor $f_i$ would not be necessary to render the expansion of $T$ finite and consequently $f$ would not be minimal. Also, note that the minimality of $f$ implies that its irreducible factors must 
\emph{all} depend non-trivially on the regulator. There are only the following two possibilities for a summand of \eq{FiniteSummands} to have a finite expansion:
\begin{equation}
\label{TwoPossibilities}
\textup{d}f_i=r_if_i\quad \vee\quad h=r_if_i,
\end{equation}
where $r_i$ denotes a rational differential form or function of the invariants and $\epsilon$ with finite expansion. However, since the left-hand sides of \eq{TwoPossibilities} are polynomial, a denominator of $r_i$ would have to be canceled by $f_i$, but this would imply that $r_i$ has an infinite expansion. Thus, $r_i$ has, in fact, to be polynomial. The first of the above possibilities implies $f_i=c_i(\epsilon)$ by an argument analogous to the one around \eq{dfargument}, where $c_i(\epsilon)$ denotes an irreducible polynomial in $\epsilon$. In the second case, $f_i$ is equal to one of the irreducible factors of $h(\varset)$. Thus, the irreducible factors of $f$ that are not given by an irreducible factor of $h$ are independent of the invariants.

\subsection{Obtaining a finite expansion with \texorpdfstring{$h$}{h}}
As mentioned before, the polynomial $f$ cannot be used to render the transformation finite, since it cannot be directly determined before the computation of $T$. In the previous section, it was shown that an irreducible factor of $f$ can either be determined from $h$ or is independent of the invariants. The latter factors may conveniently be absorbed by the remaining freedom to choose $\eta(\epsilon)$. In the following, it is outlined how this leads to a formulation of the transformation law with a finite expansion.

Let $\mathcal{S}$ denote the set of indices of the irreducible factors of $h$ that both depend non-trivially on the invariants and are equal to an irreducible factor of $f$. The product of all irreducible factors of $f$ depending only on $\epsilon$ is denoted by $c(\epsilon)$. Using this notation, $f$ can be written as follows
\begin{equation}
f=c(\epsilon)\prod_{i\in\mathcal{S}\subseteq\{1,\dots,N_h\}}h_i.
\end{equation}
From \eq{fStartsAtZero} it is clear that $c(\epsilon)$ is of the form
\begin{equation}
c(\epsilon)=c^{(0)}+\mathcal{O}(\epsilon),\quad c^{(0)}\neq 0.
\end{equation}
The remaining freedom in the choice of the overall factor $g(\epsilon)$ can be used to absorb $c(\epsilon)$ by demanding $\eta(\epsilon)=c(\epsilon)$. This choice completely fixes $g(\epsilon)$ and reduces $f$ to
\begin{equation}
f=\prod_{i\in\mathcal{S}\subseteq\{1,\dots,N_h\}}h_i.
\end{equation}
Since the set $\mathcal{S}$ is unknown prior to the computation of $T$, it is not possible to use the minimal set of factors necessary to render the expansion of $T$ finite. However, by multiplying with \emph{all} irreducible factors of $h$, the resulting transformation will have a finite expansion, though possibly contain some unnecessary factors. This amounts to defining $\hat{T}\defeq Th$, which can now easily be seen to have a finite expansion by
\begin{equation}
\hat{T}=Th=\tilde{T}\prod_{i\in\{1,\dots,N_h\}\setminus\mathcal{S}}h_i.
\end{equation}

\subsection{Solving the expanded transformation law}
\label{subsec:SolexpTrafoLaw}

The transformation law \eq{DEQTrafoAlternativ} can now be rewritten entirely in terms of quantities with finite expansion:
\begin{equation}
\label{DEQfiniteh}
-\hat{T}\textup{d}h+h\textup{d}\hat{T}-\hat{a}\hat{T}+\epsilon h\hat{T}\textup{d}\tilde{A}=0.
\end{equation}
Altogether, it was shown that for any solution $T$ of \eq{DEQTrafoAlternativ} there exists a solution $\hat{T}$ of \eq{DEQfiniteh} with finite expansion
\begin{equation}
\hat{T}=\sum_{n=l_\textup{min}}^{n_\textup{max}}\epsilon^n\hat{T}^{(n)}.
\end{equation}
Conversely, each solution $\hat{T}$ of \eq{DEQfiniteh} corresponds to a solution $T$ of \eq{DEQTrafoAlternativ} via $T=\hat{T}/h$. Thus, it can be avoided to calculate infinitely many coefficients in the expansion of $T$ by computing $\hat{T}$ instead. This can be done by expanding \eq{DEQfiniteh} in the regulator:
\begin{equation}
-\hat{T}\textup{d}h+h\textup{d}\hat{T}=\sum_{n=2l_{\textup{min}}}^{n_\textup{max}+l_\textup{max}}\epsilon^n\sum_{k=l_\textup{min}}^{\textup{min}(l_\textup{max},n-l_\textup{min})}\left(-\textup{d}h^{(k)}\hat{T}^{(n-k)}+h^{(k)}\textup{d}\hat{T}^{(n-k)}\right),
\end{equation}
\begin{align}
\epsilon h\hat{T}\textup{d}\tilde{A}&=\sum_{n=2l_{\textup{min}}}^{n_\textup{max}+l_\textup{max}}\epsilon^{n+1}\sum_{k=l_\textup{min}}^{\textup{min}(l_\textup{max},n-l_\textup{min})}h^{(k)}\hat{T}^{(n-k)}\textup{d}\tilde{A}\\
&=\sum_{n=2l_{\textup{min}}+1}^{n_\textup{max}+l_\textup{max}+1}\epsilon^{n}\sum_{k=l_\textup{min}}^{\textup{min}(l_\textup{max},n-l_\textup{min}-1)}h^{(k)}\hat{T}^{(n-k-1)}\textup{d}\tilde{A},
\end{align}
\begin{equation}
\hat{a}\hat{T}=\sum_{n=l_{\textup{min}}}^{n_\textup{max}+k_\textup{max}}\epsilon^n\sum_{k=0}^{\textup{min}(k_\textup{max},n-l_\textup{min})}\hat{a}^{(k)}\hat{T}^{(n-k)}.
\end{equation}
Note that the equation at some order $k$ only involves $\hat{T}^{(n)}$ with $n\leq k$. Therefore, the $\hat{T}^{(n)}$ can be computed successively, starting with the lowest order. Given some $a(\varset)$, the first step is to calculate $h$ and $\hat{a}$, which fixes the values of $l_\textup{min}$, $l_\textup{max}$ and $k_\textup{max}$. The value of $n_\textup{max}$ remains unknown until the solution for $\hat{T}$ is known. Therefore, it is tested at each order $k$ whether $k=n_\textup{max}$. In order to do so, it has to be checked if $\hat{T}^{(n)}=0$ for all $n>k$ solves the equations of the remaining $\textup{max}(k_\textup{max}, l_\textup{max}+1)$ subsequent orders. The algorithm stops as soon as this test is successful and returns $T=\hat{T}/h$. 

Each order in the expansion of \eq{DEQfiniteh} is a differential equation for the coefficients $\hat{T}^{(n)}$ and the resulting canonical form $\textup{d}\tilde{A}$. These differential equations do in general admit transcendental solutions for $\hat{T}^{(n)}$. In order to single out the rational solutions, it suggests itself to solve these equations with a rational ansatz of the form 
\begin{equation}
\label{ansatzThat}
\hat{T}^{(n)}=\sum_{k=1}^{|\mathcal{R}_T|}\tau_k^{(n)}r_k(\invariants),
\end{equation}
\begin{equation}
\mathcal{R}_T\defeq \left\{r_1(\invariants),\dots,r_{|\mathcal{R}_T|}(\invariants)\right\},
\end{equation}
where the $\tau_k^{(n)}$ denote unknown $m \times m$ matrices independent of the invariants and the regulator, which are to be determined by the algorithm. The choice of the set $\mathcal{R}_T$ of rational functions is discussed in detail in \sec{sec:ansatz}. Choosing the ansatz to be linear in the unknown parameters results in equations linear in the unknown parameters $\tau$ since \eq{DEQfiniteh} is linear in $\hat{T}$. The dependence of the unknown resulting canonical form $\textup{d}\tilde{A}$ on the invariants is restricted by its dlog-form, which suggests an ansatz of the form
\begin{equation}
\label{ansatzAtilde}
\tilde{A}(\invariants)=\sum_{l=1}^N\alpha_l\log(L_l(\invariants)),
\end{equation}
where the $\alpha_l$ are considered to be unknown $m\times m$ matrices independent of the invariants and the regulator. In \sec{subsec:Ansatzepsform} the set of letters in the resulting canonical form is proven to be a subset of the irreducible denominator factors of $a(\varset)$ that are independent of the regulator. Thus, the ansatz in \eq{ansatzAtilde} is guaranteed to encompass the resulting canonical form if all of the aforementioned denominator factors are included in the ansatz.

\subsection{Treatment of nonlinear parameter equations}
\label{subsec:NLequations}

In the course of applying the algorithm, the ansatz in \eqs \eqref{ansatzThat} and \eqref{ansatzAtilde} is inserted in the expansion of \eq{DEQfiniteh}. By requiring the resulting equations to hold for all allowed values of the invariants, a system of equations in the unknown parameters is obtained at each order of the expansion. Due to the term $\epsilon T\textup{d}\tilde{A}$ in \eq{DEQTrafoAlternativ} these equations can contain products of the unknown $\tau$ and $\alpha$ parameters and therefore be nonlinear in the unknowns. Instead of directly solving these nonlinear equations, it will be shown below how these equations can be reduced to linear equations by imposing appropriate constraints without compromising the generality of the algorithm.

In \sec{subsec:uniqueness}, it has been proven that the resulting canonical form is uniquely fixed up to an invertible constant transformation. Exactly this ambiguity leads to the nonlinear equations, because if $\textup{d}\tilde{A}$ was fixed, the term $\epsilon T\textup{d}\tilde{A}$ would not generate nonlinear equations. Therefore, the nonlinear equations can be turned into linear equations by fixing the degrees of freedom in the ansatz that correspond to a subsequent invertible constant transformation. In order to fix these degrees of freedom directly, they would have to be disentangled from those which are determined by the equations in the parameters. Since this would require a parameterization of the solution set of the nonlinear equations, which is essentially equivalent to solving them, a more indirect approach of fixing the freedom needs to be taken.

To this end, suppose for the moment that the parameters of the ansatz can be separated in those which are fixed by the parameter equations $\{\tau\}$ and those which correspond to the remaining freedom $\{\tau^\prime\}$. Let $T(\epsilon, \{x_j\}, \{\tau\},\{\tau^\prime\})$ be a solution of \eq{DEQTrafoAlternativ}, provided the parameters $\{\tau\}$ solve the parameter equations. According to the proof of claim \ref{claim:uniqueC}, this transformation can be thought of as the product of some fixed transformation $T_1(\varset,\{\tau\})$, which transforms the original differential equation to some canonical form $\epsilon\,\textup{d}\tilde{A}_1$, and a transformation $C(\epsilon,\{\tau^\prime\})$ parameterizing the transformation of $\textup{d}\tilde{A}_1$ to any other possible canonical form $\epsilon\,\textup{d}\tilde{A}_2(\tau^\prime)$:
\begin{equation}
\label{freeParamFactorize}
T(\epsilon, \{x_j\}, \{\tau\},\{\tau^\prime\})=T_1(\varset,\{\tau\})C(\epsilon,\{\tau^\prime\}),
\end{equation}
\begin{equation}
\label{remainingFreedomTrafo}
\textup{d}\tilde{A}_2(\{\tau^\prime\})=C(\epsilon,\{\tau^\prime\})^{-1}\textup{d}\tilde{A}_1C(\epsilon,\{\tau^\prime\}).
\end{equation}
It should be noted that $\tilde{A}_2$ does in general only depend on a subset of the parameters $\{\tau^\prime\}$ because some parameters can correspond to a non-trivial $\epsilon$ dependence of $C(\epsilon,\{\tau^\prime\})$. As mentioned above, the goal is to fix the resulting differential equation $\textup{d}\tilde{A}_2$ by fixing the corresponding parameters of $\{\tau^\prime\}$. 
This can be achieved by demanding $C(\epsilon,\{\tau^\prime\})$ to equal a fixed constant invertible transformation at some non-singular value $\epsilon=\epsilon_0$. Since the left-hand side of \eq{remainingFreedomTrafo} does not depend on $\epsilon$, this completely fixes $\textup{d}\tilde{A}_2$ irrespective of the particular value of $\epsilon_0$. However, fixing $C(\epsilon,\{\tau^\prime\})$ directly would require the computation of the factorization in \eq{freeParamFactorize}, which is only possible if the separation of the parameters into the sets $\{\tau\}$ and $\{\tau^\prime\}$ is known. Instead, $C(\epsilon,\{\tau^\prime\})$ can be fixed indirectly by demanding
\begin{equation}
\label{addconstraints}
T(\epsilon_0,\{x_{0j}\},\{\tau\},\{\tau^\prime\})=\mathbb{I}
\end{equation}
to hold at some non-singular point $\{x\}=\{x_{j0}\}$, $\epsilon=\epsilon_0$. This is equivalent to fixing $C(\epsilon,\{\tau^\prime\})$ as follows
\begin{equation}
C(\epsilon_0,\{\tau^\prime\})=T_1(\epsilon_0,\{x_{0j}\},\{\tau\})^{-1}.
\end{equation}
The constraints given by \eq{addconstraints} can be imposed without being able to separate the parameters into $\{\tau\}$ and $\{\tau^\prime\}$. Moreover, these constraints are linear in both the $\{\tau\}$ and the $\{\tau^\prime\}$, since the ansatz in \eq{ansatzThat} is linear in all parameters. Therefore, the additional constraints in \eq{addconstraints} can be used to completely fix the resulting canonical form, which turns the nonlinear parameter equations into linear equations.

Recall that the parameter equations are generated order by order in the expansion of \eq{DEQfiniteh}, and at each order it is tested whether the series terminates at the current order. The constraints in \eq{addconstraints} can only be imposed if the full $T(\epsilon, \{x_j\}, \{\tau\},\{\tau^\prime\})$ is known. Thus, the computation must have reached the order at which the series terminates. However, nonlinear equations can already occur at lower orders in the expansion, i.e., before \eq{addconstraints} can be imposed to turn them into linear equations. The strategy described in the following overcomes this point by essentially just solving the linear equations at each order and keeping the nonlinear equations until the constraints can be imposed.

At each order in the expansion of \eq{DEQfiniteh}, the linear equations are solved first, and then their solution is inserted into the nonlinear ones, which possibly turns some of them into linear equations. These newly generated linear equations can again be solved. This procedure is iterated until no further linear equations are generated. The remaining nonlinear equations are kept unsolved. It is then tested whether the series terminates at the current order $k$ by generating the additional parameter equations implied by the $\textup{max}(k_\textup{max},l_\textup{max}+1)$ following orders under the assumption $\hat{T}^{(n)}=0$ for all $n>k$. The linear equations in the parameters implied by these additional equations are then iteratively solved as described above, while the previously obtained still unsolved nonlinear equations are taken into account as well. If it turns out during this iteration that the system has no solution, the algorithm proceeds with the next order in the expansion of the transformation. If this is not the case and some nonlinear equations remain at the end of the iteration, \eq{addconstraints} is imposed. If the series does terminate at the current order, the additional constraints will turn the remaining nonlinear equations into linear ones, which then determine the transformation. If either nonlinear equations remain or the linear ones have no solution, it can be concluded that the series does not terminate at the current order and the algorithm proceeds with the next order in the expansion.

Altogether, this procedure allows to compute a transformation to a canonical form by only solving linear equations at each order without sacrificing the generality of the algorithm.

\section{Recursion over sectors}
\label{sec:OffDiagPart}
The strategy presented in \sec{sec:ExpTrafo} is applicable to differential equations $a(\varset)$ of an arbitrary number of master integrals. However, if $a(\varset)$ is comprised of more than one sector, the computational cost can be significantly reduced by making use of its block-triangular form. More precisely, the block-triangular form allows to compute the transformation to a canonical form in a recursion over the sectors of $a(\varset)$. Starting from the lowest sector, at each step of the recursion the next diagonal block is transformed into canonical form with strategy presented in \sec{sec:ExpTrafo}. The off-diagonal blocks are transformed into canonical form in a subsequent part of the recursion step, which is the topic of this section. Similar considerations have been made in \cite{Caron-Huot:2014lda, Gehrmann:2014bfa, Lee:2014ioa}.

\subsection{General structure of the recursion step}

In order to investigate the recursion step, it is assumed that the first $p$ sectors have already been transformed into a block-triangular canonical form by a transformation $t_p$. Using the general strategy from \sec{sec:ExpTrafo}, a transformation $t_{p+1}$ can be computed to transform the next diagonal block into canonical form. Thus, up to this point, the transformation
\begin{equation}
t=\left(\,
\begin{array}{|ccc|c|}
\cline{1-4}
 & & &\\
\quad & t_p & \quad & 0 \\
 & & &\\ \hline
 & 0 & & t_{p+1}\\ \hline
\end{array}
\,\right)
\end{equation}
has been applied to the original $a(\varset)$. The intermediate differential equation~$a_{I}$
\begin{equation}
\label{intermediateDEQ}
a_{I}=t^{-1}at-t^{-1}\textup{d}t
\end{equation}
is of the form
\begin{equation}
\label{aBlockinEpsForm}
a_{I}=\left(\,
\begin{array}{|ccc|c|}
\cline{1-4}
 & & &\\
\quad &   \epsilon\tilde{c} & \quad &\,\, 0 \,\,\\
 & & &\\ \hline
 & b & & \epsilon\tilde{e}\\ \hline
\end{array}
\,\right),
\end{equation}
where $\tilde{c}$ and $\tilde{e}$ are in dlog-form with $\tilde{c}$ being block-triangular. The goal of this section is to devise an algorithm to compute the remaining transformation $t_r$, such that
\begin{equation}
\label{IntTrafoEq}
a^\prime=t_r^{-1}a_{I}t_r-t_r^{-1}\textup{d}t_r
\end{equation}
attains a block-triangular canonical form
\begin{equation}
\label{aprimeincanonical}
a^\prime=\left(\,
\begin{array}{|ccc|c|}
\cline{1-4}
 & & &\\
\quad &   \epsilon\tilde{c}^\prime &\quad & 0 \\
 & & &\\ \hline
 &\epsilon\tilde{b}^\prime & & \epsilon\tilde{e}^\prime\\ \hline
\end{array}
\,\right).
\end{equation}
By assumption, there exists a transformation 
\begin{equation}
T=\left(\,
\begin{array}{|ccc|c|}
\cline{1-4}
 & & &\\
 &   T_p & & 0 \\
 & & &\\ \hline
 & T_{p+1,p} & & T_{p+1}\\ \hline
\end{array}
\,\right),
\end{equation}
transforming the original differential equation $a(\varset)$ to the canonical form $a^\prime$. Since both $t_p$ and $T_p$ transform the first $p$ sectors of $a(\varset)$ into a canonical form, they must be related by a transformation $g_p(\epsilon)$ that is independent of the invariants according to the proof of claim \ref{claim:uniqueC}:
\begin{equation}
\label{relrecursive1}
t_p=T_pg_p(\epsilon).
\end{equation}
Similarly, there exists a transformation $g_{p+1}(\epsilon)$ relating the transformations of the sector~$p+1$:
\begin{equation}
\label{relrecursive2}
t_{p+1}=T_{p+1}g_{p+1}(\epsilon).
\end{equation}
The remaining transformation $t_r$ is related to the full transformation $T$ by requiring 
\begin{equation}
t\cdot t_r=T.
\end{equation}
Solving this relation for $t_r$ and using \eq{relrecursive1} and \eq{relrecursive2} to eliminate $t$ leads to
\begin{equation}
t_r=\left(\,
\begin{array}{|ccc|c|}
\cline{1-4}
 & & &\\
 &   g_p^{-1} & & 0 \\
 & & &\\ \hline
 & g_{p+1}^{-1}T_{p+1}^{-1}T_{p+1,p} & & g_{p+1}^{-1}\\ \hline
\end{array}
\,\right).
\end{equation}
It is convenient to split the computation of $t_r$ into two consecutive steps using the following factorization
\begin{equation}
t_r=t_Dt_{g},
\end{equation}
with 
\begin{equation}
t_D=\left(\,
\begin{array}{|ccc|c|}
\cline{1-4}
 & & &\\
\quad &   \mathbb{I} & \quad &\,\, 0 \,\,\\
 & & &\\ \hline
 & D & &\mathbb{I}\\ \hline
\end{array}
\,\right),\quad 
t_{g}=\left(\,
\begin{array}{|ccc|c|}
\cline{1-4}
 & & &\\
 &  g_p^{-1}(\epsilon)  & & 0 \\
 & & &\\ \hline
 & 0 & & g_{p+1}^{-1}(\epsilon) \\ \hline
\end{array}
\,\right),
\end{equation}
and
\begin{equation}
 D=g_{p+1}^{-1}T^{-1}_{p+1}T_{p+1,p}g_p.
\end{equation}
The quantities $D$, $g_p$ and $g_{p+1}$ are determined by the following equations, which are implied by \eq{IntTrafoEq}
\begin{equation}
\label{gpgpp1eqns}
g_p\tilde{c}g_{p}^{-1}=\tilde{c}^\prime,\quad g_{p+1}\tilde{e}g_{p+1}^{-1}=\tilde{e}^\prime,
\end{equation}
\begin{equation}
\textup{d}D-\epsilon(\tilde{e}D-D\tilde{c})=b-\epsilon g_{p+1}^{-1}\tilde{b}^\prime g_p.
\end{equation}
In the latter equation, the product of three unknown quantities occurs in the term $\epsilon g_{p+1}^{-1}\tilde{b}^\prime g_p$. A linear ansatz for these quantities would result in nonlinear equations in the coefficients of the ansatz. This can be prevented by defining $b^\prime=g_{p+1}^{-1}\tilde{b}^\prime g_p$ which leads to
\begin{equation}
\label{DFullDEQ}
\textup{d}D-\epsilon(\tilde{e}D-D\tilde{c})=b-b^\prime.
\end{equation}
Note that $b^\prime$ has to be in dlog-form, since $g_p$ and $g_{p+1}$ are independent of the invariants and therefore do not spoil the dlog-form of $\tilde{b}^\prime$. Then, for a given $b$, the remaining transformation can be calculated by first solving \eq{DFullDEQ} for a rational $D$ and a $b^\prime$ in dlog-form. In a second step, the transformation $t_g$ is determined from \eq{gpgpp1eqns} and $b^\prime=g_{p+1}^{-1}\tilde{b}^\prime g_p$, which is the topic of \sec{sec:dettg}. Altogether, the above argument has shown that the calculation of a canonical basis can without loss of generality be split into the calculation of the diagonal blocks and the steps presented here. The performance gain of the recursive approach compared with transforming the differential equation all at once is mostly due to the more specific ansatzes that can be used in the various steps of the recursion. The following sections are dedicated to the solution of \eq{DFullDEQ} and the determination of $t_g$.

\subsection{Setting up a recursion over sectors for \texorpdfstring{$t_{D}$}{tD}}

The first step in the determination of the remaining transformation is the computation of $t_D$. To that end, it is necessary to solve \eq{DFullDEQ} for a rational $D$ and a $b^\prime$ in dlog-form. In this section, it is argued that the block-triangular form of $\tilde{c}$ can be used to split the computation of $D$ into a recursion over sectors. Therefore, all quantities are considered to be split into sectors according to the block-triangular structure:
\begin{equation}
D=\left(D_1,\dots,D_p\right),\quad b=\left(b_1,\dots, b_p\right),\quad b^{\prime}=\left(b^{\prime}_1,\dots,b^{\prime}_p\right),
\end{equation}
\begin{align}
\tilde{c}&=\left(\,
\begin{array}{cccc}
\tilde{c}_{1} & & &   \\
\vdots & \ddots & & \\
 & & \tilde{c}_{p-1} & \\
\tilde{c}_{p,1} & \cdots & \tilde{c}_{p,p-1} & \tilde{c}_p\\ 
\end{array}
\,\right).
\end{align}
In this notation, \eq{DFullDEQ} may equivalently be written as a system of $p$ equations of the form
\begin{equation}
\textup{d}D_{k}-\epsilon(\tilde{e}D_{k}-D_{k}\tilde{c}_{k})=\left(b_{k}-\epsilon \sum_{i=k+1}^pD_i\tilde{c}_{i,k}\right)-b_{k}^\prime,\quad k=1,\dots, p.
\end{equation}
Note that the equation for a sector $k$ only depends on the $D_n$ of higher sectors with $n \geq k$. It is, therefore, possible to solve for the $D_k$ in a recursion that starts with the highest sector. As for the recursion step, suppose that the equations for the topmost $p-k$ sectors have already been solved. The contribution of the higher sectors to the equation of sector $k$ is most naturally absorbed into the definition of
\begin{equation}
\bar{b}_k\defeq b_{k}-\epsilon  \sum_{i=k+1}^pD_i\tilde{c}_{i,k}.
\end{equation}
Thus, the following equation has to be solved 
\begin{equation}
\label{SubsecDDEQ}
\textup{d}D_k-\epsilon(\tilde{e}D_k-D_k\tilde{c}_k)=\bar{b}_k-b_k^\prime,
\end{equation}
with $\bar{b}_k$ being determined by the solution of the higher sectors. In summary, this procedure allows to split the computation of a solution of \eq{DFullDEQ} into several smaller computations of the same form, which turns out to be beneficial for the performance of the algorithm. 

\subsection{Uniqueness of the rational solution}
\label{sec:uniquenessD}
The following sections are concerned with the solution of \eq{SubsecDDEQ}. As the sector index in \eq{SubsecDDEQ} is irrelevant for these considerations, it will be suppressed from now on. In this section, the rational solution of this equation for $D$ is proven to be unique up to the addition of terms depending solely on the regulator. In practice, this result allows to exclude terms with trivial dependence on the invariants from the ansatz without losing generality.

For a given $\bar{b}$, let $D$ and $b^\prime$ satisfy 
\begin{equation}
\label{DDEQcopy2}
\textup{d}D-\epsilon(\tilde{e}D-D\tilde{c})=\bar{b}-b^\prime,
\end{equation}
with $b^\prime$ assumed to be in dlog-form. Adding a term $C(\epsilon)$ to the solution $\tilde{D}=D+C(\epsilon)$ solves the same equation
\begin{equation}
\textup{d}\tilde{D}-\epsilon(\tilde{e}\tilde{D}-\tilde{D}\tilde{c})=\bar{b}-\tilde{b}^\prime,
\end{equation}
with
\begin{equation}
\tilde{b}^\prime=b^\prime+\epsilon(\tilde{e}C(\epsilon)-C(\epsilon)\tilde{c}),
\end{equation}
which is also in dlog-form, since $\tilde{e}$ and $\tilde{c}$ are in dlog-form. This argument establishes the freedom to add terms independent of the invariants to a solution of \eq{DDEQcopy2}. The following argument proves this to be the only possible relation between two solutions of \eq{DDEQcopy2}. Let $D_1$ and $D_2$ satisfy \eq{DDEQcopy2} for a given $\bar{b}$:
\begin{align}
\textup{d}D_1-\epsilon(\tilde{e}D_1-D_1\tilde{c})&=\bar{b}-b_1^\prime,\\
\textup{d}D_2-\epsilon(\tilde{e}D_2-D_2\tilde{c})&=\bar{b}-b_2^\prime.
\end{align}
Then the difference $\bar{D}=D_1-D_2$ satisfies
\begin{equation}
\textup{d}\bar{D}-\epsilon(\tilde{e}\bar{D}-\bar{D}\tilde{c})=b_2^\prime-b_1^\prime.
\end{equation}
Let $\hat{\bar{D}}=\bar{D}\epsilon^\tau$ be defined such that the expansion of $\hat{\bar{D}}$ starts at the constant order. The equation for $\hat{\bar{D}}$ reads
\begin{equation}
\label{DhatbarDEQ}
\textup{d}\hat{\bar{D}}-\epsilon(\tilde{e}\hat{\bar{D}}-\hat{\bar{D}}\tilde{c})=B,
\end{equation}
with $B=\epsilon^\tau(b_2^\prime-b_1^\prime)$, which is in dlog-form. The first order in the expansion of \eq{DhatbarDEQ} reads
\begin{equation}
\textup{d}\hat{\bar{D}}^{(0)}=\sum_{l=1}^NB_l^{(0)}\textup{d}\log(L_l),
\end{equation}
which integrates to
\begin{equation}	
\hat{\bar{D}}^{(0)}=\sum_{l=1}^NB_l^{(0)}\log(L_l)+\mathrm{const.}
\end{equation}
As $D_1$ and $D_2$ are assumed to be rational, $\hat{\bar{D}}$ has to be rational as well and therefore
\begin{equation}
B_l^{(0)}=0,\quad l=1,\dots,N,
\end{equation}
which implies that $\hat{\bar{D}}^{(0)}$ is constant. Consider the expansion of \eq{DhatbarDEQ} at some order $n>0$
\begin{equation}
\textup{d}\hat{\bar{D}}^{(n)}=(\tilde{e}\hat{\bar{D}}^{(n-1)}-\hat{\bar{D}}^{(n-1)}\tilde{c})+\sum_{l=1}^NB_l^{(n)}\textup{d}\log(L_l).
\end{equation}
The right-hand side is in dlog-form for constant $\hat{\bar{D}}^{(n-1)}$, and therefore $\hat{\bar{D}}^{(n)}$ can only be rational if it is constant as well. This proves by induction that $\hat{\bar{D}}$ is independent of the invariants. Consequently, the difference of the solutions $\bar{D}=\hat{\bar{D}}\epsilon^{-\tau}$ has to be independent of the invariants as well. Altogether, the argument establishes the uniqueness of a rational solution for $D$ of \eq{DDEQcopy2} up to the addition of terms that are independent of the invariants. This fact can be used in practice to fix this freedom without losing generality.

\subsection{Determination of the lowest order in the expansion of \texorpdfstring{$D$}{D}}

The general strategy to solve \eq{DDEQcopy2} for a rational $D$ is to expand \eq{DDEQcopy2} in the regulator and solve the resulting equations with a rational ansatz. The previous section has revealed that for any solution $D$ there is the freedom to add a term $C(\epsilon)$ independent of the invariants and the result $D+C(\epsilon)$ is still a solution of \eq{DDEQcopy2}. This freedom may be used to remove terms in the expansion of $D$ with trivial dependence on the invariants. Let $m_\textup{min}$ denote the lowest order in the expansion of $D$ with a non-trivial dependence on the invariants, which therefore cannot be removed with the choice of $C(\epsilon)$. Note that $m_\textup{min}$ is well defined since the rational solutions of \eq{DDEQcopy2} have been shown in the previous section to be unique up to the aforementioned freedom. The ansatz for the coefficients of $D$ can without loss of generality be restricted to orders higher than or equal to $m_\textup{min}$. In the following, a lower bound on $m_\textup{min}$ will be shown to be given by the lowest order in the expansion of $\bar{b}$. Let $D$ and $\bar{b}$ have the expansions
\begin{equation}
\label{Dmindef}
D=\sum_{m=m_{\textup{min}}}^\infty\epsilon^mD^{(m)},\quad D^{(m_\textup{min})}\neq \mathrm{const.},
\end{equation}
\begin{equation}
\label{lowerBoundExists}
\bar{b}=\sum_{n=n_{\textup{min}}}^\infty\epsilon^n\bar{b}^{(n)}.
\end{equation}
The bound $m_\textup{min}\geq n_\textup{min}$ can be proven by contradiction. To that end, assume for the moment $m_\textup{min}< n_\textup{min}$ and expand \eq{DDEQcopy2} in the regulator:
\begin{align}
\textup{d}D^{(n_\textup{min})}-(\tilde{e}D^{(n_\textup{min}-1)}-D^{(n_\textup{min}-1)}\tilde{c}) &= \bar{b}^{(n_\textup{min})}-b^{\prime(n_\textup{min})}\\
\textup{d}D^{(n_\textup{min}-1)}-(\tilde{e}D^{(n_\textup{min}-2)}-D^{(n_\textup{min}-2)}\tilde{c}) &=-b^{\prime(n_\textup{min}-1)}\\
& \vdotswithin{=} \\
\textup{d}D^{(m_\textup{min}+1)}-(\tilde{e}D^{(m_\textup{min})}-D^{(m_\textup{min})}\tilde{c}) &=-b^{\prime(m_\textup{min}+1)}\\
\textup{d}D^{(m_\textup{min})} &=-b^{\prime(m_\textup{min})}.
\end{align}
Integrating the last equation yields
\begin{equation}
D^{(m_\textup{min})}=-\sum_{l=1}^N B_l^{\prime(m_\textup{min})}\log(L_l)+\mathrm{const.},
\end{equation}
for constant matrices $B_l^{\prime(m_\textup{min})}$. Since $D$ is assumed to be rational, these matrices have to vanish, and therefore it follows $D^{(m_\textup{min})}=\mathrm{const.}$, which contradicts \eq{Dmindef}. Altogether, it has been established that $m_\textup{min}\geq n_\textup{min}$, which allows to start the expansion of the ansatz for $D$ at the order $n_\textup{min}$.

\subsection{Obtaining finite expansions}

The expansion of $D$ in the regulator may have infinitely many non-vanishing terms. Using ideas similar to those in \sec{sec:ExpTrafo}, it will be shown that \eq{DDEQcopy2} can be reformulated such that a solution for $D$ can be obtained by solving only finitely many differential equations. 

Since $D$ is assumed to be rational in $\epsilon$ and the invariants, a polynomial $f(\varset)$ has to exist such that $\check{D}\defeq Df$ has a finite $\epsilon$\nobreakdash-expansion. Similarly, there exists a polynomial $k(\varset)$ such that $\check{b}\defeq \bar{b}k$ has a finite $\epsilon$\nobreakdash-expansion as well. In order to fix $f$ and $k$ up to constant factors, both are required to only contain the minimal number of irreducible factors that are necessary to satisfy the aforementioned conditions. The products of all irreducible factors of $f$ and $k$ that are independent of the invariants are subsequently denoted by $\hat{f}$ and $\hat{k}$, respectively. Then their factorizations read
\begin{equation}
f(\varset)=\hat{f}(\epsilon)\prod_{i=1}^{N_f}\bar{f}_i(\varset),
\end{equation}
\begin{equation}
k(\varset)=\hat{k}(\epsilon)\prod_{i=1}^{N_k}\bar{k}_i(\varset).
\end{equation}
Furthermore, let $\gamma(\epsilon)$ be a polynomial with a minimal number of irreducible factors, such that $b^\prime\gamma$ has a finite expansion. Note that $\gamma(\epsilon)$ does not depend on the invariants since $b^\prime$ is in dlog-form.

For a given $\bar{b}$ it is straightforward to compute $k$, but as $D$ is not known in advance, $f$ cannot be calculated directly. Therefore, the relation between $f$ and $k$ has to be investigated. In order to do so, consider the \eq{DDEQcopy2} rewritten in terms of $\check{D}$ 
\begin{equation}
\label{finite1DDEQ}
\sum_{i=1}^{N_f}\frac{-k\gamma\check{D}\textup{d}\bar{f}_i}{\bar{f}_i}=-k\gamma\textup{d}\check{D}+\epsilon k\gamma (\tilde{e}\check{D}-\check{D}\tilde{c})+f\gamma\check{b}-f\gamma k b^\prime.
\end{equation}
The right-hand side obviously has a finite expansion, and thus also the left-hand side has to have a finite expansion. By similar arguments as in \sec{sec:relationfh}, each of the summands on the left-hand side has to have a finite expansion. Note that $\textup{d}\bar{f}_i/\bar{f}_i$ cannot be equal to a rational differential form with finite expansion due to an argument analogous to the one after \eq{dfargument}. Similarly, $\check{D}/\bar{f}_i$ cannot have a finite expansion due to the minimality of $f$. Since $\gamma$ does not depend on the invariants, it follows that each $\bar{f}_i$ is equal to some $\bar{k}_j$ and thus
\begin{equation}
\label{eq:fInk}
k=\hat{k}(\epsilon)p(\varset)\bar{f}(\varset),
\end{equation}
with $p(\varset)$ being a polynomial and $\bar{f}$ denoting the product of all irreducible factors of $f$ that depend on the invariants. By applying this relation to \eq{finite1DDEQ} and dividing by $\bar{f}$, the following equation is obtained
\begin{equation}
\label{2dnformDcheckEq}
\sum_{i=1}^{N_f}\frac{-\hat{k}p\gamma\check{D}\textup{d}\bar{f}_i}{\bar{f}_i}=-\hat{k}p\gamma\textup{d}\check{D}+\hat{k}p\gamma\epsilon (\tilde{e}\check{D}-\check{D}\tilde{c})+\hat{f}\gamma\check{b}-\hat{f} k \gamma b^\prime.
\end{equation}
The same argument as above leads to $p(\varset)=r(\varset)\bar{f}(\varset)$ for some polynomial $r(\varset)$. Combining this relation with \eq{eq:fInk}, it is evident that the product $\bar{k}$ of all irreducible factors of $k$ that depend on the invariants contains two powers of $\bar{f}$
\begin{equation}
\label{kbarfsq}
\bar{k}=r\bar{f}^2.
\end{equation}
In order to learn about $\hat{f}$, the above equation is applied to \eq{2dnformDcheckEq} and subsequently divided by $\hat{f}$
\begin{equation}
\frac{\hat{k}r\gamma(\bar{f}\textup{d}\check{D}-\check{D}\textup{d}\bar{f}-\epsilon\bar{f}(\tilde{e}\check{D}-\check{D}\tilde{c}))}{\hat{f}}=\gamma\check{b}-\gamma k b^\prime.
\end{equation}
The irreducible factors of $r(\varset)$ cannot be equal to irreducible factors of $\hat{f}(\epsilon)$ because they are not independent of the invariants. The other factors in the numerator can be a product of an irreducible factor of $\hat{f}$ and a quantity with finite expansion. Since only $\hat{k}$ is known before solving the equations, some irreducible factors of $\hat{f}$ remain unknown.

\subsection{Reformulation in terms of quantities with finite expansion}
Since $f$ cannot be used in practice, as it is not computable before solving for $D$, an alternative factor 
\begin{equation}
h(\varset)=\bar{h}(\varset)\hat{h}(\epsilon),
\end{equation}
will be defined such that the expansion of 
\begin{equation}
\hat{D}\defeq Dh,
\end{equation}
is finite. The minimality of $f$ implies that all irreducible factors of $f$ need to be irreducible factors of $h$ as well. The irreducible factors of $h$ that depend on both the invariants and $\epsilon$ can be defined by
\begin{equation}
\bar{h}(\varset)^2\defeq\bar{k}(\varset)s(\varset),
\end{equation}
where the polynomial $s(\varset)$ is required to have the minimal number of irreducible factors. By virtue of \eq{kbarfsq}, this definition ensures that $\bar{h}$ captures all irreducible factors of $\bar{f}$. 
As for the irreducible factors of $\hat{f}$, it is only known that some of them may be equal to irreducible factors of $\hat{k}$. The following definition incorporates all of these factors and leaves the missing factors to a factor $g(\epsilon)$ that has to be solved for 
\begin{equation}
\hat{h}(\epsilon)\defeq\hat{k}(\epsilon)g(\epsilon).
\end{equation}
Note that the minimality of $\hat{f}$ implies that $g(\epsilon)$ has a non-vanishing constant coefficient
\begin{equation}
g^{(0)}\neq 0.
\end{equation}

With the definitions $\hat{b}\defeq \bar{b}\bar{h}^2\hat{k}$ and $\hat{b}^\prime\defeq \hat{k}gb^\prime$, the differential equation \eq{DDEQcopy2} can be rewritten entirely in terms of quantities with finite $\epsilon$\nobreakdash-expansion
\begin{equation}
\label{FinalFormDDEQ}
-\textup{d}\bar{h}\hat{D}+\bar{h}\textup{d}\hat{D}-\epsilon\bar{h}(\tilde{e}\hat{D}-\hat{D}\tilde{c})=g(\epsilon)\hat{b}-\hat{b}^\prime\bar{h}^2.
\end{equation}
All quantities on the left-hand side have finite expansions by definition. The expansion of $\hat{b}$ must be finite as well because $\hat{b}=\check{b}s$ and the expansion of $\check{b}$ is finite by definition. Together, this implies that $\hat{b}^\prime\bar{h}^2$ must have a finite expansion. Since $\hat{b}^\prime$ is in dlog-form, only factors that are independent of the invariants can render its expansion infinite. However, these factors could not be compensated by $\bar{h}^2$, which is a product of irreducible factors depending on both $\epsilon$ and the invariants, and therefore $\hat{b}^\prime$ itself has to have a finite expansion. Thus, all quantities in \eq{FinalFormDDEQ} indeed have a finite expansion. 

Altogether, the procedure is as follows: First, $\bar{k}$ and $\hat{k}$ are computed from the given $\bar{b}$, which then allows to infer $\bar{h}$ and $\hat{b}$. Subsequently, \eq{FinalFormDDEQ} can be solved for $\hat{D}$, $g$ and $\hat{b}^\prime$ all of which have a finite expansion. Finally, a solution of \eq{DDEQcopy2} is obtained via $D=\hat{D}/(\bar{h}\hat{k}g)$.

\subsection{Expansion of the reformulated equation for \texorpdfstring{$t_{D}$}{tD}}

As already mentioned above, the strategy to solve \eq{FinalFormDDEQ} is to expand it in the regulator and solve the resulting equations with a rational ansatz. The Taylor series of the polynomials $\bar{h}$, $\hat{k}$ and $g$ all start with a non-vanishing constant coefficient due to their minimality. This implies that the expansions of $\hat{D}$, $\hat{b}$ and $\hat{b}^\prime$ start at the same orders as those of $D$, $\bar{b}$ and $b^\prime$
\begin{equation}
\hat{D}=\sum_{m=n_\textup{min}}^{m_\textup{max}}\epsilon^m\hat{D}^{(m)},\quad\hat{b}=\sum_{p=n_\textup{min}}^{p_\textup{max}}\epsilon^p\hat{b}^{(p)},\quad \hat{b}^\prime=\sum_{s=n_\textup{min}}^{s_\textup{max}}\epsilon^s\hat{b}^{\prime(s)}.
\end{equation}
Let $\bar{h}_\textup{max}$, $g_\textup{max}\in\mathbb{N}$ denote the highest non-vanishing order of the Taylor expansions of $\bar{h}$ and $g$, respectively. Expanding \eq{FinalFormDDEQ} in the regulator yields
\begin{equation}
\sum_{n=n_\textup{min}}^{E_\textup{max}}\epsilon^nE^{(n)}=0,
\end{equation}
with 
\begin{align}
E^{(n)}=\,\,&\sum_{k=0}^{\textup{min}(\bar{h}_\textup{max},\,\,n-n_\textup{min})}-\textup{d}\bar{h}^{(k)}\hat{D}^{(n-k)}+\bar{h}^{(k)}\textup{d}\hat{D}^{(n-k)}\\
\,\,&-\sum_{k=0}^{\textup{min}(\bar{h}_\textup{max},\,\,n-n_\textup{min}-1)}\bar{h}^{(k)}(\tilde{e}\hat{D}^{(n-k-1)}-\hat{D}^{(n-k-1)}\tilde{c})\\
\,\,&-\sum_{k=n_\textup{min}}^{\textup{min}(p_\textup{max},\,\,n)}\hat{b}^{(k)}g^{(n-k)}\\
\,\,&+\sum_{k=0}^{\textup{min}(2\bar{h}_\textup{max},\,\,n-n_\textup{min})}(\bar{h}^2)^{(k)}\hat{b}^{\prime(n-k)},
\end{align}
\begin{equation}
E_\textup{max}=\textup{max}(m_\textup{max}+\bar{h}_\textup{max}+1, p_\textup{max}+g_\textup{max}, 2\bar{h}_\textup{max}+s_\textup{max}).
\end{equation}
The rational ansatz for the coefficients $\hat{D}^{(n)}$ is chosen to be
\begin{equation}
\label{Dansatzreal}
\hat{D}^{(n)}=\sum_{k=1}^{|\mathcal{R}_D|}\delta_k^{(n)}r_k(\invariants),
\end{equation}
\begin{equation}
\mathcal{R}_D\defeq \left\{r_1(\invariants),\dots,r_{|\mathcal{R}_D|}(\invariants)\right\},
\end{equation}
where the $\delta_k^{(n)}$ are unknown matrices of the same dimensions as $\hat{D}$ that are independent of the regulator and the invariants. The choice of the set of rational functions $\mathcal{R}_D$ is discussed in detail in \sec{sec:ansatz}. Since the unknown coefficients of $\hat{b}^\prime$ are assumed to be in dlog-form, an ansatz of the following form can be made:
\begin{equation}
\label{ansatzbhatprime}
\hat{b}^{\prime(n)}=\sum_{l=1}^N\beta^{(n)}_l\textup{d}\log\left(L_l(\invariants)\right),
\end{equation}
where the $\beta^{(n)}_l$ denote unknown matrices of the same dimensions as $\hat{b}^\prime$ that are independent of the regulator and the invariants. The polynomials $L_l(\invariants)$ are taken from set of irreducible denominator factors of $a_I$, which is shown to encompass all possible cases in \sec{subsec:Ansatzepsform}. Since the constant coefficient $g^{(0)}$ of $g(\epsilon)$ is non-zero, \eq{FinalFormDDEQ} can be divided by $g^{(0)}$. Subsequently, this factor can be absorbed into the definitions of $\hat{D}$ and $\hat{b}^\prime$. Effectively, this amounts to setting $g^{(0)}=1$ without loss of generality. All higher Taylor coefficients of $g(\epsilon)$ are treated as unknown parameters. Inserting the ansatzes above into the equations $E^{(n)}=0$ and demanding the resulting equations to hold for all allowed values of the invariants implies linear equations in the parameters of the ansatzes. These equations are then solved order by order, starting at the lowest order $n=n_\textup{min}$. Since $E_\textup{max}$ is unknown until the solution is known, it is tested at each order $n$ whether $n=E_\textup{max}$. To this end, it is checked whether the parameter equations corresponding to
\begin{align}
\hat{D}^{(i)}&=0,\quad i=n-\bar{h}_\textup{max},\dots,n,\\
g^{(i)}&=0,\quad i=n-p_\textup{max}+1,\dots,n-n_\textup{min},\\
\hat{b}^{\prime(i)}&=0,\quad i=n-2\bar{h}_\textup{max}+1,\dots,n,
\end{align}
are satisfied. Once this test has been successful, \eq{FinalFormDDEQ} is satisfied to all orders upon setting the coefficients of $\hat{D}$, $g$ and $\hat{b}^\prime$ of all, still undetermined, higher orders to zero too. Then, the algorithm stops and returns $D=\hat{D}/(\bar{h}\hat{k}g)$.

\subsection{Determination of \texorpdfstring{$t_{g}$}{tg}}
\label{sec:dettg}
After $D$ and $b^\prime$ have been calculated by solving \eq{DFullDEQ}, the second part of the recursion step is to compute the transformation 
\begin{equation}
t_{g}=\left(\,
\begin{array}{|ccc|c|}
\cline{1-4}
 & & &\\
 &  g_p^{-1}(\epsilon)  & & 0 \\
 & & &\\ \hline
 & 0 & & g_{p+1}^{-1}(\epsilon) \\ \hline
\end{array}
\,\right),
\end{equation}
which is determined by
\begin{equation}
\label{tgdetermination}
g_p\tilde{c}g_{p}^{-1}=\tilde{c}^\prime,\quad g_{p+1}\tilde{e}g_{p+1}^{-1}=\tilde{e}^\prime, \quad b^\prime=g_{p+1}^{-1}\tilde{b}^\prime g_p.
\end{equation}
Applying the transformation $t_D$ to the intermediate differential equation $a^I$ in \eq{intermediateDEQ} yields
\begin{equation}
a^D(\varset)\defeq\left(\,
\begin{array}{|ccc|c|}
\cline{1-4}
 & & &\\
 & \epsilon\tilde{c}(\invariants)  & & 0 \\
 & & &\\ \hline
 &  b^\prime(\varset)  & & \epsilon\tilde{e}(\invariants) \\ \hline
\end{array}
\,\right),
\end{equation}
which is in dlog-form. The conditions in \eq{tgdetermination} are equivalent to demanding $t_g$ to transform $a^D$ into the canonical form $a^\prime$ in \eq{aprimeincanonical}. A procedure to calculate a transformation from a differential equation in dlog-form to a canonical form has been outlined in~\cite{Lee:2014ioa} and is reproduced in the following. Since $a^D$ is in dlog-form, it can be written as
\begin{equation}
a^D=\sum_{l=1}^Na^D_l(\epsilon)\textup{d}\log(L_l(\invariants)).
\end{equation}
Every transformation $V(\epsilon)$ that transforms $a^D$ into canonical form has to satisfy
\begin{equation}
\label{tepstoconst}
V(\epsilon)^{-1}\frac{a^D_l(\epsilon)}{\epsilon}V(\epsilon)=\tilde{o}_l,\quad l=1,\dots,N,
\end{equation}
for constant matrices $\tilde{o}_l$. A necessary condition for $V(\epsilon)$ to exist is that the eigenvalues of $a^D_l(\epsilon)/\epsilon$ are constant. The following argument shows that this is indeed the case. Each of the $a^D_l$ is again of the same block-triangular form as $a^D$. The determinant of a block-triangular matrix equals the product of the determinants of its diagonal blocks. This leads to a factorization of the characteristic polynomials of the $a^D_l$:
\begin{equation}
\det(a^D_l-\lambda\mathbb{I})=\det(\epsilon\tilde{c}_l-\lambda\mathbb{I})\det(\epsilon\tilde{e}_l-\lambda\mathbb{I}).
\end{equation}
In this form, it is obvious that the eigenvalues of $a^D_l(\epsilon)$ are proportional to $\epsilon$. Therefore, the eigenvalues of $a^D_l(\epsilon)/\epsilon$ must be constant. In order to actually calculate $V(\epsilon)$, \eq{tepstoconst} would have to be solved. Since the constant matrices on the right-hand side are unknown, the components of $V(\epsilon)$ cannot be solved for directly. However, as the right-hand side of \eq{tepstoconst} is manifestly independent of $\epsilon$, the following holds
\begin{align}
V(\epsilon)^{-1}\frac{a^D_l(\epsilon)}{\epsilon}V(\epsilon)&=V(\mu)^{-1}\frac{a^D_l(\mu)}{\mu}V(\mu)\\
\Leftrightarrow\frac{a^D_l(\epsilon)}{\epsilon}V(\epsilon)V(\mu)^{-1}&=V(\epsilon)V(\mu)^{-1}\frac{a^D_l(\mu)}{\mu}\\
\label{tgLS}
\Leftrightarrow\frac{a^D_l(\epsilon)}{\epsilon}V(\epsilon,\mu)&=V(\epsilon,\mu)\frac{a^D_l(\mu)}{\mu},
\end{align}
with $V(\epsilon,\mu)\defeq V(\epsilon)V(\mu)^{-1}$. In the last form, for each $l=1,\dots,N$ there is a linear equation for $V(\epsilon,\mu)$. This set of equations can now be solved for the components of $V(\epsilon,\mu)$ subject to the constraint that the block-triangular form is preserved. Finally, a constant $\mu_0$ needs to be chosen such that $t_{g}= V(\epsilon,\mu_0)$ is non-singular. It is straightforward to check that this $t_{g}$ does indeed transform $a^D$ into canonical form:
\begin{align}
t_{g}^{-1}a^D_l(\epsilon)t_{g}
&=V(\epsilon,\mu_0)^{-1} a^D_l(\epsilon)V(\epsilon,\mu_0)\\
&=\epsilon V(\mu_0)V(\epsilon)^{-1}\frac{a^D_l(\epsilon)}{\epsilon}V(\epsilon)V(\mu_0)^{-1}\\
&=\epsilon V(\mu_0)\tilde o_lV(\mu_0)^{-1}\\
&=\epsilon\tilde{A}_l^\prime.
\end{align}
Altogether, the above procedure provides a convenient method to compute $t_g$ by solving the linear system of equations in \eq{tgLS}.

\section{Ansatz in terms of rational functions}
\label{sec:ansatz}
In the previous sections, it was shown that the computation of a rational transformation to a canonical form is equivalent to finding a rational solution of finitely many differential equations in the invariants. Since these equations do in general admit transcendental solutions as well, a rational solution is conveniently obtained with a rational ansatz. This section will investigate the type of rational functions to be used in the ansatz and develop a procedure to compute a suitable set of rational functions from the given differential equation.

\subsection{Leinartas decomposition}
\label{sec:Leinartas}

The ansatzes proposed in the previous sections depend linearly on the parameters since this will translate linear differential equations into equations in the parameters that are linear again. This raises the question whether there is a subset of rational functions that is sufficient to express any other rational function as a linear combination. An answer will be given in this section by showing that any rational function can be decomposed as a linear combination of a certain \emph{simple} type of rational functions, which allows to restrict the ansatz to functions of this type.

In the univariate case, a partial fractions decomposition of the denominator polynomial allows to decompose rational functions as a linear combination of \emph{simpler} ones. However, in the multivariate case a naive generalization of partial fractioning may run into an infinite loop. This is illustrated by the following example
\begin{align}
\frac{1}{x(x+y)}&=\frac{1}{xy}-\frac{1}{y(x+y)}\\
&=\frac{1}{xy}-\left[\frac{1}{xy}-\frac{1}{x(x+y)}\right]=\frac{1}{x(x+y)}.
\end{align}
In the first equation, the partial fractions decomposition was applied with respect to $x$, and in the second equation it was applied with respect to $y$. Apparently, this naive procedure runs into a loop, which can be avoided by a more careful generalization of the partial fractioning procedure, as was outlined in \cite{Lei78, 2012arXiv1206.4740R}. In the following, a brief account of this \emph{Leinartas decomposition} method is given, based on the above references and~\cite{Cox:2007:IVA:1204670}. The focus will be on the computational aspects, and only those proofs will be shown that are relevant for the implementation of the decomposition. For the readers convenience, some definitions and standard results about polynomial rings that are used throughout this section are collected in \app{App:PolyAlgebra}.

\subsubsection*{Denominator decomposition}

Let $K[X]$ denote the ring of polynomials in $d$ variables $X=\{x_1,\dots,x_d\}$ with coefficients in a field $K$. Again, the cases $K=\mathbb{R}$ and $K=\mathbb{C}$ are the most relevant for the present application, but there is no need to specify the field for the following considerations.
\begin{defn}[Algebraic Independence]
A set of polynomials $\{f_1,\dots,f_m\}\subset K[X]$ is called algebraically independent if there exists no non-zero polynomial $\kappa$ in $m$ variables with coefficients in $K$ such that $\kappa(f_1,\dots,f_m)=0$ in $K[X]$. $\kappa$ is called annihilating polynomial.
\end{defn}
For the Leinartas decomposition, it is necessary to compute annihilating polynomials. Let $\{f_1,\dots,f_m\}\subset K[X]$ be a set of algebraically dependent polynomials and consider the ideal $I=\langle Y_1-f_1,\dots,Y_m-f_m\rangle\subseteq K[X,Y_1,\dots,Y_m]$. It is straightforward to check that the elements of the ideal $E=I\cap K[Y_1,\dots,Y_m]$  are annihilating polynomials. The following theorem provides a means to actually compute the elements of $E$.
\begin{thm}[Elimination Theorem]
Let $I\subset  K[X,Y_1,\dots,Y_m]$ be an ideal and $G$ be a Gr\"obner basis of $I$ with respect to lexicographic order with $X>Y_1>\dots >Y_m$. Then 
\begin{equation*}
G_Y=G\cap K[Y_1,\dots,Y_m]
\end{equation*}
is a Gr\"obner basis of the ideal $I\cap K[Y_1,\dots,Y_m]$.
\end{thm}
Thus, a Gr\"obner basis of $\langle Y_1-f_1,\dots,Y_m-f_m\rangle$ can be computed with standard algorithms \cite{GBBIB706, GBBIB699, Cox:2007:IVA:1204670}, and the intersection of this basis with $K[Y_1,\dots,Y_m]$ gives a Gr\"obner basis for $E$. Every element of this basis is an annihilating polynomial.
\begin{lemma}
Any set of polynomials $\{f_1,\dots,f_m\}\subset K[X]$ with $m>d$ is algebraically dependent.
\end{lemma}
\begin{lemma}
\label{lem:anypower}
A finite set of polynomials $\{f_1,\dots,f_m\}\subset K[X]$ is algebraically dependent if and only if for all positive integers $e_1,\dots,e_m$ the set of polynomials $\{f_1^{e_1},\dots,f_m^{e_m}\}$ is algebraically dependent. 
\end{lemma}
The following considerations rely on a corollary of Hilbert's weak Nullstellensatz:
\begin{corollary}[Nullstellensatz certificate]
\label{lem:NSCertificate}
A finite set of polynomials $\{f_1,\dots, f_m\}\subset K[X]$ has no common zero in $\overline{K}^d$ if and only if there exist polynomials $h_1,\dots,h_m\in K[X]$ such that 
\begin{equation*}
1=\sum_{i=1}^mh_if_i.
\end{equation*}
The set of polynomials $\{h_1,\dots,h_m\}$ is called a Nullstellensatz certificate. 
\end{corollary}
A Nullstellensatz certificate is said to have degree $k$ if 
\begin{equation}
\max\{\textup{deg}(h_i)\,\,|\,\,i=1,\dots,m\}=k.
\end{equation}
Algorithm \ref{alg:Nullstellensatz} is a simple but sufficiently fast way to compute a Nullstellensatz certificate for a set of polynomials with no common zero.
\begin{algorithm}
\DontPrintSemicolon
\label{alg:Nullstellensatz}
 \KwIn{$\{f_1,\dots,f_m\}$ with no common zero.}
 \KwOut{Nullstellensatz certificate $\{h_1,\dots,h_m\}$ such that $\sum_{i=1}^mh_if_i=1$.}
 $k=0$\\
\Do{}{
  $\sum_{i=1}^mh_if_i=1$ with the $h_i$ being polynomials of degree $k$ with unknowns as coefficients. Extract a linear system of equations from this relation and solve it.\\
  \eIf{solution exists}{
   \KwRet{certificate}
   }{
   $k=k+1$
  }
 }
 \bigskip
 \caption{Nullstellensatz certificate.}
\end{algorithm}

The Leinartas decomposition is based on the following theorem, which provides a generalization of the partial fractions decomposition to the multivariate case.
\begin{thm}[Leinartas]
Let $f=p/q$ be a rational function with $p,q\in K[X]$ and $q=q_1^{e_1}\dots q_m^{e_m}$ be the unique factorization of $q$ in $K[X]$ and $V_i=\{x\in \overline{K}^d\, |\, q_i(x)=0\}$. Then $f$ can be written in the following form:
\begin{equation*}
f=\sum_S\frac{p_S}{\prod_{i\in S}q_i^{b_i}}, \quad b_i\in \mathbb{N}\setminus\{0\}, \quad p_S\in K[X],
\end{equation*}
with the sum running over all subsets $S\subseteq \{1, \dots, m\}$ with $\cap_{i\in S}V_i\neq\emptyset$ and $\{q_i \,|\, i\in S\}$ being algebraically independent.\footnote{Throughout this thesis, the number zero is understood to be included in the natural numbers.}
\end{thm}
The proof of this theorem will be presented because it directly translates to an algorithm that decomposes rational functions into the above form. The decomposition can be separated into two consecutive steps. In the first step, a form is attained that satisfies $\cap_{i\in S}V_i\neq\emptyset$ for each summand. This step is called \textit{Nullstellensatz decomposition}. Let $f=p/q$ be a rational function. In the case $\cap_{i=1}^mV_i\neq\emptyset$, the Nullstellensatz decomposition is already complete. Thus, it remains to consider the case $\cap_{i=1}^mV_i=\emptyset$. As $q_i$ has the same zero-set as $q_i^{e_i}$, it follows that $\{q_1^{e_1},\dots,q_m^{e_m}\}$ has no common zero in $\overline{K}^d$. According to corollary \ref{lem:NSCertificate}, a Nullstellensatz certificate $1=\sum_{i=1}^mh_iq^{e_i}_i$ exists in this situation. Multiplying $f$ with this factor of one yields
\begin{equation}
f=\frac{p\sum_{i=1}^mh_iq^{e_i}_i}{q}=\sum_{i=1}^m\frac{ph_i}{q_1^{e_1}\cdots \widehat{q_i^{e_i}}\cdots q_m^{e_m}}.
\end{equation}
This step is applied repeatedly until the denominator factors of each term have a common zero. Note that this procedure will eventually stop since single irreducible factors always have a zero, i.e., $V_i\neq \emptyset$. In the second step, the goal is to achieve that $\{q_1,\dots,q_m\}$ is algebraically independent for each summand. Let $f=p/q$ be a summand of the Nullstellensatz decomposition. If $\{q_1,\dots,q_m\}$ is algebraically independent, then this term is already in the desired form. If this is not the case, the set $\{q_1^{e_1},\dots,q_m^{e_m}\}$ is also algebraically dependent by virtue of lemma \ref{lem:anypower}. Therefore, an annihilating polynomial $\kappa=\sum_{\nu\in S}c_\nu Y^\nu\in K[Y_1,\dots,Y_m]$ exists, which has been written in multi-index notation with $S\subset\mathbb{N}^m$. Let $\mu\in S$ refer to the powers of the monomial with the smallest norm $\|\mu\|=\sum_{i=1}^m\mu_i$. The annihilating polynomial vanishes on $Q=(q_1^{e_1},\dots,q_m^{e_m})$:
\begin{align}
\kappa(Q)&=0,\\
\Rightarrow \quad c_\mu Q^\mu&=-\sum_{\nu\in S\setminus\{\mu\}}c_\nu Q^\nu,\\
\Rightarrow \quad 1&=\frac{-\sum_{\nu\in S\setminus\{\mu\}}c_\nu Q^\nu}{c_\mu Q^\mu}.
\end{align}
This factor of one can be used to decompose $f$
\begin{equation}
f=\frac{p}{q}= \sum_{\nu\in S\setminus\{\mu\}}\frac{-pc_\nu Q^\nu}{c_\mu Q^{\mu+1}}=\sum_{\nu\in S\setminus\{\mu\}}\frac{-pc_\nu}{c_\mu}\prod_{i=1}^m\frac{q_i^{e_i\nu_i}}{q_i^{e_i(\mu_i+1)}}.
\end{equation}
As $\mu$ has the smallest norm in $S$, there has to exist some $j$ for each $\nu\in S$ such that $\mu_j+1\leq \nu_j$ and therefore $e_j(\mu_j+1)\leq e_j\nu_j$. So in each summand at least one factor in the denominator cancels. Again, this step is applied repeatedly to all summands whose denominator factors are algebraically dependent. Eventually, this procedure will stop, since a single irreducible factor is obviously algebraically independent. This completes the proof of the Leinartas theorem. Following this proof, a recursive algorithm can be built that computes the above decomposition of rational functions.

\subsubsection*{Numerator decomposition}

The Leinartas decomposition as presented in \cite{Lei78, 2012arXiv1206.4740R} leaves the numerator polynomial untouched. However, by employing multivariate polynomial division, the above decomposition can be extended to the numerator polynomial as well, which results in summands with simpler numerator polynomials. The precise meaning of \emph{simple} in this context will be stated below. 

Consider a summand $f=p/(q_1^{e_1}\dots q_m^{e_m})$ of the above decomposition, i.e., with $\cap_{i\in S}V_i\neq\emptyset$ and the $q_i$ being algebraically independent. The numerator polynomial $p$ can be decomposed according to the following theorem (cf.~\cite{Cox:2007:IVA:1204670}).
\begin{thm}[Division Algorithm]
\label{thm:DivAlg}
Fix some monomial ordering on $\mathbb{N}^d$ and let $(f_1,\dots,f_m)$ be an ordered m-tuple of polynomials in $K[X]$. Then every $p\in K[X]$ can be written as
\begin{equation*}
p=\beta_1f_1+\dots+\beta_mf_m+r,
\end{equation*}
with $\beta_1,\dots,\beta_m,r\in K[X]$ and either $r=0$ or $r$ is a linear combination of monomials with coefficients in $K$ such that no monomial is divisible by any of the $\textup{LT}(f_1),\dots,\textup{LT}(f_m)$. Moreover, for all $\beta_if_i\neq 0$ the following holds
\begin{equation*}
\textup{multideg}(p)\geq\textup{multideg}(\beta_if_i).
\end{equation*}
\end{thm}
It should be noted that the resulting decomposition depends on both the ordering of the $(f_1, \dots, f_m)$ and the monomial ordering. Let the ordered tuple of polynomials be given by the set of denominator polynomials $(q_1,\dots,q_m)$ and apply the above theorem to the numerator polynomial
\begin{equation}
p=\beta_1q_1+\dots+\beta_mq_m+r,
\end{equation}
to arrive at
\begin{equation}
\label{fNumDecomposition}
f=\frac{r}{q_1^{e_1}\dots q_m^{e_m}}+\sum_{i=1}^m\frac{\beta_i}{q_1^{e_1}\dots q_i^{e_i-1}\dots q_m^{e_m}}.
\end{equation}
The denominator factors of the resulting summands are still algebraically independent since every subset of an algebraically independent set of polynomials is algebraically independent. Moreover, every subset of a set of polynomials that share a common zero has a common zero as well. So after decomposing the numerator as above, the denominator polynomials of the resulting summands still have a common zero and are algebraically independent. Therefore, this decomposition can be applied recursively. The recursion stops at a summand whenever there is no monomial of the numerator polynomial that is divisible by the leading term of any of the denominator polynomials. It has to be shown that the recursion will always stop after a finite number of steps. For the first summand in \eq{fNumDecomposition} the recursion trivially stops. Concerning the other terms, it is sufficient to show that the multidegree strictly decreases
\begin{equation}
\label{eq:multidegStrictlydec}
\textup{multideg}(p)>\textup{multideg}(\beta_i)
\end{equation}
at each step, due to property \ref{def:monOrd:3} of definition \ref{def:monOrd} given in \app{App:PolyAlgebra}. Lemma \ref{lem:lowerboundmonoordering} implies $\textup{multideg}(q_i)\geq 0$ with respect to any monomial ordering. However, the case $\textup{multideg}(q_i)=0$ cannot occur, since it implies $q_i=\mathrm{const.}$ Thus, $\textup{multideg}(q_i)$ is strictly greater than zero. Using property \ref{def:monOrd:2} of definition \ref{def:monOrd} and lemma \ref{lem:multidegProperties} it follows
\begin{equation}
\textup{multideg}(\beta_iq_i)=\textup{multideg}(q_i)+\textup{multideg}(\beta_i)>\textup{multideg}(\beta_i).
\end{equation}
Theorem \ref{thm:DivAlg} implies $\textup{multideg(p)}\geq\textup{multideg}(\beta_iq_i)$, which together with the above inequality proves \eq{eq:multidegStrictlydec}. This completes the decomposition of the numerator polynomial. 

The terms in such a decomposition are not necessarily linearly independent over $K$, as the following example illustrates
\begin{equation}
\frac{1}{x+y}+\frac{y}{x(x+y)}-\frac{1}{x}=0.
\end{equation}
In the last step, this redundancy is removed by eliminating all such relations from the set of summands. Altogether, it has been demonstrated that every multivariate rational function can be decomposed into $K$-linearly independent summands such that the denominator polynomials of each summand share a common zero and are algebraically independent, and the numerator polynomial is not divisible by the leading term of any of its denominator polynomials. In the following, the individual summands of this decomposition are referred to as \emph{Leinartas functions}.

\subsection{Ansatz for diagonal blocks}
\label{subsec:AnsatzDB}
In this section, the choice of the set $\mathcal{R}_T$ of rational functions used in the ansatz for diagonal blocks is discussed. The basic compromise with the ansatz is to choose it large enough to encompass the solution and as small as possible to keep the resulting number of equations small and therefore allow the algorithm to perform well. The goal of this section is to present a procedure to generate a finite set of rational functions for a given differential form $a(\varset)$, which can then be used as an ansatz. The first step towards this goal is to determine the set of possible denominator factors of the transformation to a canonical basis. A natural guess is to consider the set of irreducible denominator factors of $\hat{a}$, which is proven in the following to contain all possible factors.

It is useful to first define some notation. Let $f(\invariants)$ be an irreducible polynomial and $S(\varset)$ some matrix-valued rational differential form or function. Then, the notation 
\begin{equation}
S\sim\frac{1}{f^n}
\end{equation}
indicates $n\in\mathbb{N}$ to be the maximal number for which $S$ can be written as
\begin{equation}
S=R\frac{1}{f^n}.
\end{equation}
Here $R$ is required to be non-zero and not to be the product of $f$ and a quantity, which is finite on the set of all zeros of $f$. The set $\mathcal{I}(S)$ of irreducible denominator factors of $S$ is then given by those factors $f$ with $S\sim1/f^k$ and $k\geq 1$.

\begin{claim}
\label{claim:factorsT}
Each irreducible denominator factor $f(\invariants)$ of a rational solution $\hat{T}$ of \eq{DEQfiniteh} is an irreducible denominator factor of $\hat{a}$.
\end{claim}
The assertion in the claim is equivalent to $\mathcal{I}(\hat{T})\subseteq\mathcal{I}(\hat{a})$, which will be proven by showing that
\begin{equation}
\label{Thatdenfactor}
\hat{T}\sim\frac{1}{f^n}, \quad n\geq 1
\end{equation}
implies 
\begin{equation}
\hat{a}\sim\frac{1}{f^k},\quad k\geq 1.
\end{equation}
To this end, \eq{Thatdenfactor} is assumed to hold. It is instructive to rearrange the terms in \eq{DEQfiniteh}
\begin{equation}
\label{rearrangedTL}
-\hat{T}\textup{d}h+h(\textup{d}\hat{T}+\epsilon \hat{T}\textup{d}\tilde{A})=\hat{a}\hat{T}.
\end{equation}
Since $\textup{d}h$ is polynomial, the term $\hat{T}\textup{d}h$ behaves as
\begin{equation}
\label{Tdhorder}
\hat{T}\textup{d}h\sim \frac{1}{f^k},\quad k\leq n.
\end{equation}
Given \eq{Thatdenfactor}, there must be a lowest order $s\in\mathbb{Z}$ in the $\epsilon$\nobreakdash-expansion of $\hat{T}$ with
\begin{equation}
\hat{T}^{(s)}\sim\frac{1}{f^n},
\end{equation}
and consequently
\begin{equation}
\label{thatleqn}
\hat{T}^{(s-1)}\sim\frac{1}{f^k},\quad k<n.
\end{equation}
The derivative raises the power of $f$ by one, which implies
\begin{equation}
\textup{d}\hat{T}^{(s)}\sim\frac{1}{f^{n+1}}.
\end{equation}
Note that $\textup{d}\tilde{A}$ is in dlog-form, and therefore 
\begin{equation}
\label{aTildeleq1}
\textup{d}\tilde{A}\sim\frac{1}{f^k},\quad k\leq 1.
\end{equation}
Taking both \eq{thatleqn} and \eq{aTildeleq1} into account, it follows
\begin{equation}
\hat{T}^{(s-1)}\textup{d}\tilde{A}\sim\frac{1}{f^k},\quad k\leq n.
\end{equation}
This implies for the order $s$ of the expansion of $(\textup{d}\hat{T}+\epsilon \hat{T}\textup{d}\tilde{A})$
\begin{equation}
\textup{d}\hat{T}^{(s)}+\hat{T}^{(s-1)}\textup{d}\tilde{A}\sim\frac{1}{f^{n+1}},
\end{equation}
which also holds for the full expression $(\textup{d}\hat{T}+\epsilon \hat{T}\textup{d}\tilde{A})$. Due to the minimality requirement, $h$ does not contain any irreducible factors independent of $\epsilon$, and therefore the multiplication of the term in brackets with $h$ in \eq{rearrangedTL} cannot cancel any power of $f$, because $f$ is independent of $\epsilon$. Since \eq{Tdhorder} shows that the term $\hat{T}\textup{d}h$ is of lower order in $f$ than $h(\textup{d}\hat{T}+\epsilon \hat{T}\textup{d}\tilde{A})$, the whole left-hand side of \eq{rearrangedTL} is of order $1/f^{n+1}$ and consequently the right-hand side as well:
\begin{equation}
\hat{a}\hat{T}\sim\frac{1}{f^{n+1}}.
\end{equation}
Since $\hat{T}$ is only of order $1/f^n$, it can be concluded  
\begin{equation}
\hat{a}\sim\frac{1}{f^k},\quad k\geq 1,
\end{equation}
which proves the claim. Thus, the ansatz can without loss of generality be restricted to the set 
\begin{equation}
\mathcal{Q}=\left\{\frac{x_1^{p_1}\cdots x_M^{p_M}}{f_1^{q_1}\dots f_U^{q_U}}\,\,\,\Bigg|\,\,\, p_1,\dots, p_M, q_1,\dots, q_U \in \mathbb{N}\right\}
\end{equation}
of rational functions with the denominator factors drawn from the set $\mathcal{I}(\hat{a})=\{f_1,\dots,f_U\}$ of irreducible denominator factors of $\hat{a}$.

As was argued in \sec{sec:Leinartas}, rational functions may be decomposed in terms of Leinartas functions. Let $\mathcal{L}(\mathcal{Q})$ denote a basis of the $K$-span of $\mathcal{Q}$ in terms of Leinartas functions. While $\mathcal{L}(\mathcal{Q})$ is guaranteed to contain the correct ansatz, it is still an infinite set. Therefore, a constructive procedure is needed to generate a finite subset of $\mathcal{L}(\mathcal{Q})$ for a given $a(\varset)$. This procedure should be inexpensive to compute while yielding a correct ansatz for most practical examples. Since the procedure outlined in the following is not proven to generate a correct ansatz, it is important to be able to systematically enlarge the ansatz in a way that is guaranteed to eventually encompass the solution.

The strategy to define a finite subset of $\mathcal{L}(\mathcal{Q})$ is to set restrictions on the powers of the invariants in the numerator as well as on the powers of the denominator factors.

While the powers of those factors occurring in $a(\varset)$ may be suspected to be a good indicator for the powers in the transformation, the following simple example demonstrates this to be false. Consider the differential form 
\begin{equation}
a(\epsilon,\{x\})=\left(-\frac{\alpha}{x}+\frac{\epsilon}{x}\right)\textup{d}x,\quad \alpha\in\mathbb{Z},
\end{equation}
which contains the factor $x$ with the negative power one. However, for any given integer $\alpha$, the rational transformation to the canonical form reads
\begin{equation}
T(\epsilon,\{x\})=\frac{1}{x^\alpha}.
\end{equation}
Consequently, the transformation can contain any power of the factor $x$, while the power of the same factor in the differential equation remains fixed. A much better predictor is given by the determinant of the transformation, which in the one-dimensional example above is identical to the transformation itself and therefore always yields the correct power of the factor $x$. For higher-dimensional differential equations, the determinant does not fix the transformation but still carries information on the powers of the irreducible denominator factors of the transformation. Let the determinant of the transformation read
\begin{equation}
\det(T)=F(\varset)\prod_{i=1}^U f_i^{-\lambda_i}, \quad \lambda_i\in\mathbb{Z},\,\, f_i\in\mathcal{I}(\hat{a}),
\end{equation}
where $F(\varset)$ denotes the product of all irreducible factors with a non-trivial dependence on the regulator. Then, for each factor $f_i$ with $\lambda_i>0$, there has to be a component $T_{jl}$ of the transformation satisfying
\begin{equation}
\label{lowerBoundsfi}
T_{jl}\sim \frac{1}{f_i^k}, \quad k\geq \left\lceil{\frac{\lambda_i}{m}}\right\rceil,
\end{equation}
where $m$ denotes the dimension of the differential equation. Thus, the determinant sets lower bounds on the maximal powers of the denominator factors in the transformation, which have to be taken into account in the construction of the ansatz.

In the following, a finite subset of $\mathcal{Q}$ will be constructed, which then leads to a finite subset of $\mathcal{L}(\mathcal{Q})$ by taking a basis of its $K$-span in terms of Leinartas functions. The powers $\lambda_i$ are used to define a set of denominators
\begin{equation}
\mathcal{D}(\delta_D)=\left\{\frac{1}{f^{p_{i_1}}_{i_1}\cdots f^{p_{i_M}}_{i_M}} \,\,\Bigg|\,\, f_{i_j}\in\mathcal{I}(\hat{a}), 0\leq p_{i}\leq\Theta(\lambda_{i})\lambda_i+\delta_D, i_j \neq i_k\,\, \text{for}\,\, j \neq k\right\},
\end{equation}
which has been restricted to at most $M$ denominator factors with $M$ denoting the number of invariants. Any higher number of polynomials in $M$ invariants is algebraically dependent and therefore reducible in terms of Leinartas functions with $M$ or fewer denominator polynomials. The parameter $\delta_D\in\mathbb{N}$ has been introduced to define a way to enlarge the set of $\mathcal{D}(\delta_D)$ systematically. The default value is going to be $\delta_D=0$. The lower bounds in \eq{lowerBoundsfi} are satisfied for all allowed values of $\delta_D$. For the numerators, consider the set of all possible monomials up to a fixed bound on their total degree
\begin{equation}
\mathcal{N}(\delta_N)=\left\{x_1^{\nu_1}\cdots x_M^{\nu_M} \,\,\,\Bigg|\,\,\, \nu_1,\dots,\nu_M\in\mathbb{N},\,\,\, \sum_{i=1}^M\nu_i\leq 3+\delta_N\right\},
\end{equation}
where the parameter $\delta_N$ has been introduced to control the highest total degree of the monomials in $\mathcal{N}(\delta_N)$. For the default value $\delta_N=0$, the highest total degree of the numerator monomials is three. This choice is made based on practical examples and is intended to make the default value $\delta_N=0$ work for most cases and at the same time yield a rather small ansatz. Furthermore, it has proven useful to also include the following sets of monomials
\begin{equation}
\mathcal{N}_\textup{det}=\left\{\text{numerator monomials of}\,\, \det(T) \right\},
\end{equation}
\begin{equation}
\mathcal{N}_a=\left\{ \text{numerator monomials of the}\,\,  \hat{a}^{(k)}(\{x_j\}) \right\},
\end{equation}
in order to capture the correct ansatz in more cases already with the default value $\delta_N=0$. Usually, the inclusion of $\mathcal{N}_\textup{det}$ and $\mathcal{N}_a$ does not significantly enlarge the ansatz, while making the default value work for more examples. Finally, the ansatz $\mathcal{R}_T$ is obtained by computing a basis of Leinartas functions of the $K$-span of the set of rational functions drawing their numerators and denominators from the sets defined above:
\begin{equation}
\mathcal{R}_T(\delta_D,\delta_N)=\mathcal{L}\left(\left\{ \frac{p}{f}  \,\,\,\Bigg|\,\,\, f\in \mathcal{D}(\delta_D),\,\,\, p\in \mathcal{N}(\delta_N)\cup \mathcal{N}_a\cup\mathcal{N}_\textup{det}  \right\}\right).
\end{equation}
The set $\mathcal{R}_T(\delta_D,\delta_N)$ is finite and contains all elements of $\mathcal{L}(\mathcal{Q})$ necessary to represent the elements of $\mathcal{Q}$ with denominators from $\mathcal{D}(\delta_D)$ and numerators from $\mathcal{N}(\delta_N)$. Therefore, by increasing the values of $\delta_D$ and $\delta_N$, the set $\mathcal{R}_T(\delta_D,\delta_N)$ can be systematically extended to the whole set of $\mathcal{L}(\mathcal{Q})$, which contains the correct ansatz. While the correct ansatz is necessarily contained in $\mathcal{L}(\mathcal{Q})$, the choice of the finite subset $\mathcal{R}_T\subset\mathcal{L}(\mathcal{Q})$ presented here is a heuristic procedure. However, the knowledge of upper bounds on $\delta_D$ and $\delta_N$ would be enough to turn the algorithm into a computable criterion for the existence of a rational transformation that transforms a given differential equation into a canonical form.

\subsection{Ansatz for the resulting canonical form}
\label{subsec:Ansatzepsform}
The ansatz for the resulting canonical form in \eq{ansatzAtilde} requires the knowledge of a set of polynomials in the invariants that encompasses the set of letters of the resulting canonical form. In this section, these letters will be shown to be a subset of the set $\mathcal{I}(a)$ of irreducible denominator factors of the original differential equation with trivial dependence on the regulator. This result immediately implies a similar statement for the ansatz in \eq{ansatzbhatprime} used in the calculation of the remaining transformation in \sec{sec:OffDiagPart}.

Consider the transformation law in \eq{aprimeINeps}:
\begin{equation}
\label{trafoLawbracket}
\epsilon\,\textup{d}\tilde{A}=T^{-1}(aT-\textup{d}T).
\end{equation}
Since the transformation $T$ is rational, the derivative does not alter the set of its denominator factors, i.e., $\mathcal{I}(\textup{d}T)=\mathcal{I}(T)$. Claim \ref{claim:factorsT} implies $\mathcal{I}(T)\subseteq\mathcal{I}(a)$ and thus
\begin{equation}
\mathcal{I}(aT-\textup{d}T)\subseteq\mathcal{I}(a).
\end{equation}
The denominator factors of $T^{-1}$ can be deduced by writing the inverse as
\begin{equation}
T^{-1}=\det(T)^{-1}\textup{adj}(T).
\end{equation}
The cofactors in the adjugate of $T$ are a sum of products of components of $T$, which implies 
\begin{equation}
\mathcal{I}(\textup{adj}(T))\subseteq\mathcal{I}(T)\subseteq\mathcal{I}(a).
\end{equation}
According to \eq{detFactorization}, the determinant of $T$ can be written in the form
\begin{equation}
\det(T)=F(\varset)\prod_{i=1}^U f_i(\invariants)^{-\lambda_i}, \quad \lambda_i\in\mathbb{Z},\,\,\, f_i\in\mathcal{I}(a),
\end{equation}
which leads to 
\begin{equation}
\mathcal{I}(\det(T)^{-1})\subseteq\mathcal{I}(a).
\end{equation}
Thus, the irreducible denominator factors of the right-hand side of \eq{trafoLawbracket} have been shown to be a subset of $\mathcal{I}(a)$, hence the same holds for those of the left-hand side:
\begin{equation}
\mathcal{I}(\textup{d}\tilde{A})\subseteq\mathcal{I}(a).
\end{equation}
Since $\mathcal{I}(\textup{d}\tilde{A})$ is equal to the set of letters of $\tilde{A}$, this allows to restrict the set of polynomials in the ansatz in \eq{ansatzAtilde} to the set $\mathcal{I}(a)$. 

The ansatz in \eq{ansatzbhatprime} for the dlog-form achieved by the transformation $t_D$ also requires the choice of a set of letters. The set of letters occurring in this dlog-form and in particular in $\hat{b}^\prime$ cannot be changed by the subsequent transformation $t_g(\epsilon)$, and therefore $\mathcal{I}(\hat{b}^\prime)\subseteq\mathcal{I}(\textup{d}\tilde{A})$ must hold. Applying the above argument to this situation then leads to $\mathcal{I}(\hat{b}^\prime)\subseteq\mathcal{I}(a^I)$. Thus, the letters in the ansatz may safely be restricted to $\mathcal{I}(a^I)$.

\subsection{Ansatz for off-diagonal blocks}
\label{subsec:AnsatzOD}
The transformation $t_D$, which transforms the off-diagonal blocks into dlog-form, is determined by \eq{DFullDEQ}. A rational solution of this equation for $D$ has been shown to be computable by solving the reformulated \eq{FinalFormDDEQ} for $\hat{D}$ by making a rational ansatz in \eq{Dansatzreal} in terms of Leinartas functions. The goal of this section is to develop a procedure to construct a set of Leinartas functions $\mathcal{R}_D$ to be used as an ansatz. First, the set of irreducible denominator factors of $D$ will be shown to be a subset of the irreducible denominator factors of $b$. Here and in the following, these factors are assumed not to depend on the regulator unless stated otherwise. The argument proceeds similarly to the one in the proof of claim \ref{claim:factorsT}. In this case, it will even be possible to derive upper bounds on the powers of the irreducible denominator factors of $D$. In a second step, these global upper bounds will be refined to upper bounds for the individual components of $D$, which reduces the number of rational functions in the ansatz considerably. Note that the bounds derived for $D$ apply to $\hat{D}$ as well.

The set of possible irreducible denominator factors occurring in a rational solution $D$ of 
\begin{equation}
\label{DDEQcopy}
\textup{d}D-\epsilon(\tilde{e}D-D\tilde{c})=b-b^\prime
\end{equation}
can be determined from the denominator factors of $b$. In order to demonstrate this, assume
\begin{equation}
D\sim\frac{1}{f^n},\quad n\geq 1,
\end{equation}
where the same notation as in \sec{subsec:AnsatzDB} is used. Then, there exists a lowest order $s\in\mathbb{Z}$ in the expansion of $D$ with
\begin{equation}
D^{(s)}\sim\frac{1}{f^n},
\end{equation}
and therefore 
\begin{equation}
D^{(s-1)}\sim\frac{1}{f^k},\quad k<n.
\end{equation}
The derivative raises the order of $f$ by one:
\begin{equation}
\textup{d}D^{(s)}\sim\frac{1}{f^{n+1}}.
\end{equation}
Since $\tilde{e}$ and $\tilde{c}$ are in dlog-form and thus at most of order $1/f$, it follows
\begin{equation}
\textup{d}D^{(s)}-(\tilde{e}D^{(s-1)}-D^{(s-1)}\tilde{c})\sim\frac{1}{f^{n+1}},
\end{equation}
which in turn implies the left-hand side and therefore also the right-hand side of \eq{DDEQcopy} to be of order $1/f^{n+1}$:
\begin{equation}
b-b^\prime\sim\frac{1}{f^{n+1}}.
\end{equation}
As $b^\prime$ is in dlog-form and consequently at most of order $1/f$, it can be concluded 
\begin{equation}
b\sim\frac{1}{f^{n+1}}.
\end{equation}
This result allows to extract upper bounds on the order of the irreducible denominator factors of $D$ from a given $b$. Let $\mathcal{I}(b)=\{f_1,\dots,f_U\}$ denote the set of irreducible denominator factors of $b$ and $\lambda_i$ the order of the denominator factor $f_i$
\begin{equation}
b\sim\frac{1}{f_i^{\lambda_i}}, \quad i=1,\dots,U.
\end{equation}
According to the argument above, the upper bounds $\mu_i$ 
\begin{equation}
D\sim\frac{1}{f^{k_i}_i}, \quad 0\leq k_i \leq \mu_i, \quad i=1,\dots,U,
\end{equation}
are given by
\begin{equation}
\label{globalUpperBounds}
\mu_i=\lambda_i-1,\quad i=1,\dots, U.
\end{equation}
Rather than using these bounds to make an ansatz, it is beneficial to reduce the combinatorics of the ansatz by refining the above bounds. The idea is to infer bounds on the powers of the denominator factors of individual components of the solution $D$ rather than for all components at once. To this end, assume 
\begin{equation}
D_{ij}\sim\frac{1}{f^n}, \quad n\geq 1,
\end{equation}
and let $s\in\mathbb{Z}$ denote the lowest order in the expansion of $D_{ij}$ with
\begin{equation}
D^{(s)}_{ij}\sim\frac{1}{f^n}.
\end{equation}
The derivative raises the order by one:
\begin{equation}
\textup{d}D^{(s)}_{ij}\sim\frac{1}{f^{n+1}}.
\end{equation}
Consider a component of the order $s$ in the expansion of \eq{DDEQcopy}
\begin{equation}
\textup{d}D_{ij}^{(s)}-(\tilde{e}D^{(s-1)}-D^{(s-1)}\tilde{c})_{ij}=b^{(s)}_{ij}-b^{\prime(s)}_{ij}.
\end{equation}
Since $b^\prime$ is in dlog-form, this term cannot cancel the order $1/f^{n+1}$ of the derivative term. Therefore at least one of the following cases must be true:
\newline
\\\textbf{case 1:}
\begin{equation}
b^{(s)}_{ij}\sim\frac{1}{f^{k}},\quad k\geq n+1,
\end{equation}
\\\textbf{case 2:}
\begin{equation}
(\tilde{e}D^{(s-1)}-D^{(s-1)}\tilde{c})_{ij}\sim\frac{1}{f^{k}},\quad k\geq n+1.
\end{equation}
In case 2, there has to be at least one index $\alpha$ with either
\begin{equation}
\label{eialphaone}
\tilde{e}_{i\alpha}\sim\frac{1}{f^1}\quad \text{and}\quad D_{\alpha j}^{(s-1)}\sim\frac{1}{f^k},\quad k\geq n
\end{equation}
or 
\begin{equation}
\label{eialphazero}
\tilde{e}_{i\alpha}\sim\frac{1}{f^0}\quad \text{and}\quad D_{\alpha j}^{(s-1)}\sim\frac{1}{f^{k}},\quad k\geq n+1,
\end{equation}
or there exists at least one index $\beta$ with either
\begin{equation}
\label{cbetajone}
\tilde{c}_{\beta j}\sim\frac{1}{f^1}\quad \text{and}\quad D_{i\beta}^{(s-1)}\sim\frac{1}{f^k},\quad k\geq n
\end{equation}
or 
\begin{equation}
\label{cbetajzero}
\tilde{c}_{\beta j}\sim\frac{1}{f^0}\quad \text{and}\quad D_{i\beta}^{(s-1)}\sim\frac{1}{f^{k}},\quad k\geq n+1.
\end{equation}
So far, the assumption $D^{(s)}_{ij}\sim 1/f^n$ has been demonstrated to imply either $b^{(s)}_{ij}\sim 1/f^{k}$ for some $k\geq n+1$ (case 1) or that some other component of $D^{(s-1)}$ is of order $1/f^k$ with $k\geq n$ (case 2). In case 2, the whole argument can be applied again to the respective components of $D^{(s-1)}$. This can be repeated until either the lowest order in the expansion is reached and therefore case 2 is not possible anymore or at some point only case 1 is possible due to the structure of $\tilde{e}$ and $\tilde{c}$. Thus, all possible chains of this argument necessarily end with case 1. Since $\tilde{e}$, $\tilde{c}$ and $b$ are known prior to the computation of $D$, the chains can be followed backwards in order to derive upper bounds on the powers of the denominator factors of the components of $D$. The idea is to consider all chains at once and start at the last step by reversing case 1 for all components of $D$. Using the powers of the denominator factors of $b$
\begin{equation}
b_{ij}\sim\frac{1}{f^{\lambda_{k,ij}}_{k}},
\end{equation}
case 1 is reversed for all components by setting the upper bounds $\mu_{k,ij}$ of $D_{ij}$ on $f_k$
\begin{equation}
D_{ij}\sim\frac{1}{f_k^p},\quad 0\leq p\leq \mu_{k,ij},
\end{equation}
to
\begin{equation}
\mu_{k,ij}=\lambda_{k,ij}-1, \quad \forall k,i,j.
\end{equation}
It can then be deduced from $\tilde{e}$ and $\tilde{c}$ for each component which other components could have implied the current bounds via case 2. For instance, if there exists an $\alpha$ with 
\begin{equation}
\tilde{e}_{i\alpha}\sim\frac{1}{f^1},
\end{equation}
and the current bound for the order in $1/f$ of $D_{\alpha j}$ is $n$, case 2 is reversed by setting the bound on the order of $D_{ij}$ to $n$ as well unless it is already higher. At each step it is checked for all components $D_{ij}$, whether there is an $\tilde{e}_{i\alpha}$ as in \eq{eialphaone} or \eq{eialphazero} or a $\tilde{c}_{\beta j}$ as in \eq{cbetajone} or \eq{cbetajzero}. If this is the case, the bounds are updated accordingly. This is repeated until the bounds do not change anymore, and therefore they incorporate all possible cases. Algorithm \ref{alg:upperBounds} summarizes the procedure.
\begin{algorithm}[h!]
\DontPrintSemicolon
\label{alg:upperBounds}
 \KwIn{$\{\lambda_{k,ij}\}$, $\tilde{e}$, $\tilde{c}$.}
 \KwOut{Set of upper bounds $\mu_{k,ij}$ with $D_{ij}\sim 1/f^k$ and $0\leq k\leq \mu_{k,ij}$.}
 $\mu_{k,ij}=\lambda_{k,ij}-1$\\
\Repeat{bounds $\mu$ do not change anymore}{
\ForEach{$k$, $i$, $j$, $\alpha$, $\beta$}{
\lIf{ $\tilde{e}_{i\alpha}\sim\frac{1}{f^0}$}{$\mu_{k,ij}=\textup{max}\left(\mu_{k,ij},\mu_{k,\alpha j}-1\right)$}
\lIf{ $\tilde{e}_{i\alpha}\sim\frac{1}{f^1}$}{$\mu_{k,ij}=\textup{max}\left(\mu_{k,ij},\mu_{k,\alpha j}\right)$}
\lIf{ $\tilde{c}_{\beta j}\sim\frac{1}{f^0}$}{$\mu_{k,ij}=\textup{max}\left(\mu_{k,ij},\mu_{k,i \beta}-1\right)$}
\lIf{ $\tilde{c}_{\beta j}\sim\frac{1}{f^1}$}{$\mu_{k,ij}=\textup{max}\left(\mu_{k,ij},\mu_{k,i \beta}\right)$}
}
}
\Return $\{\mu_{k,ij}\}$\;
\bigskip
\caption{Determination of upper bounds on the powers of the denominator factors of the components of $D$.}
\end{algorithm}
Since the values of the bounds $\mu_{k,ij}$ can only increase during each iteration in algorithm \ref{alg:upperBounds} and the overall bounds $\mu_k$ given in \eq{globalUpperBounds} are upper bounds on the bounds of the components
\begin{equation}
\mu_{k,ij}\leq \mu_k,\quad \forall k,i,j,
\end{equation}
it is clear that the algorithm terminates after a finite number of steps. The bounds computed with algorithm \ref{alg:upperBounds} hold for $\hat{D}=\bar{h}\hat{k}gD$ as well, since $\bar{h}$, $\hat{k}$ and $g$ do not contain any irreducible factors with trivial dependence on the regulator. To obtain an ansatz for the coefficients of $\hat{D}$ define
\begin{equation}
\mathcal{R}_{ij}(\delta_N)=\left\{ \frac{p}{f_1^{q_1}\dots f_U^{q_U}} \,\,\,\Bigg|\,\,\,p\in\mathcal{N}(\delta_N),\,0\leq q_k\leq\mu_{k,ij}\,\, \forall k\right\},
\end{equation}
with
\begin{equation}
\mathcal{N}(\delta_N)=\left\{x_1^{\nu_1}\cdots x_M^{\nu_M} \,\,\,\Bigg|\,\,\, \nu_1,\dots,\nu_M\in\mathbb{N},\,\,\, \sum_{i=1}^M\nu_i\leq 3+\delta_N\right\}.
\end{equation}
The above argument shows that for high enough $\delta_N$, the coefficients in the expansion of the component $\hat{D}_{ij}$ are an element of the $K$-span of $\mathcal{R}_{ij}(\delta_N)$. For the ansatz, a basis of Leinartas functions of the $K$-span of the union of all $\mathcal{R}_{ij}(\delta_N)$ is taken:
\begin{equation}
\mathcal{R}_D(\delta_N)=\mathcal{L}\left(\bigcup_{i,j}\mathcal{R}_{ij}(\delta_N)\right).
\end{equation}
It would be more efficient to make a different ansatz for each component of $\hat{D}$ using $\mathcal{L}(\mathcal{R}_{ij}(\delta_N))$. However, this functionality has not been implemented yet.

       \chapter{The \textit{CANONICA} package}
\label{chap:package}

The algorithm presented in \chap{chap:algorithm} has been implemented in the \texttt{\justify Mathematica} package \textit{CANONICA}~\cite{Meyer:2017joq}, which represents the first publicly available program to compute transformations to a canonical form for differential equations depending on multiple variables. This chapter aims to give an overview of the package and its capabilities. The most frequently used functions and data structures of \textit{CANONICA} are introduced in \sec{subsec:UsageEx} along with simple examples of their usage. A detailed description of all functions and options can be found in the interactive manual notebook of the package. Also note that a quick reference guide in \app{App:Lists} provides details on the installation, the files contained in the package and a short description of all available functions and options. The limitations of the algorithm and its implementation are discussed in \sec{sec:TestsNLimits}.

The material presented in this chapter is based on the publication~\cite{Meyer:2017joq}, which accompanies the \textit{CANONICA} package.

%
%

\section{Usage examples}
\label{subsec:UsageEx}

The most common input required by \textit{CANONICA} is a differential equation of the form in \eq{DEQDifferentialForm}, which is determined by the differential form $a(\varset)$. A simple example depending on the invariants $x$ and $y$ is given by
\begin{align}
a(\epsilon,\{x,y\})=&\left(\,
\begin{array}{cc}
-\frac{2+\epsilon}{x} & 0 \\
0 & -\frac{1+\epsilon}{x}  \\
\end{array}
\,\right)\textup{d}x \nonumber \\
&+\left(\,
\begin{array}{cc}
0 & 0 \\
\frac{(-1+\epsilon)x}{(-1+y)y} & \frac{1-\epsilon(1+y)}{(-1+y)y}  \\
\end{array}
\,\right)\textup{d}y.
\end{align}
The differential form $a(\varset)$ is represented in \textit{CANONICA} as a list of the matrix-valued coefficients of the differentials of the invariants. For the dimensional regulator $\epsilon$, the protected symbol \texttt{\justify eps} must be used. The above differential equation is assigned to the symbol \texttt{a} with 
\begin{verse}
\begin{verbatim}
a = {
  {{-(2+eps)/x, 0}, {0, -(1+eps)/x}}
  ,
   {{0, 0}, {((-1+eps)x)/((-1+y)y), (1-eps(1+y))/((-1+y)y)}}
 };
\end{verbatim}
\end{verse}
Most functions in \textit{CANONICA} that require a differential equation as input, also require a list of the invariants indicating the order of the corresponding coefficient matrices in the differential equation argument. In the case of the example, this list is given by
\begin{verse}
\begin{verbatim}
invariants = {x, y};
\end{verbatim}
\end{verse}
The algorithm to compute a transformation to a canonical form for diagonal blocks as outlined in \sec{sec:ExpTrafo} is implemented in the function \texttt{\justify TransformDiagonalBlock}. For the above example, this function is called as follows:
\begin{verse}
\begin{verbatim}
res=TransformDiagonalBlock[a, invariants]
\end{verbatim}
\end{verse}
which returns the output
\begin{verbatim}
      {
       {{(1-2eps)/x^2, (1-2eps)/x^2}, {(1-eps)/x, (1-eps)/(xy)}}
       ,
       {
        {{-(eps/x), 0}, {0, -(eps/x)}}
        ,
        {{-eps/(-1+y), -eps/(y-y^2)}, {eps/(-1+y), eps/(y-y^2)}}
       }
      }
\end{verbatim}
\texttt{\justify TransformDiagonalBlock} returns a list with two entries. The first entry contains the transformation 
\begin{equation}
T=\left(\begin{array}{cc}
\frac{(1-2\epsilon)}{x^2} & \frac{(1-2\epsilon)}{x^2} \\
\frac{(1-\epsilon)}{x} & \frac{(1-\epsilon)}{xy}\end{array}\right),
\end{equation}
and the second entry contains the differential equation in canonical form
\begin{equation}
\epsilon\,\textup{d}\tilde{A}=\left(\begin{array}{cc}
-\frac{\epsilon}{x} & 0 \\
0 & -\frac{\epsilon}{x}\end{array}\right)\textup{d}x
+
\left(\begin{array}{cc}
-\frac{\epsilon}{-1+y} & -\frac{\epsilon}{y-y^2} \\
\frac{\epsilon}{-1+y} & -\frac{\epsilon}{y-y^2}\end{array}\right)\textup{d}y.
\end{equation}
The resulting canonical form is, of course, redundant information since it can be computed by applying the transformation to the original differential equation. However, the resulting differential equation is generated in the course of the computation of the transformation and applying the transformation can be a costly operation in itself for larger differential equations.

The application of a transformation to a differential equation, according to the transformation law \eq{DEQTrafo}, is implemented in the function \texttt{\justify TransformDE}, which for some transformation 
\begin{verse}
\begin{verbatim}
trafo = res[[1]];
\end{verbatim}
\end{verse}
is called as
\begin{verse}
\begin{verbatim}
TransformDE[a, invariants, trafo]
\end{verbatim}
\end{verse}
and returns the resulting differential equation. In order to apply \eq{DEQTrafo}, the inverse of the transformation needs to be computed. With the build-in \texttt{\justify Mathematica} command, this can consume significant computation time for larger matrices. A considerably better performance is achieved in \texttt{\justify TransformDE} by exploiting the block-triangular structure of the transformations.

The function \texttt{\justify TransformDiagonalBlock} is in principle applicable to differential equations comprising several sectors. However, it has been argued in \sec{sec:OffDiagPart} that the performance can be improved significantly by splitting the computation according to the block-triangular structure of the differential equation and performing the computation in a recursion over the sectors of the differential equation. In \textit{CANONICA}, this recursive approach is implemented in the function \texttt{\justify RecursivelyTransformSectors}. In addition to the two arguments related to the differential equation itself, this function expects an argument that defines the boundaries of the diagonal blocks. The differential equation in the example actually splits into two blocks of dimension one leading to the boundaries
\begin{verse}
\begin{verbatim}
boundaries = {{1, 1}, {2, 2}};
\end{verbatim}
\end{verse}
Each entry of the boundaries list corresponds to one diagonal block, which is specified by the position of its lowest and highest integral. Instead of using \texttt{\justify TransformDiagonalBlock} to transform \texttt{\justify a} into canonical form all at once, the following command
\begin{verse}
\begin{verbatim}
RecursivelyTransformSectors[a, invariants, boundaries, {1, 2}]
\end{verbatim}
\end{verse}
computes the transformation in a recursion over the sectors, as described in \sec{sec:OffDiagPart}. The last argument determines the sectors at which the computation starts and ends. The output is of the same format as described above for \texttt{\justify TransformDiagonalBlock}. If some lower sectors have already been transformed into canonical form and the computation should therefore not start at the first sector, the differential equation of the lower sectors in canonical form and the transformation leading to it need to be provided as two additional arguments.

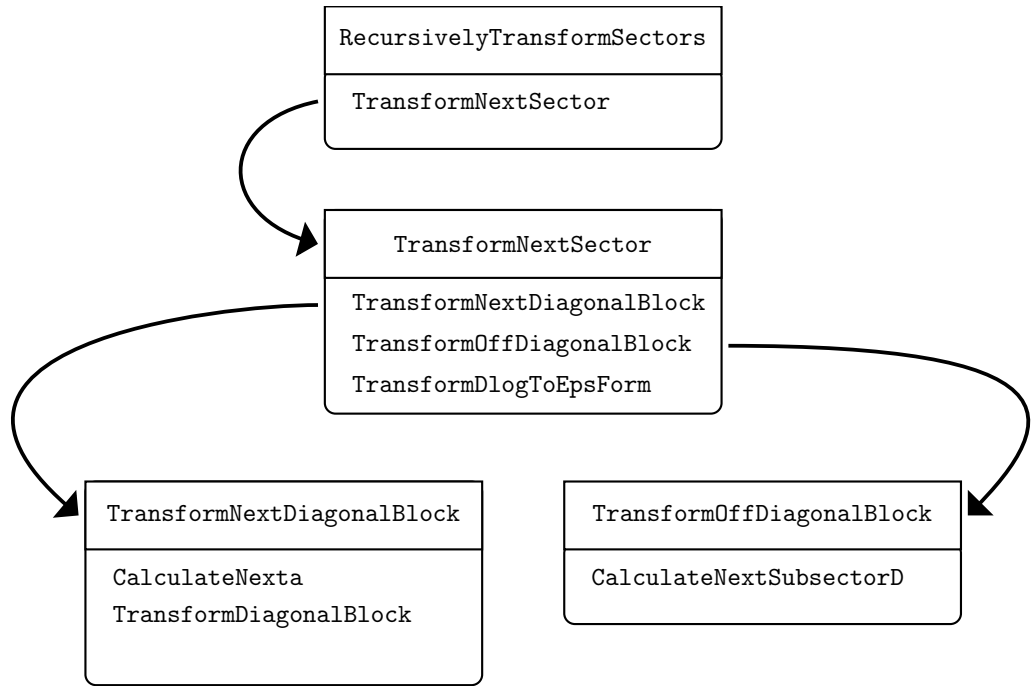
\begin{figure}[h!]
\centering
\begin{tikzpicture}[thick,scale=0.9, every node/.style={scale=0.9}]
\clip(-4,-2) rectangle (11,8);

\def\frow{0}
\def\srow{7}
\def\mrow{\frow*0.5+\srow*0.5}

\def\vadd{-4}
\def\vvadd{-4}
\def\vvvadd{3}

\small
\tikzset{
bignodeone/.style={rectangle,rounded corners, draw=black, top color=white, inner    sep=1em,minimum width=5.80cm, minimum height=1.2cm, text centered},
regnodeone/.style={rectangle,draw=black, top color=white, inner sep=1pt,minimum width=5.8cm, minimum height=1cm, text centered},
bignode/.style={rectangle,rounded corners, draw=black, top color=white, inner    sep=1em,minimum width=5.80cm, minimum height=3cm, text centered},
regnode/.style={rectangle,draw=black, top color=white, inner sep=1pt,minimum width=5.8cm, minimum height=1cm, text centered},
nregnode/.style={rectangle,draw=white, top color=white, inner sep=1pt,minimum width=1.2cm, minimum height=0.4cm,text width=5cm}
}
\node[bignodeone] (axiif) at (\mrow,3.5+\vvvadd) {};
\node[regnode] (RecursivelyTransformSectors) at (\mrow, 4.5+\vvvadd) {\texttt{RecursivelyTransformSectors}};
\node[nregnode] (pTransformNextSector) at (\mrow,3.6+\vvvadd) {\texttt{TransformNextSector}}; 

\node[bignode] (axiif) at (\mrow,3.5) {};
\node[regnode] (sTransformNextSector) at (\mrow, 4.5) {\texttt{TransformNextSector}};
\node[nregnode] (pTransformNextDiagonalBlock) at (\mrow,3.6) {\texttt{TransformNextDiagonalBlock}}; 
\node[nregnode] (pTransformOffDiagonalBlock) at (\mrow,3.0) {\texttt{TransformOffDiagonalBlock}};
\node[nregnode] (pTransformDlogToEpsForm) at (\mrow,2.4) {\texttt{TransformDlogToEpsForm}};

\node[bignode] (axiif) at (\frow,3.5+\vadd) {};
\node[regnode] (sTransformNextDiagonalBlock) at (\frow, 4.5+\vadd) {\texttt{TransformNextDiagonalBlock}};
\node[nregnode] (pCalculateNexta) at (\frow,3.6+\vadd) {\texttt{CalculateNexta}}; 
\node[nregnode] (pTransformDiagonalBlock) at (\frow,3.0+\vadd) {\texttt{TransformDiagonalBlock}};

\node[bignodeone] (axiif) at (\srow,3.5+\vvadd) {};
\node[regnode] (sTransformOffDiagonalBlock) at (\srow, 4.5+\vvadd) {\texttt{TransformOffDiagonalBlock}};
\node[nregnode] (pCalculateNextSubsectorD) at (\srow,3.6+\vvadd) {\texttt{CalculateNextSubsectorD}}; 

\draw[draw=black,line width=0.5mm,solid, -triangle 90] (\mrow-3,3.6+\vvvadd) .. controls (\mrow-4.5,3.6+\vvvadd-0.3) and (\mrow-4.5,3.6+\vvvadd-1.7) .. (\mrow-3, 4.5);

\draw[draw=black,line width=0.5mm,solid, -triangle 90] (\mrow-3,3.6) .. controls (\mrow-4.5,3.6) and (\frow-6,3.6+\vvvadd-3.6) .. (\frow-3, 4.5+\vadd);

\draw[draw=black,line width=0.5mm,solid, -triangle 90] (\mrow+3,3.0) .. controls (\srow+4.5,3.0) and (\srow+4.5,\vvvadd-1) .. (\srow+3, 4.5+\vadd);

\end{tikzpicture} 
\medskip
\caption{Hierarchy of the main functions in \textit{CANONICA}. Each block lists the public functions called by the function in the blocks title.} 
\label{fig:hierarchy}
\end{figure}

\textit{CANONICA} also has functionality to extract the boundaries of the diagonal blocks from the differential equation. The function \texttt{\justify SectorBoundariesFromDE} extracts the most fine-grained boundaries compatible with the differential equation. For instance, in the example above
\begin{verse}
\begin{verbatim}
SectorBoundariesFromDE[a]
\end{verbatim}
\end{verse}
returns 
\begin{verse}
\begin{verbatim}
{{1, 1}, {2, 2}}
\end{verbatim}
\end{verse}
The boundaries obtained in this way may be too fine for the algorithm to find the solution because the solution space could be over-constrained by splitting the transformation into smaller blocks. It is safer to choose the boundaries according to the sector-ids of the integrals, which in general yields coarser-grained boundaries. For a given list of integrals specified by their propagator powers
\begin{verse}
\begin{verbatim}
masterIntegrals={Int["T1", {0, 1, 0, 1, 0, 1, 0, 0, 0}],
                 Int["T1", {0, 1, 0, 1, 1, 1, 0, 0, 0}], 
                 Int["T1", {0, 0, 1, 1, 1, 1, 0, 0, 0}],
                 Int["T1", {0, 1, 1, 1, 1, 1, 0, 0, 0}], 
                 Int["T1", {1, 1, 0, 0, 0, 0, 1, 0, 0}], 
                 Int["T1", {1, 1, -1, 0, 0, 0, 1, 0, 0}]};
\end{verbatim}
\end{verse}
the boundaries for the corresponding differential equation can be computed via
\begin{verse}
\begin{verbatim}
SectorBoundariesFromID[masterIntegrals]
\end{verbatim}
\end{verse}
provided the integrals are ordered with respect to their sector-ids. \texttt{\justify SectorBoundariesFromID} then groups integrals with identical sector-ids together in a sector and returns
\begin{verse}
\begin{verbatim}
{{1, 1}, {2, 2}, {3, 3}, {4, 4}, {5, 6}}
\end{verbatim}
\end{verse}
While the main function of \textit{CANONICA} is \texttt{\justify RecursivelyTransformSectors}, it is in some cases useful to be able to perform only certain steps of the algorithm. For this reason, there is a hierarchy of functions available in \textit{CANONICA} allowing to break the calculation of the transformation into smaller steps. The hierarchy of these lower-level functions is illustrated in \fig{fig:hierarchy}. For more information on specific functions see the manual notebook included in the package or \app{App:Lists}.

\section{Tests and limitations}
\label{sec:TestsNLimits}
\textit{CANONICA} has been successfully tested on a variety of non-trivial single- and multi-scale problems, some of which are presented in the following \chap{chap:applications}. Additional examples are included in the \textit{CANONICA} package. All tests have been performed with the \texttt{\justify Mathematica} versions 10 and 11 on a Linux operating system. 

The algorithm is limited to differential equations for which a \emph{rational} transformation to a canonical form exists. This represents a limitation of the algorithm since it is well known that rational differential equations can require non-rational transformations to attain a canonical form. The following example illustrates this behavior:
\begin{equation}
a(\epsilon,\{x\})=\left(\frac{1}{2x}+\frac{\epsilon}{x}\right)\textup{d}x,
\end{equation}
where the transformation to a canonical form is given by
\begin{equation}
T(\epsilon,\{x\})=\sqrt{x}.
\end{equation}
In such a case it may be possible to render the transformation rational with a change of coordinates. For instance, in the above example, the differential form transforms under the change of variables 
\begin{equation}
x=y^2
\end{equation}
into
\begin{equation}
a(\epsilon,\{y\})=\left(\frac{1}{y}+\frac{2\epsilon}{y}\right)\textup{d}y,
\end{equation}
which admits the rational transformation to a canonical form
\begin{equation}
T(\epsilon,\{y\})=y.
\end{equation}
While a change of coordinates can remove non-rational letters in more complicated examples as well \cite{Henn:2013woa, Henn:2014lfa, Gehrmann:2014bfa, Caola:2014lpa}, this has neither been proven to be always possible nor is a general method to construct such coordinate changes known. In fact, the existence of such a procedure appears to be unlikely, given that the number of independent roots can largely outgrow the number of variables in a problem~\cite{Bonciani:2016qxi}. It is therefore desirable to extend the algorithm to transformations with algebraic dependence on the invariants.

A further limitation stems from the heuristic nature of the ansatz used to solve the transformation law. The algorithm is able to compute a rational transformation of a given differential equation into canonical form whenever such a transformation exists, and it is decomposable in terms of the ansatz that is used. However, if no such transformation exists for the given ansatz, the equations in the parameters of the ansatz will not have a solution. In this case, either the ansatz is not general enough, or a rational transformation to a canonical form does not exist at all. A sufficient condition for the latter case is the presence of non-rational factors in the determinant of the transformation, which can be computed via \eq{DetIsFixed}. However, if a canonical form does not exist at all or requires a non-rational transformation with rational determinant, \textit{CANONICA} has no functionality to distinguish these cases from the case where a rational transformation exists but the ansatz that is used is insufficient. 



In practice, \textit{CANONICA} is limited by the size of the available memory, which imposes limits on the size of the systems of linear parameter equations that can be handled. The main factors determining the sizes of these systems of linear equations are the size of the differential equation itself and the size of the ansatz. Thus, the size of the ansatz and thereby the range of feasible problems is typically limited by the available memory. In general, the run time and memory consumption are highly problem dependent, which makes it complicated to specify the scaling behavior of the algorithm. For instance, the most complex example discussed in \chap{chap:applications} has a run time of about 20 minutes and a memory consumption of less than 8 GB. This topology is a two-loop double box topology depending on three dimensionless scales, which contributes to the NNLO QCD corrections to the production of single top-quarks.

More generally, it should be noted that the strategy of using a canonical basis of master integrals is known to be limited, as there exist Feynman integrals which do not admit such a basis. However, typically only a few sectors (cf. e.g., \cite{Adams:2016xah, Bonciani:2016qxi}) of such integral topologies do not admit a canonical basis, while all lower sectors can still be cast in a canonical form. Thus, \textit{CANONICA} can still be useful for the treatment of integral topologies not evaluating to Chen iterated integrals.

       \chapter{Applications}
\label{chap:applications}

In this chapter, \textit{CANONICA} is applied to state-of-the-art multi-loop topologies encountered in phenomenologically relevant calculations. For most of the presented topologies, the transformation and the resulting differential equation are too large to be displayed completely. Therefore, some simpler univariate problems are included as well, for which the resulting differential equations are compact enough to be shown in full detail. In all other cases, only some samples of the result are shown. However, all topologies are included in full detail as examples in the \textit{CANONICA} package~\cite{Meyer:2017joq}. 

The reduction to master integrals and the corresponding differential equations have for all topologies been computed with Reduze \cite{Studerus:2009ye, vonManteuffel:2012np} and are included in the package as well. The cited run times of the examples are obtained on a single core machine.

\section{Massless planar double box}
\begin{figure}[H]
\begin{center}
\includegraphics[scale=1]{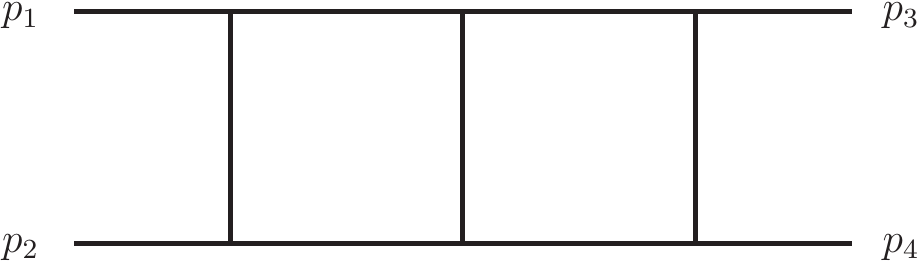}
\end{center}
\caption{Massless planar double box.}
\label{fig:mldb}
\end{figure}
The massless planar double box topology in \fig{fig:mldb} contributes, for instance, to the NNLO QCD corrections to the production of two jets~\cite{Ridder:2013mf}. This topology has first been computed in~\cite{Smirnov:1999gc}, and a treatment with the differential equations approach can be found in~\cite{Henn:2013pwa}. The topology is given by the following set of inverse propagators and irreducible scalar products:
\begin{equation}
\begin{array}{lll}
P_1=l_1^2, & P_{4}=(l_2-p_3-p_4)^2, & P_{7}=(l_1-l_2)^2, \\
P_2=(l_1-p_3)^2, & P_{5}=(l_2-p_1)^2, &  P_{8}=(l_2-p_3)^2, \\
P_3=(l_1-p_3-p_4)^2, &P_{6}=l_2^2, & P_{9}=(l_1-p_1)^2. \\  
\end{array}
\end{equation}
The momenta $p_1$ and $p_2$ are incoming, and $p_3$ and $p_4$ are outgoing with the kinematics given by
\begin{equation}
p_1+p_2=p_3+p_4,
\end{equation}
\begin{equation}
p_1^2=0,\quad p_2^2=0,\quad p_3^2=0,\quad p_4^2=0,
\end{equation}
\begin{equation}
s\defeq(p_1+p_2)^2,\quad t\defeq(p_1-p_3)^2.
\end{equation}
The topology has a basis of 8 master integrals:
\small
\begin{equation}
\begin{array}{lll}
\vec{g}^{\,\textup{DBP}}(\epsilon,s,t)\,\,\,=&\big(I_\textup{DBP}(1,0,1,1,0,1,0,0,0),&I_\textup{DBPx124}(1,0,0,1,0,0,1,0,0),\\
&I_\textup{DBP}(1,0,0,1,0,0,1,0,0),&I_\textup{DBP}(1,0,1,0,1,0,1,0,0),\\
&I_\textup{DBP}(1,1,1,0,1,0,1,0,0),&I_\textup{DBP}(1,1,0,1,1,0,1,0,0),\\
&I_\textup{DBP}(1,1,1,1,1,1,1,-1,0),&I_\textup{DBP}(1,1,1,1,1,1,1,0,0)\big).
\end{array}
\end{equation}
\normalsize
The notation $\textup{x124}$ in the topology name refers to the topology with the external momenta $p_1$, $p_2$ and $p_4$ cyclically permuted. A vector of dimensionless master integrals is obtained by factoring out the mass-dimension
\begin{equation}
f^{\textup{DBP}}_i(\epsilon,x)=(t)^{-\textup{dim}(g^{\textup{DBP}}_i)/2}g^{\textup{DBP}}_i(\epsilon,s,t),
\end{equation}
with 
\begin{equation}
x=\frac{s}{t}.
\end{equation}
After loading the \textit{CANONICA} package with 
\begin{verse}
\begin{verbatim}
Get["CANONICA.m"];
\end{verbatim}
\end{verse}
and assigning the differential equation with respect to the master integrals $\vec{f}$ to the symbol \texttt{\justify a}, the sector boundaries are extracted from the differential equation by calling
\begin{verse}
\begin{verbatim}
boundaries=SectorBoundariesFromDE[a]
\end{verbatim}
\end{verse}
which returns
\begin{verse}
\begin{verbatim}
{{1, 1}, {2, 2}, {3, 3}, {4, 4}, {5, 5}, {6, 6}, {7, 8}}
\end{verbatim}
\end{verse}
Using the extracted boundaries, the recursive strategy presented in \sec{sec:OffDiagPart} can be applied by calling the function \texttt{\justify RecursivelyTransformSectors} as follows:
\begin{verse}
\begin{verbatim}
RecursivelyTransformSectors[a, {x}, boundaries, {1, 7}]
\end{verbatim}
\end{verse}
After a run time of about two seconds, this function returns the transformation and the resulting differential equation in canonical form. Below, the transformation is shown by expressing the original basis of master integrals $\vec{f}$ in terms of the integrals of the canonical basis $\vec{f}^\prime$:

\footnotesize
\begin{equation}
f^\textup{DBP}_{1}=f^{\textup{DBP}\prime}_{1}, \quad f^\textup{DBP}_{2}=\left(\frac{2\epsilon (2\epsilon-1)}{(3\epsilon-2) (3\epsilon-1)}\right)f^{\textup{DBP}\prime}_{2},
\end{equation}
\normalsize
\footnotesize
\begin{align}
f^\textup{DBP}_{3}=&
\left(\frac{2\epsilon (2\epsilon-1) x}{(3\epsilon-2) (3\epsilon-1)}\right)f^{\textup{DBP}\prime}_{1}
+\left(\frac{2\epsilon (2\epsilon-1) x}{(3\epsilon-2) (3\epsilon-1)}\right)f^{\textup{DBP}\prime}_{3},
\end{align}
\normalsize
\footnotesize
\begin{equation}
f^\textup{DBP}_{4}=\left(\frac{2-4\epsilon}{1-3\epsilon}\right)f^{\textup{DBP}\prime}_{4}, \quad f^\textup{DBP}_{5}=\left(\frac{2\epsilon-1}{\epsilon x}\right)f^{\textup{DBP}\prime}_{5}, \quad f^\textup{DBP}_{6}=\left(\frac{(1-2\epsilon)^2}{\epsilon^2 (x+1)}\right)f^{\textup{DBP}\prime}_{6},
\end{equation}
\normalsize
\footnotesize
\begin{equation}
f^\textup{DBP}_{7}=\left(\frac{(1-2\epsilon)^2}{\epsilon^2 x^2}\right)f^{\textup{DBP}\prime}_{7}, \quad f^\textup{DBP}_{8}=\left(\frac{(1-2\epsilon)^2}{\epsilon^2 x^2}\right)f^{\textup{DBP}\prime}_{8}.
\end{equation}
\normalsize
The resulting canonical form can be written as
\begin{equation}
\textup{d}A=\epsilon\left[\frac{\tilde{A}_1}{x}+\frac{\tilde{A}_2}{1+x}\right]\textup{d}x,
\end{equation}
where the constant matrices $\tilde{A}_1$ and $\tilde{A}_2$ are given by
\begin{equation}
\tilde{A}_1=\left(\,
\begin{array}{cccccccc}
-2 & 0 & 0 & 0 & 0 & 0 & 0 & 0\\
0 & 0 & 0 & 0 & 0 & 0 & 0 & 0\\
0 & 0 & -2 & 0 & 0 & 0 & 0 & 0\\
0 & 0 & 0 & -2 & 0 & 0 & 0 & 0\\
0 & 0 & 0 & 0 & -1 & 0 & 0 & 0\\
-2 & 2 & -2 & 0 & 0 & -2 & 0 & 0\\
2 & 0 & 0 & 0 & 0 & 0 & -2 & 0\\
-10 & 0 & -12 & 6 & 0 & 0 & 2 & -2\\
\end{array}
\,\right)
\end{equation}
and 
\begin{equation}
\tilde{A}_2=\left(\,
\begin{array}{cccccccc}
0 & 0 & 0 & 0 & 0 & 0 & 0 & 0\\
0 & 0 & 0 & 0 & 0 & 0 & 0 & 0\\
0 & 0 & 0 & 0 & 0 & 0 & 0 & 0\\
0 & 0 & 0 & 0 & 0 & 0 & 0 & 0\\
0 & 2 & 0 & 2 & 1 & 0 & 0 & 0\\
0 & 0 & 0 & 0 & 0 & 2 & 0 & 0\\
11 & 18 & 12 & 15 & 12 & -18 & -1 & 1\\
22 & 12 & 24 & 6 & 12 & -12 & -2 & 2\\
\end{array}
\,\right).
\end{equation}
The alphabet of the canonical form is given by
\begin{equation}
\mathcal{A}=\{x,\, 1+x\}.
\end{equation}
In the following applications, the usage of \textit{CANONICA} is very similar and will, therefore, be omitted.

\section{Massless non-planar double box}
\begin{figure}[H]
\begin{center}
\includegraphics[scale=1]{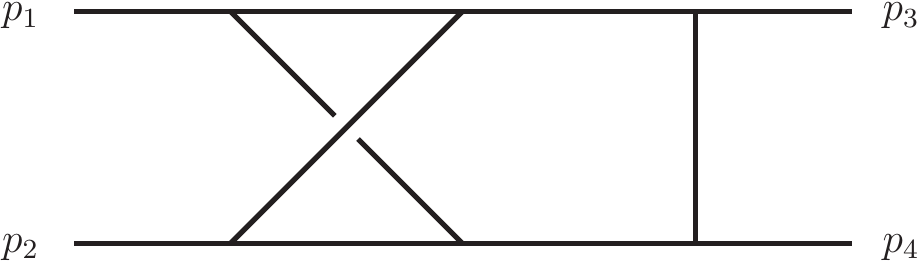}
\end{center}
\caption{Massless non-planar double box.}
\label{fig:mldbnp}
\end{figure}
The massless non-planar double box topology in \fig{fig:mldbnp} contributes, for instance, to the NNLO QCD corrections to the production of two jets~\cite{Ridder:2013mf}. This topology has first been computed in~\cite{Tausk:1999vh}, and a treatment using differential equations can be found in~\cite{Argeri:2014qva}. The topology is given by the following set of inverse propagators and irreducible scalar products:
\begin{equation}
\begin{array}{lll}
P_1=l_1^2, & P_{4}=l_2^2, & P_{7}=(l_1-l_2+p_3-p_1)^2, \\
P_2=(l_1-p_4)^2, & P_{5}=(l_2-l_1-p_3)^2, &  P_{8}=(l_1+p_2)^2, \\
P_3=(l_2-p_2)^2, &P_{6}=(l_1+p_3)^2, & P_{9}=(l_2-p_3)^2. \\  
\end{array}
\end{equation}
The momenta $p_1$ and $p_2$ are incoming, and $p_3$ and $p_4$ are outgoing with the kinematics given by
\begin{equation}
p_1+p_2=p_3+p_4,
\end{equation}
\begin{equation}
p_1^2=0,\quad p_2^2=0,\quad p_3^2=0,\quad p_4^2=0,
\end{equation}
\begin{equation}
s\defeq(p_1+p_2)^2,\quad t\defeq(p_1-p_3)^2.
\end{equation}
The topology has a basis of 12 master integrals:
\small
\begin{equation}
\label{OldBasisDBP}
\begin{array}{lll}
\vec{g}^{\,\textup{DBNP}}(\epsilon,s,t)\,\,\,=&\big(I_\textup{DBNPx124}(1,0,1,0,1,0,0,0,0),&I_\textup{DBNPx12}(1,0,1,0,1,0,0,0,0),\\
&I_\textup{DBNP}(1,0,1,0,1,0,0,0,0),&I_\textup{DBNPx123}(1,1,1,1,1,0,0,0,0),\\
&I_\textup{DBNPx12}(1,1,1,1,1,0,0,0,0),&I_\textup{DBNP}(1,1,1,1,1,0,0,0,0),\\
&I_\textup{DBNP}(0,1,1,0,1,1,0,0,0),&I_\textup{DBPNPx12}(1,1,1,0,1,1,0,0,0),\\
&I_\textup{DBNP}(1,1,1,0,1,1,0,0,0),&I_\textup{DBNP}(0,1,1,1,1,1,1,0,0),\\
&I_\textup{DBNP}(1,1,1,1,1,1,1,0,0),&I_\textup{DBNP}(1,1,1,1,1,1,1,-1,0)\big).\end{array}
\end{equation}
\normalsize
A vector of dimensionless master integrals is obtained by factoring out the mass-dimension
\begin{equation}
f^{\textup{DBNP}}_i(\epsilon,x)=(t)^{-\textup{dim}(g^{\textup{DBNP}}_i)/2}g^{\textup{DBNP}}_i(\epsilon,s,t),
\end{equation}
with 
\begin{equation}
x=\frac{s}{t}.
\end{equation}
Recursively computing the transformation to a canonical basis with \textit{CANONICA} takes about four seconds. In the following, the original basis of master integrals $\vec{f}$ is expressed in terms of the canonical basis $\vec{f}^\prime$:

\footnotesize
\begin{align}
f^\textup{DBNP}_{1}=&
\left(\frac{2\epsilon^3 x}{(2\epsilon-1) (3\epsilon-2) (3\epsilon-1)}\right)f^{\textup{DBNP}\prime}_{1},
\end{align}
\normalsize
\footnotesize
\begin{align}
f^\textup{DBNP}_{2}=&
\left(\frac{2\epsilon^3}{(2\epsilon-1) (3\epsilon-2) (3\epsilon-1)}\right)f^{\textup{DBNP}\prime}_{2},
\end{align}
\normalsize
\footnotesize
\begin{align}
f^\textup{DBNP}_{3}=&
\left(\frac{2\epsilon^3 (x+1)}{(2\epsilon-1) (3\epsilon-2) (3\epsilon-1)}\right)f^{\textup{DBNP}\prime}_{3},
\end{align}
\normalsize
\footnotesize
\begin{equation}
f^\textup{DBNP}_{4}=\left(\frac{1}{x}\right)f^{\textup{DBNP}\prime}_{4},\quad f^\textup{DBNP}_{5}=\left(\frac{1}{x+1}\right)f^{\textup{DBNP}\prime}_{5},\quad f^\textup{DBNP}_{6}=f^{\textup{DBNP}\prime}_{6},
\end{equation}
\normalsize
\footnotesize
\begin{equation}
f^\textup{DBNP}_{7}=\left(\frac{2\epsilon^2}{(2\epsilon-1) (3\epsilon-1)}\right)f^{\textup{DBNP}\prime}_{7},\quad f^\textup{DBNP}_{8}=\left(\frac{\epsilon}{(2\epsilon-1) x}\right)f^{\textup{DBNP}\prime}_{8},
\end{equation}
\normalsize
\footnotesize
\begin{equation}
f^\textup{DBNP}_{9}=\left(\frac{\epsilon}{(2\epsilon-1) x}\right)f^{\textup{DBNP}\prime}_{9}, \quad f^\textup{DBNP}_{10}=\left(\frac{1}{x^2}\right)f^{\textup{DBNP}\prime}_{10},
\end{equation}
\normalsize
\footnotesize
\begin{align}
f^\textup{DBNP}_{11}=&
\left(-\frac{6 \left(\epsilon \left(x^2+14 x+12\right)+3 (x+1)\right)}{(4\epsilon+1) x^2 (x+1)}\right)f^{\textup{DBNP}\prime}_{1}
+\left(-\frac{6 (\epsilon (5 x+6)+x+1)}{(4\epsilon+1) x^2 (x+1)}\right)f^{\textup{DBNP}\prime}_{2}\nonumber\\
&+\left(-\frac{6 (\epsilon (x+6)+1)}{(4\epsilon+1) x^2}\right)f^{\textup{DBNP}\prime}_{3}
+\left(\frac{6\epsilon (x+2)}{(4\epsilon+1) x (x+1)}\right)f^{\textup{DBNP}\prime}_{4}\nonumber\\
&+\left(\frac{6 (\epsilon (x+6)+1)}{(4\epsilon+1) x^2}\right)f^{\textup{DBNP}\prime}_{5}
+\left(\frac{6 (\epsilon (5 x+6)+x+1)}{(4\epsilon+1) x^2 (x+1)}\right)f^{\textup{DBNP}\prime}_{6}\nonumber\\
&+\left(\frac{2}{x^2}\right)f^{\textup{DBNP}\prime}_{10}
+\left(\frac{1}{x^2}\right)f^{\textup{DBNP}\prime}_{11},
\end{align}
\normalsize
\footnotesize
\begin{align}
f^\textup{DBNP}_{12}=&
\left(\frac{3 (24\epsilon (2 x+1)+11 x+6)}{(4\epsilon+1) x^2 (x+1)}\right)f^{\textup{DBNP}\prime}_{1}
+\left(-\frac{3 (4\epsilon (x-3)+x-2)}{(4\epsilon+1) x^2 (x+1)}\right)f^{\textup{DBNP}\prime}_{2}\nonumber\\
&+\left(\frac{12\epsilon (4 x+3)+9 x+6}{(4\epsilon+1) x^2 (x+1)}\right)f^{\textup{DBNP}\prime}_{3}
+\left(-\frac{12\epsilon}{(4\epsilon+1) x (x+1)}\right)f^{\textup{DBNP}\prime}_{4}\nonumber\\
&+\left(-\frac{3 (4\epsilon (4 x+3)+3 x+2)}{(4\epsilon+1) x^2 (x+1)}\right)f^{\textup{DBNP}\prime}_{5}
+\left(\frac{3 (4\epsilon (x-3)+x-2)}{(4\epsilon+1) x^2 (x+1)}\right)f^{\textup{DBNP}\prime}_{6}\nonumber\\
&+\left(-\frac{1}{x^2}\right)f^{\textup{DBNP}\prime}_{10}
+\left(-\frac{1}{x^2}\right)f^{\textup{DBNP}\prime}_{11}
+\left(-\frac{1}{x (x+1)}\right)f^{\textup{DBNP}\prime}_{12}.
\end{align}
\normalsize
The resulting canonical form can be written as
\begin{equation}
\textup{d}A=\epsilon\left[\frac{\tilde{A}_1}{x}+\frac{\tilde{A}_2}{1+x}\right]\textup{d}x,
\end{equation}
where the constant matrices $\tilde{A}_1$ and $\tilde{A}_2$ are given by
\begin{equation}
\tilde{A}_1=\left(\,
\begin{array}{cccccccccccc}
-2 & 0 & 0 & 0 & 0 & 0 & 0 & 0 & 0 & 0 & 0 & 0\\
0 & 0 & 0 & 0 & 0 & 0 & 0 & 0 & 0 & 0 & 0 & 0\\
0 & 0 & 0 & 0 & 0 & 0 & 0 & 0 & 0 & 0 & 0 & 0\\
0 & 0 & 0 & 2 & 0 & 0 & 0 & 0 & 0 & 0 & 0 & 0\\
-2 & 2 & 0 & 0 & -2 & 0 & 0 & 0 & 0 & 0 & 0 & 0\\
2 & 0 & 2 & 0 & 0 & -2 & 0 & 0 & 0 & 0 & 0 & 0\\
0 & 0 & 0 & 0 & 0 & 0 & -2 & 0 & 0 & 0 & 0 & 0\\
0 & 0 & 0 & 0 & 0 & 0 & 0 & -1 & 0 & 0 & 0 & 0\\
0 & 0 & 0 & 0 & 0 & 0 & 0 & 0 & -1 & 0 & 0 & 0\\
0 & 0 & 0 & 0 & 0 & 0 & 0 & 0 & 0 & -2 & 0 & 0\\
0 & 0 & 0 & 0 & 0 & 0 & 0 & 0 & 0 & 0 & -2 & 0\\
0 & 0 & 0 & 0 & 0 & 0 & 12 & 6 & 6 & 0 & 1 & 0\\
\end{array}
\,\right)
\end{equation}
and
\begin{equation}
\tilde{A}_2=\left(\,
\begin{array}{cccccccccccc}
0 & 0 & 0 & 0 & 0 & 0 & 0 & 0 & 0 & 0 & 0 & 0\\
0 & 0 & 0 & 0 & 0 & 0 & 0 & 0 & 0 & 0 & 0 & 0\\
0 & 0 & -2 & 0 & 0 & 0 & 0 & 0 & 0 & 0 & 0 & 0\\
0 & -2 & -2 & -2 & 0 & 0 & 0 & 0 & 0 & 0 & 0 & 0\\
0 & 0 & 0 & 0 & 2 & 0 & 0 & 0 & 0 & 0 & 0 & 0\\
-2 & 0 & -2 & 0 & 0 & -2 & 0 & 0 & 0 & 0 & 0 & 0\\
0 & 0 & 0 & 0 & 0 & 0 & 0 & 0 & 0 & 0 & 0 & 0\\
0 & 2 & 0 & 0 & 0 & 0 & 2 & 1 & 0 & 0 & 0 & 0\\
0 & 0 & 2 & 0 & 0 & 0 & -2 & 0 & -2 & 0 & 0 & 0\\
0 & 0 & 0 & 0 & 0 & 0 & 0 & 0 & 0 & 0 & 0 & 0\\
-36 & 0 & 0 & -12 & 0 & 0 & 0 & 0 & 0 & 2 & 1 & 2\\
0 & -12 & -12 & 6 & -12 & -12 & -6 & 0 & -6 & 0 & 0 & -2\\
\end{array}
\,\right).
\end{equation}
The alphabet of the resulting canonical form is given by
\begin{equation}
\mathcal{A}=\{x,\, 1+x\}.
\end{equation}

\section{K4 integral}
\begin{figure}[H]
\begin{center}
\includegraphics[scale=1]{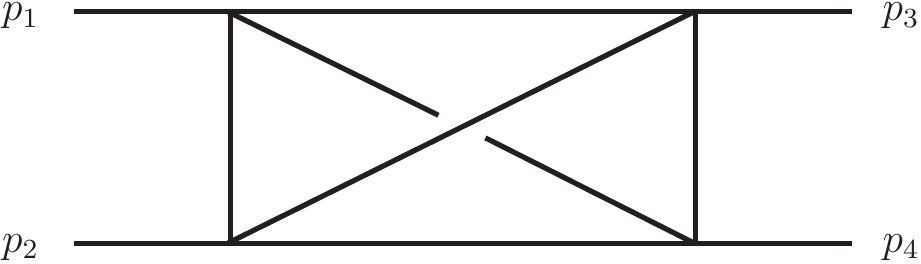}
\end{center}
\caption{K4 integral.}
\label{fig:K4}
\end{figure}
The K4 integral topology in \fig{fig:K4} contributes, for instance, to the NNNLO QCD corrections to the production of two jets, which have yet to be calculated. This integral topology was first evaluated with the differential equations method in~\cite{Henn:2013nsa}. The topology is given by the following set of inverse propagators and irreducible scalar products:
\begin{equation}
\begin{array}{lll}
P_1=(l_1+l_3)^2, & P_{6}=(l_1+l_2+l_3+p_3)^2, & P_{11}=(l_1+l_2)^2, \\
P_2=(l_1+l_2+p_1+p_2)^2, & P_{7}=l_1^2, &  P_{12}=(l_3+p_1)^2, \\
P_3=l_3^2, &P_{8}=(l_1+p_1+p_2)^2, & P_{13}=(l_2+p_1)^2, \\  
P_4=l_2^2, &P_{9}=(l_1+l_2+l_3)^2, & P_{14}=(l_1-p_3)^2, \\  
P_5=(l_1+p_1)^2, &P_{10}=(l_1+l_2+l_3+p_1+p_2)^2, & P_{15}=(l_3-p_3)^2. \\  
\end{array}
\end{equation}
The momenta $p_1$ and $p_2$ are incoming, and $p_3$ and $p_4$ are outgoing with the kinematics given by
\begin{equation}
p_1+p_2=p_3+p_4,
\end{equation}
\begin{equation}
p_1^2=0,\quad p_2^2=0,\quad p_3^2=0,\quad p_4^2=0,
\end{equation}
\begin{equation}
s\defeq(p_1+p_2)^2,\quad t\defeq(p_1-p_3)^2.
\end{equation}
The topology has a basis of 10 master integrals:
\small
\begin{equation}
\label{OldBasisK4}
\begin{array}{ll}
\vec{g}^{\,\textup{K4}}(\epsilon,s,t)\,\,\,=&\big(I_\textup{K4x124}(1,1,1,1,0,0,0,0,0,0,0,0,0,0,0),\\
&I_\textup{K4x1234}(1,1,1,1,0,0,0,0,0,0,0,0,0,0,0),\\
&I_\textup{K4}(1,1,1,1,0,0,0,0,0,0,0,0,0,0,0),\\
&I_\textup{K4}(2,1,1,1,1,1,0,0,0,0,0,0,0,0,0),\\
&I_\textup{K4}(1,1,2,1,1,1,0,0,0,0,0,0,0,0,0),\\
&I_\textup{K4}(1,1,1,1,2,1,0,0,0,0,0,0,0,0,0),\\
&I_\textup{K4}(1,1,1,1,1,1,-1,0,0,0,0,0,0,0,0),\\
&I_\textup{K4}(1,1,1,1,1,1,0,0,0,0,-1,0,0,0,0),\\
&I_\textup{K4}(1,1,1,1,1,1,0,0,0,0,0,-1,0,0,0),\\
&I_\textup{K4}(1,1,1,1,1,1,0,0,0,0,0,0,0,0,0)\big).
\end{array}
\end{equation}
\normalsize
A vector of dimensionless master integrals is obtained by factoring out the mass-dimension
\begin{equation}
f^\textup{K4}_i(\epsilon,x)=(t)^{-\textup{dim}(g^\textup{K4}_i)/2}g^\textup{K4}_i(\epsilon,s,t),
\end{equation}
with 
\begin{equation}
x=\frac{s}{t}.
\end{equation}
\textit{CANONICA} computes the transformation to a canonical basis in about 10 minutes. For brevity, only a sample of the transformation is shown in the following:

\footnotesize
\begin{align}
f^\textup{K4}_{6}=&
\left(\frac{(2\epsilon-1) (4\epsilon-1) \left(36\epsilon^3-63\epsilon^2+35\epsilon-6\right)}{9\epsilon^4 x}\right)f^{\textup{K4}\prime}_{6},\nonumber\\
\end{align}
\normalsize
\footnotesize
\begin{align}
f^\textup{K4}_{7}=&
\left(\frac{(3\epsilon-2) (3\epsilon-1) (4\epsilon-3) \left(2\epsilon^3 (9 x+47)+5\epsilon^2 (x-9)+\epsilon (5-6 x)+x\right)}{4\epsilon^3 (2\epsilon-1) (5\epsilon-2) (5\epsilon-1)}\right)f^{\textup{K4}\prime}_{1}\nonumber\\
&+\left(-\frac{(3\epsilon-2) (3\epsilon-1) (4\epsilon-3) \left(\epsilon^3 (76 x+94)-5\epsilon^2 (10 x+9)+\epsilon (11 x+5)-x\right)}{4\epsilon^3 (2\epsilon-1) (5\epsilon-2) (5\epsilon-1)}\right)f^{\textup{K4}\prime}_{2}\nonumber\\
&+\left(-\frac{(3\epsilon-2) (4\epsilon-3) \left(148\epsilon^5-184\epsilon^4+128\epsilon^3-55\epsilon^2+12\epsilon-1\right) x}{4\epsilon^3 (2\epsilon-1)^2 (5\epsilon-2) (5\epsilon-1)}\right)f^{\textup{K4}\prime}_{3}\nonumber\\
&+\left(\frac{2 \left(36\epsilon^3-63\epsilon^2+35\epsilon-6\right) \left(x (x+1)-\epsilon \left(4 x^2+7 x+2\right)\right)}{9\epsilon (2\epsilon-1) (5\epsilon-2) (5\epsilon-1)}\right)f^{\textup{K4}\prime}_{4}\nonumber\\
&+\left(\frac{2 \left(36\epsilon^3-63\epsilon^2+35\epsilon-6\right) \left(\epsilon \left(x^2+3 x-2\right)-x\right)}{9\epsilon (2\epsilon-1) (5\epsilon-2) (5\epsilon-1) (x+1)}\right)f^{\textup{K4}\prime}_{5}\nonumber\\
&+\left(\frac{2 \left(36\epsilon^3-63\epsilon^2+35\epsilon-6\right) \left(\epsilon \left(x^2+8 x+8\right)-2 (x+1)\right)}{9\epsilon (2\epsilon-1) (5\epsilon-2) (5\epsilon-1) x}\right)f^{\textup{K4}\prime}_{6}\nonumber\\
&+\left(-\frac{2 \left(36\epsilon^3-63\epsilon^2+35\epsilon-6\right) \left(2\epsilon^2 (x+9)-9\epsilon+1\right)}{9\epsilon^2 (2\epsilon-1) (5\epsilon-2) (5\epsilon-1)}\right)f^{\textup{K4}\prime}_{7}\nonumber\\
&+\left(\frac{4 \left(36\epsilon^3-63\epsilon^2+35\epsilon-6\right) (x+2)}{9 (2\epsilon-1) (5\epsilon-2) (5\epsilon-1)}\right)f^{\textup{K4}\prime}_{8}\nonumber\\
&+\left(-\frac{2 (7\epsilon-1) \left(36\epsilon^3-63\epsilon^2+35\epsilon-6\right) x}{9\epsilon^2 (5\epsilon-2) (5\epsilon-1)}\right)f^{\textup{K4}\prime}_{9}\nonumber\\
&+\left(-\frac{2 \left(36\epsilon^3-63\epsilon^2+35\epsilon-6\right) \left(2\epsilon^3 (79 x-30)-112\epsilon^2 x+7\epsilon x+x\right)}{9\epsilon^3 (2\epsilon-1) (5\epsilon-2) (5\epsilon-1)}\right)f^{\textup{K4}\prime}_{10},
\end{align}
\normalsize
\footnotesize
\begin{align}
f^\textup{K4}_{10}=&
\left(-\frac{(2\epsilon-1) (3\epsilon-2) (3\epsilon-1) (4\epsilon-3)}{2\epsilon^3 (5\epsilon-1)}\right)f^{\textup{K4}\prime}_{1}
+\left(-\frac{(2\epsilon-1) (3\epsilon-2) (3\epsilon-1) (4\epsilon-3)}{2\epsilon^3 (5\epsilon-1)}\right)f^{\textup{K4}\prime}_{2}\nonumber\\
&+\left(-\frac{(2\epsilon-1) (3\epsilon-2) (3\epsilon-1) (4\epsilon-3)}{2\epsilon^3 (5\epsilon-1)}\right)f^{\textup{K4}\prime}_{3}
+\left(\frac{4 (2\epsilon-1) \left(36\epsilon^3-63\epsilon^2+35\epsilon-6\right)}{9\epsilon^3 (5\epsilon-1)}\right)f^{\textup{K4}\prime}_{10}.
\end{align}
\normalsize
The resulting canonical form can be written as
\begin{equation}
\textup{d}A=\epsilon\left[\frac{\tilde{A}_1}{x}+\frac{\tilde{A}_2}{1+x}\right]\textup{d}x,
\end{equation}
where the constant matrices $\tilde{A}_1$ and $\tilde{A}_2$ are given by
\begin{equation}
\tilde{A}_1=\left(\,
\begin{array}{cccccccccc}
0 & 0 & 0 & 0 & 0 & 0 & 0 & 0 & 0 & 0\\
0 & 0 & 0 & 0 & 0 & 0 & 0 & 0 & 0 & 0\\
0 & 0 & -3 & 0 & 0 & 0 & 0 & 0 & 0 & 0\\
0 & 0 & 0 & -2 & 1 & 1 & -2 & -2 & 4 & 30\\
0 & 0 & 0 & 1 & -2 & -1 & 2 & -2 & -4 & -90\\
0 & 0 & 0 & 0 & 0 & 1 & 0 & 0 & 0 & 0\\
0 & 0 & 0 & 0 & 0 & 0 & 0 & 0 & 0 & 0\\
0 & 0 & 0 & 1 & 1 & 0 & -16 & -5 & 30 & 390\\
0 & 0 & -\frac{27}{4} & 0 & 0 & 1 & -16 & 0 & 32 & 496\\
0 & 0 & 0 & 0 & 0 & 0 & 1 & 0 & -2 & -31\\
\end{array}
\,\right)
\end{equation}
and
\begin{equation}
\tilde{A}_2=\left(\,
\begin{array}{cccccccccc}
0 & 0 & 0 & 0 & 0 & 0 & 0 & 0 & 0 & 0\\
0 & -3 & 0 & 0 & 0 & 0 & 0 & 0 & 0 & 0\\
0 & 0 & 0 & 0 & 0 & 0 & 0 & 0 & 0 & 0\\
0 & 0 & 0 & -2 & -1 & -1 & 2 & 2 & -4 & -30\\
0 & 0 & 0 & 0 & 1 & 0 & 0 & 0 & 0 & 0\\
-\frac{27}{2} & \frac{27}{2} & 0 & -1 & -1 & -2 & 2 & 2 & 0 & 30\\
\frac{27}{8} & \frac{27}{8} & \frac{27}{4} & 1 & -1 & -1 & 2 & 0 & -7 & -104\\
-54 & \frac{189}{4} & 0 & 0 & -1 & 0 & 16 & 16 & -16 & 16\\
-\frac{351}{8} & \frac{189}{4} & \frac{27}{4} & 1 & 0 & -1 & 16 & 14 & -21 & -90\\
\frac{27}{8} & -\frac{27}{8} & 0 & 0 & 0 & 0 & -1 & -1 & 1 & -1\\
\end{array}
\,\right).
\end{equation}
The alphabet of the resulting canonical form is given by
\begin{equation}
\mathcal{A}=\{x, 1+x\}.
\end{equation}

\section{Triple box}
\begin{figure}[H]
\begin{center}
\includegraphics[scale=1]{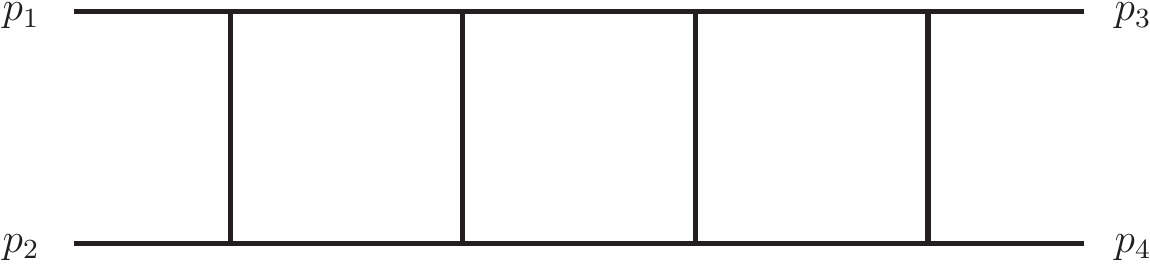}
\end{center}
\caption{Triple box.}
\label{fig:tbox}
\end{figure}
The triple box topology in \fig{fig:tbox} contributes, for instance, to the NNNLO QCD corrections to the production of two jets. Analytical results for the triple box integral have first been obtained in~\cite{Smirnov:2003vi} and more recently in~\cite{Henn:2013fah} by using differential equations. The topology is given by the following set of inverse propagators and irreducible scalar products:
\begin{equation}
\begin{array}{lll}
P_1=l_1^2, & P_{6}=(l_3+p_1+p_2)^2, & P_{11}=(l_1+p_3)^2, \\
P_2=(l_1+p_1+p_2)^2, & P_{7}=(l_1+p_1)^2, &  P_{12}=(l_2+p_1)^2, \\
P_3=l_2^2, &P_{8}=(l_1-l_2)^2, & P_{13}=(l_2+p_3)^2, \\  
P_4=(l_2+p_1+p_2)^2, &P_{9}=(l_2-l_3)^2, & P_{14}=(l_3+p_1)^2, \\  
P_5=l_3^2, &P_{10}=(l_3+p_3)^2, & P_{15}=(l_1-l_3)^2. \\  
\end{array}
\end{equation}
The momenta $p_1$ and $p_2$ are incoming, and $p_3$ and $p_4$ are outgoing with the kinematics given by
\begin{equation}
p_1+p_2=p_3+p_4,
\end{equation}
\begin{equation}
p_1^2=0,\quad p_2^2=0,\quad p_3^2=0,\quad p_4^2=0,
\end{equation}
\begin{equation}
s\defeq(p_1+p_2)^2,\quad t\defeq(p_1-p_3)^2.
\end{equation}
The topology has a basis of 26 master integrals:
\small
\begin{equation}
\label{OldBasisTriplebox}
\begin{array}{lll}
\vec{g}^{\,\textup{TB}}(\epsilon,s,t)\,\,\,=&\\
\big(I_\textup{TB}(1,1,1,1,1,1,0,0,0,0,0,0,0,0,0),&I_\textup{TB}(0,1,1,0,1,1,0,1,0,0,0,0,0,0,0),\\
I_\textup{TB}(0,0,1,1,1,1,1,1,0,0,0,0,0,0,0),&I_\textup{TBx124}(0,1,0,0,1,0,0,1,1,0,0,0,0,0,0),\\
I_\textup{TB}(0,1,0,0,1,0,0,1,1,0,0,0,0,0,0),&I_\textup{TB}(1,0,0,1,1,0,0,1,1,0,0,0,0,0,0),\\
I_\textup{TB}(1,1,0,0,1,1,0,1,1,0,0,0,0,0,0),&I_\textup{TB}(0,0,0,1,1,0,1,1,1,0,0,0,0,0,0),\\
I_\textup{TB}(0,0,0,0,1,1,1,1,1,0,0,0,0,0,0),&I_\textup{TB}(0,1,1,0,1,1,1,1,1,0,0,0,0,0,0),\\
I_\textup{TB}(1,1,0,0,0,0,1,1,1,1,0,0,0,0,0),&I_\textup{TB}(0,1,1,0,0,0,1,1,1,1,0,0,0,0,0),\\
I_\textup{TB}(-1,0,1,1,0,0,1,1,1,1,0,0,0,0,0),&I_\textup{TB}(0,0,1,1,0,0,1,1,1,1,0,0,0,0,0),\\
I_\textup{TB}(1,1,1,1,-1,0,1,1,1,1,0,0,0,0,0),&I_\textup{TB}(1,1,1,1,0,0,1,1,1,1,0,0,0,0,0),\\
I_\textup{TB}(0,1,0,0,1,0,1,1,1,1,0,0,0,0,0),&I_\textup{TB}(1,-1,0,1,1,0,1,1,1,1,0,0,0,0,0),\\
I_\textup{TB}(1,0,0,1,1,0,1,1,1,1,0,0,0,0,0),&I_\textup{TB}(1,1,-1,1,1,0,1,1,1,1,0,0,0,0,0),\\
I_\textup{TB}(1,1,0,1,1,0,1,1,1,1,0,0,0,0,0),&I_\textup{TB}(1,1,-1,0,1,1,1,1,1,1,0,0,0,0,0),\\
I_\textup{TB}(1,1,0,0,1,1,1,1,1,1,0,0,0,0,0),&I_\textup{TB}(1,1,1,1,1,1,1,1,1,1,-1,0,0,0,0),\\
I_\textup{TB}(1,1,1,1,1,1,1,1,1,1,0,-1,0,0,0),&I_\textup{TB}(1,1,1,1,1,1,1,1,1,1,0,0,0,0,0)\big).
\end{array}
\end{equation}
\normalsize
A vector of dimensionless master integrals is obtained by factoring out the mass-dimension
\begin{equation}
f^\textup{TB}_i(\epsilon,x)=(t)^{-\textup{dim}(g^\textup{TB}_i)/2}g^\textup{TB}_i(\epsilon,s,t),
\end{equation}
with 
\begin{equation}
x=\frac{s}{t}.
\end{equation}
The transformation to a canonical basis of which a sample is shown below is computed with \textit{CANONICA} in about two minutes.

\footnotesize
\begin{align}
f^\textup{TB}_{7}=&
\left(\frac{15\epsilon^3 \left(814\epsilon^2-679\epsilon+97\right)}{4 (2\epsilon-1)^2 (3\epsilon-1)^2 (4\epsilon-1)}\right)f^{\textup{TB}\prime}_{1}
+\left(\frac{15\epsilon^3 \left(62\epsilon^2-77\epsilon+11\right)}{4 (2\epsilon-1)^2 (3\epsilon-1)^2 (4\epsilon-1)}\right)f^{\textup{TB}\prime}_{2}\nonumber\\
&+\left(\frac{75\epsilon^3 \left(236\epsilon^2-203\epsilon+29\right)}{8 (2\epsilon-1)^2 (3\epsilon-1)^2 (4\epsilon-1)}\right)f^{\textup{TB}\prime}_{3}
+\left(\frac{15\epsilon^3 \left(664\epsilon^2-469\epsilon+67\right)}{8 (2\epsilon-1)^2 (3\epsilon-1)^2 (4\epsilon-1)}\right)f^{\textup{TB}\prime}_{5}\nonumber\\
&+\left(\frac{315\epsilon^3}{8 (2\epsilon-1)^2 (3\epsilon-1)}\right)f^{\textup{TB}\prime}_{6}
+\left(\frac{45\epsilon^3}{8 (2\epsilon-1)^2 (3\epsilon-1)}\right)f^{\textup{TB}\prime}_{7},
\end{align}
\normalsize
\footnotesize
\begin{align}
f^\textup{TB}_{15}=&
\left(\frac{243 (\epsilon-2)\epsilon^2}{4 (2\epsilon-1) (3\epsilon-1)^2 x}\right)f^{\textup{TB}\prime}_{1}
+\left(\frac{621 (\epsilon-2)\epsilon^2}{8 (2\epsilon-1) (3\epsilon-1)^2 x}\right)f^{\textup{TB}\prime}_{2}
+\left(\frac{711 (\epsilon-2)\epsilon^2}{8 (2\epsilon-1) (3\epsilon-1)^2 x}\right)f^{\textup{TB}\prime}_{3}\nonumber\\
&+\left(-\frac{189\epsilon \left(\epsilon^2 (6 x-7)-2\epsilon (x-3)-3\right)}{16 (2\epsilon-1) (3\epsilon-1)^2 x^2}\right)f^{\textup{TB}\prime}_{4}
+\left(\frac{189 (\epsilon-2)\epsilon^2}{2 (2\epsilon-1) (3\epsilon-1)^2 x}\right)f^{\textup{TB}\prime}_{5}\nonumber\\
&+\left(\frac{189 (\epsilon-2)\epsilon^2}{2 (2\epsilon-1) (3\epsilon-1)^2 x}\right)f^{\textup{TB}\prime}_{6}
+\left(\frac{189 (\epsilon-2)\epsilon^2}{2 (2\epsilon-1) (3\epsilon-1)^2 x}\right)f^{\textup{TB}\prime}_{7}\nonumber\\
&+\left(\frac{189 (\epsilon-2)\epsilon^2}{2 (2\epsilon-1) (3\epsilon-1)^2 x}\right)f^{\textup{TB}\prime}_{8}
+\left(\frac{189 (\epsilon-2)\epsilon^2}{4 (2\epsilon-1) (3\epsilon-1)^2 x}\right)f^{\textup{TB}\prime}_{11}\nonumber\\
&+\left(\frac{189\epsilon \left(\epsilon^2 (3 x-5)-\epsilon (x-8)-3\right)}{8 (2\epsilon-1) (3\epsilon-1)^2 x^2}\right)f^{\textup{TB}\prime}_{12}
+\left(-\frac{189 (\epsilon-3)\epsilon}{8 (2\epsilon-1) (3\epsilon-1) x^2}\right)f^{\textup{TB}\prime}_{13}\nonumber\\
&+\left(-\frac{189\epsilon^3}{(2\epsilon-1) (3\epsilon-1)^2 x}\right)f^{\textup{TB}\prime}_{14}
+\left(\frac{75 (\epsilon-1)\epsilon^2}{2 (2\epsilon-1) (3\epsilon-1)^2 x}\right)f^{\textup{TB}\prime}_{15}\nonumber\\
&+\left(-\frac{75 (\epsilon-1)^2\epsilon}{2 (2\epsilon-1) (3\epsilon-1)^2 x^2}\right)f^{\textup{TB}\prime}_{16},
\end{align}
\normalsize
\footnotesize
\begin{align}
f^\textup{TB}_{17}=&
\left(\frac{2625\epsilon^3 x}{4 (3\epsilon-1)^2 (4\epsilon-1) (x+1)}\right)f^{\textup{TB}\prime}_{1}
+\left(\frac{525\epsilon^3 x}{4 (3\epsilon-1)^2 (4\epsilon-1) (x+1)}\right)f^{\textup{TB}\prime}_{2}\nonumber\\
&+\left(\frac{525\epsilon^3 x}{(3\epsilon-1)^2 (4\epsilon-1) (x+1)}\right)f^{\textup{TB}\prime}_{3}
+\left(\frac{63\epsilon^2 (\epsilon (6 x+3)-x)}{2 (3\epsilon-1)^2 (4\epsilon-1) (x+1)^2}\right)f^{\textup{TB}\prime}_{4}\nonumber\\
&+\left(\frac{525\epsilon^3 x}{4 (3\epsilon-1)^2 (4\epsilon-1) (x+1)}\right)f^{\textup{TB}\prime}_{5}
+\left(\frac{35\epsilon^2 x}{4 (3\epsilon-1) (4\epsilon-1) (x+1)^2}\right)f^{\textup{TB}\prime}_{17}.
\end{align}
\normalsize
The alphabet of the resulting canonical form is given by
\begin{equation}
\mathcal{A}=\big\{x,\,1+x\big\}.
\end{equation}
For brevity, the resulting canonical form is not presented here but is given in the \textit{CANONICA} package.

\section{Drell--Yan with one internal mass}
\begin{figure}[H]
\begin{center}
\includegraphics[scale=1]{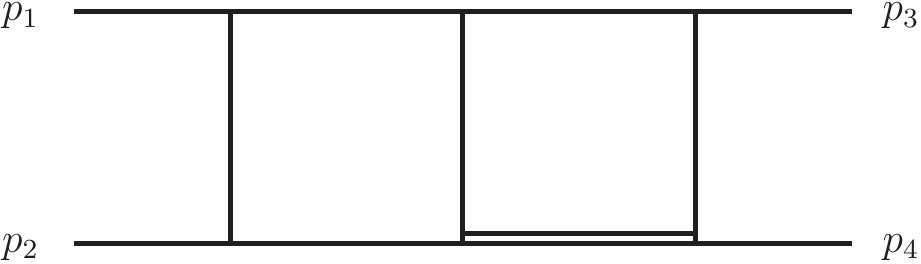}
\end{center}
\caption{Drell--Yan with one massive propagator indicated by the double line.}
\label{fig:DY}
\end{figure}
The Drell--Yan topology in \fig{fig:DY} contributes to the mixed electroweak and QCD corrections to the Drell--Yan process. This topology has been calculated with the differential equations approach in~\cite{Bonciani:2016ypc}. The topology is given by the following set of inverse propagators and irreducible scalar products:
\begin{equation}
\begin{array}{lll}
P_1=l_1^2, & P_{4}=(l_2-p_3-p_4)^2, & P_{7}=(l_1-l_2)^2, \\
P_2=(l_1-p_3)^2, & P_{5}=(l_2-p_1)^2, &  P_{8}=(l_2-p_3)^2, \\
P_3=(l_1-p_3-p_4)^2-m^2, &P_{6}=l_2^2, & P_{9}=(l_1-p_1)^2. \\  
\end{array}
\end{equation}
The momenta $p_1$ and $p_2$ are incoming, and $p_3$ and $p_4$ are outgoing with the kinematics given by
\begin{equation}
p_1+p_2=p_3+p_4,
\end{equation}
\begin{equation}
p_1^2=0,\quad p_2^2=0,\quad p_3^2=0,\quad p_4^2=0,
\end{equation}
\begin{equation}
s\defeq(p_1+p_2)^2,\quad t\defeq(p_1-p_3)^2.
\end{equation}
The topology has a basis of 25 master integrals:
\small
\begin{equation}
\begin{array}{lll}
\vec{g}^{\,\textup{DYOM}}(\epsilon,s,t,m)\,\,\,=&\big(I_\textup{DYOM}(0,0,1,1,0,1,0,0,0),&I_\textup{DYOM}(1,0,1,1,0,1,0,0,0),\\
&I_\textup{DYOMx124}(1,0,0,1,0,0,1,0,0),&I_\textup{DYOM}(1,0,0,1,0,0,1,0,0),\\
&I_\textup{DYOM}(0,0,1,1,0,0,1,0,0),&I_\textup{DYOM}(1,0,1,1,0,0,1,0,0),\\
&I_\textup{DYOM}(1,0,1,0,1,0,1,0,0),&I_\textup{DYOM}(0,1,1,0,1,0,1,0,0),\\
&I_\textup{DYOM}(1,1,1,0,1,0,1,0,0),&I_\textup{DYOM}(1,1,0,1,1,0,1,0,0),\\
&I_\textup{DYOM}(1,0,1,1,1,0,1,0,0),&I_\textup{DYOM}(0,1,1,1,1,0,1,0,0),\\
&I_\textup{DYOM}(1,1,1,1,1,0,1,0,0),&I_\textup{DYOM}(0,0,1,0,0,1,1,0,0),\\
&I_\textup{DYOM}(-1,0,1,0,0,1,1,0,0),&I_\textup{DYOM}(0,1,0,1,0,1,1,0,0),\\
&I_\textup{DYOM}(0,1,1,1,0,1,1,0,0),&I_\textup{DYOM}(-1,1,1,1,0,1,1,0,0),\\
&I_\textup{DYOM}(0,1,1,0,1,1,1,0,0),&I_\textup{DYOM}(-1,1,1,0,1,1,1,0,0),\\
&I_\textup{DYOM}(0,1,0,1,1,1,1,0,0),&I_\textup{DYOM}(0,1,1,1,1,1,1,0,0),\\
&I_\textup{DYOM}(-1,1,1,1,1,1,1,0,0),&I_\textup{DYOM}(1,1,1,1,1,1,1,0,0),\\
&I_\textup{DYOM}(1,1,1,1,1,1,1,-1,0)\big).&
\end{array}
\end{equation}
\normalsize
A vector of dimensionless master integrals is obtained by factoring out the mass-dimension
\begin{equation}
f^\textup{DYOM}_i(\epsilon,x,y)=(m)^{-\textup{dim}(g^\textup{DYOM}_i)}g^\textup{DYOM}_i(\epsilon,s,t,m),
\end{equation}
with 
\begin{equation}
x=\frac{s}{m^2}, \quad y=\frac{t}{m^2}.
\end{equation}
The transformation to a canonical basis of which a sample is shown in the following is computed with \textit{CANONICA} in about one minute.

\footnotesize
\begin{align}
f^\textup{DYOM}_{15}=&
\left(-\frac{15\epsilon^3 \left(x^2-1\right)}{2 (2\epsilon-1) (3\epsilon-2) (3\epsilon-1) x}\right)f^{\textup{DYOM}\prime}_{14}\nonumber\\
&+\left(-\frac{45\epsilon^2 \left(\epsilon \left(3 x^2-7 x-4\right)+2 x\right)}{4 (2\epsilon-1) (3\epsilon-2) (3\epsilon-1) x}\right)f^{\textup{DYOM}\prime}_{15},
\end{align}
\normalsize
\footnotesize
\begin{align}
f^\textup{DYOM}_{18}=&
\left(-\frac{9}{16 x}\right)f^{\textup{DYOM}\prime}_{1}
+\left(-\frac{63\epsilon^5-99\epsilon^4-25\epsilon^3+85\epsilon^2-44\epsilon+8}{8 (\epsilon-1)\epsilon (2\epsilon-1) (3\epsilon-1) x}\right)f^{\textup{DYOM}\prime}_{5}\nonumber\\
&+\left(\frac{45}{8 x}\right)f^{\textup{DYOM}\prime}_{15}
+\left(\frac{45}{16 x}\right)f^{\textup{DYOM}\prime}_{16}
+\left(\frac{3}{2 x}\right)f^{\textup{DYOM}\prime}_{18},
\end{align}
\normalsize
\footnotesize
\begin{align}
f^\textup{DYOM}_{23}=&
\left(\frac{3}{2 x y}\right)f^{\textup{DYOM}\prime}_{3}
+\left(-\frac{189\epsilon^4-477\epsilon^3+519\epsilon^2-289\epsilon+62}{8 (\epsilon-1) (2\epsilon-1) (3\epsilon-1) x y}\right)f^{\textup{DYOM}\prime}_{5}\nonumber\\
&+\left(\frac{3}{2 x y}\right)f^{\textup{DYOM}\prime}_{8}
+\left(\frac{3}{2 x y}\right)f^{\textup{DYOM}\prime}_{12}
+\left(-\frac{135}{8 x y}\right)f^{\textup{DYOM}\prime}_{15}\nonumber\\
&+\left(-\frac{3}{2 x y}\right)f^{\textup{DYOM}\prime}_{18}
+\left(-\frac{3}{x y}\right)f^{\textup{DYOM}\prime}_{20}
+\left(-\frac{1}{x y}\right)f^{\textup{DYOM}\prime}_{23}.
\end{align}
\normalsize
The alphabet of the resulting canonical form is given by
\begin{equation}
\mathcal{A}=\big\{-1+x,\,x,\,y,\,1+y,\,x+y,\,xy-x-y\big\}.
\end{equation}
For brevity, the resulting canonical form is not presented here but is given in the \textit{CANONICA} package.

\section{Vector boson pair production}
\begin{figure}[H]
\begin{center}
\includegraphics[scale=1]{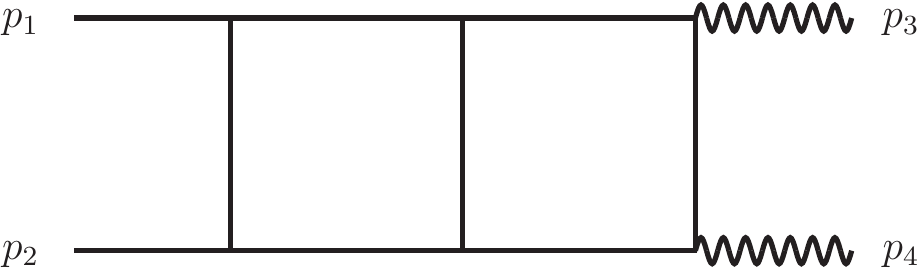}
\end{center}
\caption{Vector boson pair production topology 1.}
\label{fig:VVT1}
\end{figure}
\begin{figure}[H]
\begin{center}
\includegraphics[scale=1]{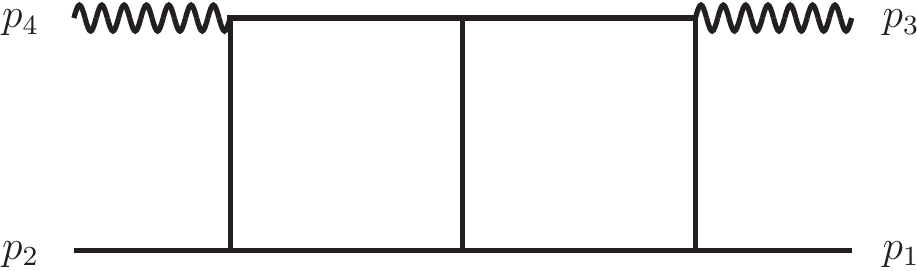}
\end{center}
\caption{Vector boson pair production topology 2.}
\label{fig:VVT2}
\end{figure}
The topologies in Figs. \ref{fig:VVT1} and \ref{fig:VVT2} occur in the computation of the NNLO QCD corrections to the production of two massive vector bosons \cite{Cascioli:2014yka, Gehrmann:2014fva, Grazzini:2016swo}. These topologies have been considered in \cite{Gehrmann:2013cxs, Henn:2014lfa, Gehrmann:2014bfa, Caola:2014lpa, Papadopoulos:2014hla, Gehrmann:2015ora} with the differential equations approach and are given by the following sets of inverse propagators and irreducible scalar products:
\paragraph*{Topology 1:}
\begin{equation}
\begin{array}{lll}
P_1=l_1^2, & P_4=(l_2-p_3-p_4)^2, & P_7=(l_2-p_1)^2, \\
P_2=(l_1-p_3-p_4)^2, & P_5=(l_1-p_3)^2, &  P_8=(l_2-p_3)^2, \\
P_3=l_2^2, &P_6=(l_1-l_2)^2, & P_9=(l_1-p_1)^2. \\  
\end{array}
\end{equation}
\paragraph*{Topology 2:}
\begin{equation}
\begin{array}{lll}
P_1=l_1^2, & P_4=(l_2+p_1-p_3)^2, & P_7=(l_2+p_4)^2, \\
P_2=(l_1+p_1-p_3)^2, & P_5=(l_1-p_3)^2, &  P_8=(l_2-p_3)^2, \\
P_3=l_2^2, &P_6=(l_1-l_2)^2, & P_9=(l_1+p_4)^2. \\  
\end{array}
\end{equation}
The momenta $p_1$ and $p_2$ are incoming, and $p_3$ and $p_4$ are outgoing, and the kinematics of both topologies are given by
\begin{equation}
p_1+p_2=p_3+p_4,
\end{equation}
\begin{equation}
p_1^2=0,\quad p_2^2=0,\quad p_3^2=m_3^2,\quad p_4^2=m_4^2,
\end{equation}
\begin{equation}
s=(p_1+p_2)^2,\quad t=(p_1-p_3)^2.
\end{equation}
Topology 1 has a basis of 31 master integrals:
\small
\begin{equation}
\label{OldBasisVVT1}
\begin{array}{lll}
\vec{g}^{\,\textup{VV1}}(\epsilon,s,t,m_3,m_4)\,\,\,=&\big(I_\textup{VV1}(1,1,1,1,0,0,0,0,0),&I_\textup{VV1}(1,0,1,1,1,0,0,0,0),\\
&I_\textup{VV1}(0,1,1,1,1,0,0,0,0),&I_\textup{VV1}(1,1,1,1,1,0,0,0,0),\\
&I_\textup{VV1}(0,1,1,0,0,1,0,0,0),&I_\textup{VV1}(0,0,1,0,1,1,0,0,0),\\
&I_\textup{VV1}(-1,1,1,0,1,1,0,0,0),&I_\textup{VV1}(0,1,1,0,1,1,0,0,0),\\
&I_\textup{VV1}(0,0,0,1,1,1,0,0,0),&I_\textup{VV1}(1,-1,0,1,1,1,0,0,0),\\
&I_\textup{VV1}(1,0,0,1,1,1,0,0,0),&I_\textup{VV1}(-1,0,1,1,1,1,0,0,0),\\
&I_\textup{VV1}(0,0,1,1,1,1,0,0,0),&I_\textup{VV1}(1,0,1,1,1,1,0,0,0),\\
&I_\textup{VV1}(0,1,1,1,1,1,0,0,0),&I_\textup{VV1}(1,1,0,0,0,1,1,0,0),\\
&I_\textup{VV1}(0,0,0,0,1,1,1,0,0),&I_\textup{VV1}(1,0,0,0,1,1,1,0,0),\\
&I_\textup{VV1}(0,1,0,0,1,1,1,0,0),&I_\textup{VV1}(1,1,-1,0,1,1,1,0,0),\\
&I_\textup{VV1}(1,1,0,0,1,1,1,0,0),&I_\textup{VV1}(-1,1,1,0,1,1,1,0,0),\\
&I_\textup{VV1}(0,1,1,0,1,1,1,0,0),&I_\textup{VV1}(1,-1,0,1,1,1,1,0,0),\\
&I_\textup{VV1}(1,0,0,1,1,1,1,0,0),&I_\textup{VV1}(0,0,1,1,1,1,1,0,0),\\
&I_\textup{VV1}(1,0,1,1,1,1,1,0,0),&I_\textup{VV1}(0,1,1,1,1,1,1,0,0),\\
&I_\textup{VV1}(1,1,1,1,1,1,1,-1,0),&I_\textup{VV1}(1,1,1,1,1,1,1,0,-1),\\
&I_\textup{VV1}(1,1,1,1,1,1,1,0,0)\big).&\end{array}
\end{equation}
\normalsize
Topology 2 has a basis of 29 master integrals:
\small
\begin{equation}
\label{OldBasisVVT2}
\begin{array}{lll}
\vec{g}^{\,\textup{VV2}}(\epsilon,s,t,m_3,m_4)\,\,\,=&\big(I_\textup{VV2}(1,1,1,1,0,0,0,0,0),&I_\textup{VV2}(1,0,1,1,1,0,0,0,0),\\
&I_\textup{VV2}(0,1,1,0,0,1,0,0,0),&I_\textup{VV2}(0,0,1,0,1,1,0,0,0),\\
&I_\textup{VV2}(1,0,0,1,1,1,0,0,0),&I_\textup{VV2}(0,0,1,1,1,1,0,0,0),\\
&I_\textup{VV2}(1,0,1,1,1,1,0,0,0),&I_\textup{VV2}(1,1,1,0,0,0,1,0,0),\\
&I_\textup{VV2}(1,0,1,0,1,0,1,0,0),&I_\textup{VV2}(1,0,0,0,0,1,1,0,0),\\
&I_\textup{VV2}(1,1,0,0,0,1,1,0,0),&I_\textup{VV2}(0,1,1,0,0,1,1,0,0),\\
&I_\textup{VV2}(1,1,1,0,0,1,1,0,0),&I_\textup{VV2}(0,0,0,0,1,1,1,0,0),\\
&I_\textup{VV2}(1,-1,0,0,1,1,1,0,0),&I_\textup{VV2}(1,0,0,0,1,1,1,0,0),\\
&I_\textup{VV2}(1,1,0,0,1,1,1,0,0),&I_\textup{VV2}(-1,0,1,0,1,1,1,0,0),\\
&I_\textup{VV2}(0,0,1,0,1,1,1,0,0),&I_\textup{VV2}(1,0,1,0,1,1,1,0,0),\\
&I_\textup{VV2}(-1,1,1,0,1,1,1,0,0),&I_\textup{VV2}(0,1,1,0,1,1,1,0,0),\\
&I_\textup{VV2}(1,1,1,0,1,1,1,0,0),&I_\textup{VV2}(1,-1,0,1,1,1,1,0,0),\\
&I_\textup{VV2}(1,0,0,1,1,1,1,0,0),&I_\textup{VV2}(0,0,1,1,1,1,1,0,0),\\
&I_\textup{VV2}(1,0,1,1,1,1,1,0,0),&I_\textup{VV2}(1,1,1,1,1,1,1,-1,0),\\
&I_\textup{VV2}(1,1,1,1,1,1,1,0,0)\big).&\end{array}
\end{equation}
\normalsize
A vector of dimensionless master integrals is obtained by factoring out the mass-dimension
\begin{equation}
f^\textup{VV1}_i(\epsilon,x,y,z)=(m_3)^{-\textup{dim}(g^\textup{VV1}_i)}g^\textup{VV1}_i(\epsilon,s,t,m_3,m_4),
\end{equation}
\begin{equation}
f^\textup{VV2}_i(\epsilon,x,y,z)=(m_3)^{-\textup{dim}(g^\textup{VV2}_i)}g^\textup{VV2}_i(\epsilon,s,t,m_3,m_4),
\end{equation}
where the set of dimensionless parameters is taken to be the same as in~\cite{Henn:2014lfa}:
\begin{equation}
(1+x)(1+xy)=\frac{s}{m_3^2},\quad -xz=\frac{t}{m_3^2},\quad x^2y=\frac{m_4^2}{m_3^2}.
\end{equation}
The calculation of the transformation to a canonical basis with \textit{CANONICA} takes about 8 minutes for topology 1 and about 15 minutes for topology 2. Samples of the transformations are presented in the following:
\paragraph*{Topology 1:}

\footnotesize
\begin{align}
f^\textup{VV1}_{8}=&
\left(\frac{2\epsilon^2 (x+1)}{(2\epsilon-1) (3\epsilon-1) x (y-1)}\right)f^{\textup{VV1}\prime}_{5}
+\left(-\frac{2\epsilon^2 (x y+1)}{(2\epsilon-1) (3\epsilon-1) x (y-1)}\right)f^{\textup{VV1}\prime}_{6}\nonumber\\
&+\left(-\frac{2\epsilon^2 (x+1)}{(2\epsilon-1) (3\epsilon-1) x (y-1)}\right)f^{\textup{VV1}\prime}_{7}
+\left(-\frac{2\epsilon^2}{(2\epsilon-1) (3\epsilon-1)}\right)f^{\textup{VV1}\prime}_{8},
\end{align}
\normalsize
\footnotesize
\begin{align}
f^\textup{VV1}_{10}=&
\left(\frac{2\epsilon^2 \left(\epsilon \left(5 x^2 y+3 x (y+1)+1\right)-x (2 x y+y+1)\right)}{(2\epsilon-1) (3\epsilon-2) (3\epsilon-1)}\right)f^{\textup{VV1}\prime}_{5}\nonumber\\
&+\left(\frac{2\epsilon^3 x^2 y}{(2\epsilon-1) (3\epsilon-2) (3\epsilon-1)}\right)f^{\textup{VV1}\prime}_{9}
+\left(-\frac{2\epsilon^2 x (x y+1) (\epsilon ((x-1) y+2)-x y-1)}{(2\epsilon-1) (3\epsilon-2) (3\epsilon-1) (y-1)}\right)f^{\textup{VV1}\prime}_{10}\nonumber\\
&+\left(-\frac{2\epsilon^2 x (2 x y+y+1)}{(3\epsilon-2) (3\epsilon-1)}\right)f^{\textup{VV1}\prime}_{11},
\end{align}
\normalsize
\footnotesize
\begin{align}
f^\textup{VV1}_{31}=&
\left(\frac{17}{x (x+1)^2 z (x y+1)^2}\right)f^{\textup{VV1}\prime}_{1}
+\left(\frac{24}{x (x+1)^2 z (x y+1)^2}\right)f^{\textup{VV1}\prime}_{5}\nonumber\\
&+\left(-\frac{1}{2 x (x+1)^2 z (x y+1)^2}\right)f^{\textup{VV1}\prime}_{9}
+\left(\frac{32}{7 x (x+1)^2 z (x y+1)^2}\right)f^{\textup{VV1}\prime}_{10}\nonumber\\
&+\left(-\frac{64}{7 x (x+1)^2 z (x y+1)^2}\right)f^{\textup{VV1}\prime}_{11}
+\left(\frac{9}{x (x+1)^2 z (x y+1)^2}\right)f^{\textup{VV1}\prime}_{16}\nonumber\\
&+\left(\frac{11}{15 x (x+1)^2 z (x y+1)^2}\right)f^{\textup{VV1}\prime}_{17}
+\left(\frac{2}{3 x (x+1)^2 z (x y+1)^2}\right)f^{\textup{VV1}\prime}_{19}\nonumber\\
&+\left(-\frac{6}{x (x+1)^2 z (x y+1)^2}\right)f^{\textup{VV1}\prime}_{22}
+\left(-\frac{36}{x (x+1)^2 z (x y+1)^2}\right)f^{\textup{VV1}\prime}_{23}\nonumber\\
&+\left(-\frac{5}{x (x+1)^2 z (x y+1)^2}\right)f^{\textup{VV1}\prime}_{24}
+\left(-\frac{23}{x (x+1)^2 z (x y+1)^2}\right)f^{\textup{VV1}\prime}_{25}\nonumber\\
&+\left(\frac{2}{x (x+1)^2 z (x y+1)^2}\right)f^{\textup{VV1}\prime}_{31}.
\end{align}
\normalsize
\paragraph*{Topology 2:}

\footnotesize
\begin{align}
f^\textup{VV2}_{16}=&
\left(\frac{2\epsilon^2}{(2\epsilon-1) (3\epsilon-1)}\right)f^{\textup{VV2}\prime}_{14}
+\left(\frac{2\epsilon^2}{(2\epsilon-1) (3\epsilon-1)}\right)f^{\textup{VV2}\prime}_{15}\nonumber\\
&+\left(\frac{\epsilon (\epsilon x y+\epsilon)}{2 (2\epsilon-1) (3\epsilon-1) (y-1)}\right)f^{\textup{VV2}\prime}_{16},
\end{align}
\normalsize
\footnotesize
\begin{align}
f^\textup{VV2}_{18}=&
\left(\frac{\epsilon^2 \left(\epsilon \left(x^2 y (2 y-1)-x y-2\right)-x^2 y^2+1\right)}{(2\epsilon-1) (3\epsilon-2) (3\epsilon-1) x (y-1)}\right)f^{\textup{VV2}\prime}_{4}\nonumber\\
&+\left(\frac{2\epsilon^2 (x+1) (\epsilon (x (y-2)+2)+x-1)}{(2\epsilon-1) (3\epsilon-2) (3\epsilon-1) x (y-1)}\right)f^{\textup{VV2}\prime}_{18}\nonumber\\
&+\left(-\frac{2\epsilon^2 (\epsilon (2 x (y+1)+1)-x (y+1))}{(2\epsilon-1) (3\epsilon-2) (3\epsilon-1)}\right)f^{\textup{VV2}\prime}_{19},
\end{align}
\normalsize
\footnotesize
\begin{align}
f^\textup{VV2}_{28}=&
\left(\frac{y (z-2)+z}{x z (x y+z) (x y z-x y+x z+z)}\right)f^{\textup{VV2}\prime}_{23}
+\left(\frac{1}{x^2 z (x y+z)}\right)f^{\textup{VV2}\prime}_{28}.\nonumber\\
\end{align}
\normalsize
The resulting canonical forms, which are omitted here for brevity, exhibit the alphabets
\begin{align}
\mathcal{A}^\textup{VV1}=\,\,&\big\{x,\,1+x,\,1-y,\,y,\,1+xy,\,1+x(1+y-z),\,1-z,\nonumber\\
\,\,&\,\,\,\,1+(1+x)y-z,\,z,\,z-y,\,z+xy,\,1+xz\big\},
\end{align}
\begin{align}
\mathcal{A}^\textup{VV2}=\,\,&\big\{x,\,1+x,\,1-y,\,y,\,1+xy,\,1+x(1+y-z),\,1-z,\nonumber\\
\,\,&\,\,\,\,1+(1+x)y-z,\,z,\,z-y,\,z+xy,\,1+xz,\nonumber\\
\,\,&\,\,\,\,z-y+yz+xyz,\,z-xy+xz+xyz\big\}.
\end{align}

\section{Single top-quark production}
\begin{figure}[H]
\begin{center}
\includegraphics[scale=1]{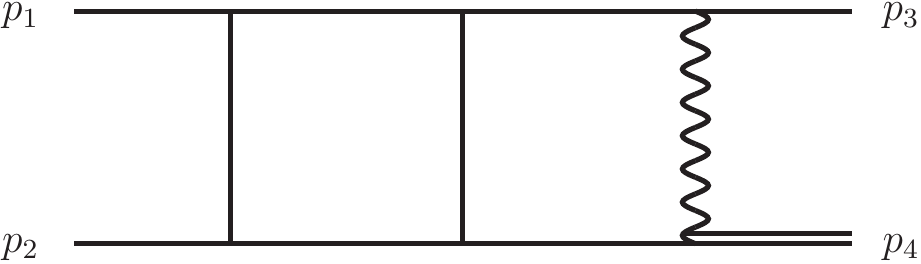}
\end{center}
\caption{Single top-quark production topology 1.}
\label{fig:STT1}
\end{figure}
\begin{figure}[H]
\begin{center}
\includegraphics[scale=1]{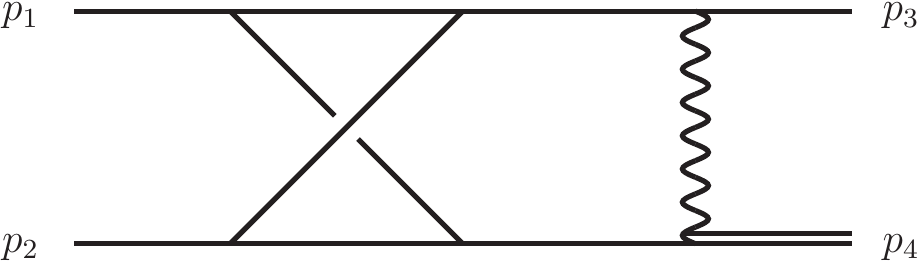}
\end{center}
\caption{Single top-quark production topology 2.}
\label{fig:STT2}
\end{figure}
The integral topologies in Figs. \ref{fig:STT1} and \ref{fig:STT2} are necessary to include certain color-suppressed contributions in the NNLO QCD corrections to single top-quark production~\cite{Assadsolimani:2014oga}, which have been neglected in other calculations~\cite{Brucherseifer:2014ama, Berger:2017zof}. These integrals have not been considered before and therefore represent a new result. The topologies are given by the following sets of inverse propagators and irreducible scalar products:
\paragraph*{Topology 1:}
\begin{equation}
\begin{array}{lll}
P_1=l_2^2, & P_4=(l_2+p_2)^2, & P_7=(l_1+l_2-p_1+p_3)^2, \\
P_2=l_1^2-m_W^2, & P_5=(l_1-p_4)^2, &  P_8=(l_1-p_2)^2, \\
P_3=(l_1+p_3)^2, &P_6=(l_2-p_1)^2, & P_9=(l_2+p_3+p_1)^2. \\  
\end{array}
\end{equation}
\paragraph*{Topology 2:}
\begin{equation}
\begin{array}{lll}
P_1=l_2^2, & P_4=(l_2-p_2)^2, & P_7=(l_1-l_2-p_1+p_3)^2, \\
P_2=l_1^2-m_W^2, & P_5=(l_1-p_4)^2, &  P_8=(l_1+p_2)^2, \\
P_3=(l_1+p_3)^2, &P_6=(l_2-l_1-p_3)^2, & P_9=(l_2-p_3)^2. \\  
\end{array}
\end{equation}
The momenta $p_1$ and $p_2$ are incoming, and $p_3$ and $p_4$ are outgoing, and the kinematics of both topologies are given by
\begin{equation}
p_1+p_2=p_3+p_4,
\end{equation}
\begin{equation}
p_1^2=0,\quad p_2^2=0,\quad p_3^2=0,\quad p_4^2=m_t^2,
\end{equation}
\begin{equation}
s=(p_1+p_2)^2,\quad t=(p_2-p_3)^2.
\end{equation}
Topology 1 has a basis of 31 master integrals: 
\small
\begin{equation}
\label{OldBasisSTT1}
\begin{array}{lll}
\vec{g}^{\,\textup{ST1}}(\epsilon,s,t,m_t^2,m_W^2)\,\,\,=&\big(I_\textup{ST1}(0,1,0,1,0,1,0,0,0),&I_\textup{ST1}(0,1,0,1,1,1,0,0,0),\\
&I_\textup{ST1}(0,0,1,1,1,1,0,0,0),&I_\textup{ST1}(0,1,1,1,1,1,0,0,0),\\
&I_\textup{ST1}(1,1,0,0,0,0,1,0,0),&I_\textup{ST1}(1,1,-1,0,0,0,1,0,0),\\
&I_\textup{ST1}(0,1,0,1,0,0,1,0,0),&I_\textup{ST1}(-1,1,0,1,0,0,1,0,0),\\
&I_\textup{ST1}(0,0,1,1,0,0,1,0,0),&I_\textup{ST1}(0,1,1,1,0,0,1,0,0),\\
&I_\textup{ST1}(1,1,1,1,0,0,1,0,0),&I_\textup{ST1}(1,1,1,1,-1,0,1,0,0),\\
&I_\textup{ST1}(1,1,0,0,1,0,1,0,0),&I_\textup{ST1}(1,0,1,0,1,0,1,0,0),\\
&I_\textup{ST1}(1,1,1,0,1,0,1,0,0),&I_\textup{ST1}(1,1,1,-1,1,0,1,0,0),\\
&I_\textup{ST1}(0,1,0,0,0,1,1,0,0),&I_\textup{ST1}(0,1,0,1,0,1,1,0,0),\\
&I_\textup{ST1}(-1,1,0,1,0,1,1,0,0),&I_\textup{ST1}(1,1,0,1,0,1,1,0,0),\\
&I_\textup{ST1}(1,1,-1,1,0,1,1,0,0),&I_\textup{ST1}(0,1,1,1,0,1,1,0,0),\\
&I_\textup{ST1}(0,1,0,0,1,1,1,0,0),&I_\textup{ST1}(-1,1,0,0,1,1,1,0,0),\\
&I_\textup{ST1}(1,1,0,0,1,1,1,0,0),&I_\textup{ST1}(1,1,-1,0,1,1,1,0,0),\\
&I_\textup{ST1}(0,1,0,1,1,1,1,0,0),&I_\textup{ST1}(1,1,0,1,1,1,1,0,0),\\
&I_\textup{ST1}(1,1,1,1,1,1,1,0,0),&I_\textup{ST1}(1,1,1,1,1,1,1,0,-1),\\
&I_\textup{ST1}(1,1,1,1,1,1,1,-1,0)\big).&
\end{array}
\end{equation}
\normalsize
Topology 2 has a basis of 35 master integrals: 
\small
\begin{equation}
\label{OldBasisSTT2}
\begin{array}{lll}
\vec{g}^{\,\textup{ST2}}(\epsilon,s,t,m_t^2,m_W^2)\,\,\,=&\big(I_\textup{ST2}(1,1,0,0,0,1,0,0,0),&I_\textup{ST2x12}(-1,1,0,1,0,1,0,0,0),\\
&I_\textup{ST2x12}(0,1,0,1,0,1,0,0,0),&I_\textup{ST2}(-1,1,0,1,0,1,0,0,0),\\
&I_\textup{ST2}(0,1,0,1,0,1,0,0,0),&I_\textup{ST2}(1,0,0,0,1,1,0,0,0),\\
&I_\textup{ST2}(1,1,-1,0,1,1,0,0,0),&I_\textup{ST2}(1,1,0,0,1,1,0,0,0),\\
&I_\textup{ST2x12}(0,1,0,1,1,1,0,0,0),&I_\textup{ST2}(0,1,0,1,1,1,0,0,0),\\
&I_\textup{ST2x12}(1,1,-1,1,1,1,0,0,0),&I_\textup{ST2x12}(1,1,0,1,1,1,0,0,0),\\
&I_\textup{ST2}(1,1,-1,1,1,1,0,0,0),&I_\textup{ST2}(1,1,0,1,1,1,0,0,0),\\
&I_\textup{ST2}(0,0,1,1,1,1,0,0,0),&I_\textup{ST2x12}(-1,1,1,1,1,1,0,0,0),\\
&I_\textup{ST2x12}(0,1,1,1,1,1,0,0,0),&I_\textup{ST2}(-1,1,1,1,1,1,0,0,0),\\
&I_\textup{ST2}(0,1,1,1,1,1,0,0,0),&I_\textup{ST2}(-1,1,0,1,0,0,1,0,0),\\
&I_\textup{ST2}(0,1,-1,1,0,0,1,0,0),&I_\textup{ST2}(0,1,1,1,0,0,1,0,0),\\
&I_\textup{ST2x12}(1,1,1,1,-1,0,1,0,0),&I_\textup{ST2x12}(1,1,1,1,0,0,1,0,0),\\
&I_\textup{ST2}(1,1,1,1,-1,0,1,0,0),&I_\textup{ST2}(1,1,1,1,0,0,1,0,0),\\
&I_\textup{ST2}(1,1,-1,1,0,1,1,0,0),&I_\textup{ST2}(1,1,0,1,0,1,1,0,0),\\
&I_\textup{ST2}(1,1,1,1,0,1,1,0,0),&I_\textup{ST2}(1,1,-1,1,1,1,1,0,0),\\
&I_\textup{ST2}(1,1,0,1,1,1,1,0,0),&I_\textup{ST2}(1,0,1,1,1,1,1,0,0),\\
&I_\textup{ST2}(1,1,1,1,1,1,1,-2,0),&I_\textup{ST2}(1,1,1,1,1,1,1,-1,0),\\
&I_\textup{ST2}(1,1,1,1,1,1,1,0,0)\big).&
\end{array}
\end{equation}
\normalsize
A vector of dimensionless master integrals is obtained by factoring out the mass-dimension
\begin{equation}
f^\textup{ST1}_i(\epsilon,x,y,z)=(m_W)^{-\textup{dim}(g^\textup{ST1}_i)}g^\textup{ST1}_i(\epsilon,s,t,m_t,m_W),
\end{equation}
\begin{equation}
f^\textup{ST2}_i(\epsilon,x,y,z)=(m_W)^{-\textup{dim}(g^\textup{ST2}_i)}g^\textup{ST2}_i(\epsilon,s,t,m_t,m_W),
\end{equation}
with
\begin{equation}
x=\frac{s}{m_W^2},\quad y=\frac{t}{m_W^2},\quad z=\frac{m_t^2}{m_W^2}.
\end{equation}
The calculation of the transformation to a canonical basis with \textit{CANONICA} takes about 12 minutes for topology 1 and about 20 minutes for topology 2. Samples of the transformations are presented in the following:

\paragraph*{Topology 1:}

\footnotesize
\begin{align}
f^\textup{ST1}_{10}=&
\left(-\frac{3\epsilon^2 \left((3\epsilon-1) x^2+x (-5\epsilon (z-1)+z-1)+2\epsilon (z-1) z\right)}{4 (2\epsilon-1)^2 (3\epsilon-1) (x-z)^2}\right)f^{\textup{ST1}\prime}_{7}\nonumber\\
&+\left(-\frac{3\epsilon^2 ((3\epsilon-1) x-4\epsilon z+z)}{2 (2\epsilon-1)^2 (3\epsilon-1) (x-z)}\right)f^{\textup{ST1}\prime}_{8}\nonumber\\
&+\left(\frac{3\epsilon^3 x}{4 (2\epsilon-1)^2 (3\epsilon-1) (x-z)}\right)f^{\textup{ST1}\prime}_{9}
+\left(\frac{9\epsilon^2 x (x-z+1)}{40 (2\epsilon-1)^2 (x-z)^2}\right)f^{\textup{ST1}\prime}_{10},
\end{align}
\normalsize
\footnotesize
\begin{align}
f^\textup{ST1}_{18}=&
\left(\frac{\epsilon^2}{20 (2\epsilon-1) (3\epsilon-1)}\right)f^{\textup{ST1}\prime}_{1}
+\left(\frac{3\epsilon^2 (x-z+1)}{2 (2\epsilon-1) (3\epsilon-1) (x-z)}\right)f^{\textup{ST1}\prime}_{7}\nonumber\\
&+\left(\frac{3\epsilon (\epsilon (x-z-3)+1)}{2 (2\epsilon-1) (3\epsilon-1) (x-z)}\right)f^{\textup{ST1}\prime}_{8}
+\left(-\frac{\epsilon}{20 (2\epsilon-1) (x-z)}\right)f^{\textup{ST1}\prime}_{17}\nonumber\\
&+\left(-\frac{\epsilon^2 (x-z+1)}{(2\epsilon-1) (3\epsilon-1) (x-z)}\right)f^{\textup{ST1}\prime}_{18}
+\left(-\frac{\epsilon (3\epsilon (3 x-3 z+5)-2)}{2 (2\epsilon-1) (3\epsilon-1) (x-z)}\right)f^{\textup{ST1}\prime}_{19},
\end{align}
\normalsize
\footnotesize
\begin{align}
f^\textup{ST1}_{31}=&
\left(\frac{6 x+6 y-3 z}{20 x^2 (x+y-z+1)}\right)f^{\textup{ST1}\prime}_{3}
+\left(-\frac{3}{8 x (x-z)}\right)f^{\textup{ST1}\prime}_{7}
+\left(\frac{3 (-x+y+z)}{8 x (x-z) (x+y-z)}\right)f^{\textup{ST1}\prime}_{8}\nonumber\\
&+\left(-\frac{9 (2 x+2 y-z)}{4 x^2 (x+y-z+1)}\right)f^{\textup{ST1}\prime}_{9}
+\left(\frac{9 y}{40 x (x-z) (x+y-z)}\right)f^{\textup{ST1}\prime}_{10}\nonumber\\
&+\left(-\frac{7 y}{4 x (x-z) (x+y-z)}\right)f^{\textup{ST1}\prime}_{11}
+\left(-\frac{y}{4 x (x-z) (x+y-z)}\right)f^{\textup{ST1}\prime}_{12}\nonumber\\
&+\left(\frac{6 x+6 y-3 z}{4 x^2 (x+y-z+1)}\right)f^{\textup{ST1}\prime}_{14}
+\left(\frac{-x+y+z}{80 x (x-z) (x+y-z)}\right)f^{\textup{ST1}\prime}_{17}\nonumber\\
&+\left(\frac{1}{x (x+y-z)}\right)f^{\textup{ST1}\prime}_{18}
+\left(\frac{9}{2 x (x+y-z)}\right)f^{\textup{ST1}\prime}_{19}
+\left(\frac{2 y}{x (x-z) (x+y-z)}\right)f^{\textup{ST1}\prime}_{21}\nonumber\\
&+\left(\frac{y}{x (x-z) (x+y-z)}\right)f^{\textup{ST1}\prime}_{22}
+\left(\frac{64 (2 x+(y-1) z)}{15 x^2 (x-z) (x+y-z+1)}\right)f^{\textup{ST1}\prime}_{23}\nonumber\\
&+\left(\frac{32 (x+y-1)}{15 x^2 (x+y-z+1)}-\frac{14}{5 x (x-z)}\right)f^{\textup{ST1}\prime}_{24}
+\left(\frac{8 (x+y-1)}{x^2 (x+y-z+1)}\right)f^{\textup{ST1}\prime}_{25}\nonumber\\
&+\left(\frac{16 (x+y-1)}{5 x^2 (x+y-z+1)}\right)f^{\textup{ST1}\prime}_{26}
+\left(\frac{x+y-1}{x^2 (x+y-z+1)}\right)f^{\textup{ST1}\prime}_{29}
+\left(\frac{1}{x (x-z)}\right)f^{\textup{ST1}\prime}_{31}.
\end{align}
\normalsize

\paragraph*{Topology 2:}

\footnotesize
\begin{align}
f^\textup{ST2}_{8}=&
\left(\frac{\epsilon^2 (\epsilon (x (2 z-1)+(5-2 z) z)+z (-x+z-2))}{2 (2\epsilon-1) (3\epsilon-2) (3\epsilon-1) z}\right)f^{\textup{ST2}\prime}_{1}
+\left(\frac{5\epsilon^3 x (z-1)}{(2\epsilon-1) (3\epsilon-2) (3\epsilon-1) z}\right)f^{\textup{ST2}\prime}_{6}\nonumber\\
&+\left(-\frac{5\epsilon^2 (x-z+1) (\epsilon (x (2 z-1)-2 (z-1) z)+z (-x+z-1))}{(2\epsilon-1) (3\epsilon-2) (3\epsilon-1) z (x-z)}\right)f^{\textup{ST2}\prime}_{7}\nonumber\\
&+\left(-\frac{5\epsilon^2 (z-1) ((z-1) z-x (z+1))}{(3\epsilon-2) (3\epsilon-1) z^2}\right)f^{\textup{ST2}\prime}_{8},
\end{align}
\normalsize
\footnotesize
\begin{align}
f^\textup{ST2}_{21}=&
\left(\frac{10\epsilon^3 (z-1) \left(\epsilon \left(2 z^2+7 z-1\right)-2 z^2-5 z+1\right) (x+y)}{9 (\epsilon-1) (2\epsilon-1) (3\epsilon-2) (3\epsilon-1) z^2}\right)f^{\textup{ST2}\prime}_{20}\nonumber\\
&+\left(\frac{20\epsilon^2 \left(\epsilon^2 \left(z^2+14 z+1\right)-\epsilon z (z+15)+3 z\right) (x+y)}{9 (\epsilon-1) (2\epsilon-1) (3\epsilon-2) (3\epsilon-1) z}\right)f^{\textup{ST2}\prime}_{21},
\end{align}
\normalsize
\footnotesize
\begin{align}
f^\textup{ST2}_{34}=&
\left(-\frac{5 (x+y) (x+y+1)}{3 x^2 (z-1) (x+y-z+1)}\right)f^{\textup{ST2}\prime}_{7}
+\left(-\frac{5 (x+y) (x+y+1)}{3 x^2 (z-1) (x+y-z+1)}\right)f^{\textup{ST2}\prime}_{8}\nonumber\\
&+\left(-\frac{5 (x+y+1) (23 x+23 y-32 z+32)}{6 x^2 (z-1) (x+y-z+1)}\right)f^{\textup{ST2}\prime}_{11}\nonumber\\
&+\left(-\frac{(x+y+1) (25 x+25 y-32 z+32)}{3 x^2 (z-1) (x+y-z+1)}\right)f^{\textup{ST2}\prime}_{12}\nonumber\\
&+\left(-\frac{5 (x+y+1) (37 x+37 y-4 z+4)}{6 x^2 (z-1) (x+y-z+1)}-\frac{20 (y+1)}{3 x^2 (y-1)}\right)f^{\textup{ST2}\prime}_{13}\nonumber\\
&+\left(\frac{5 (x+y+1)}{3 x^2 (z-1)}+\frac{5 (y+1)}{x^2 (y-1)}\right)f^{\textup{ST2}\prime}_{14}
+\left(\frac{43 (y+1)}{3 x^2 (y-1)}-\frac{29 (x+y+1)}{12 x^2 (x+y-z+1)}\right)f^{\textup{ST2}\prime}_{23}\nonumber\\
&+\left(\frac{-x-y-1}{x^2 (x+y-z+1)}+\frac{55 (y+1)}{12 x^2 (y-1)}\right)f^{\textup{ST2}\prime}_{24}
+\left(\frac{-x-y-1}{x^2 (x+y-z+1)}+\frac{y}{x^2}+\frac{1}{x^2}\right)f^{\textup{ST2}\prime}_{29}\nonumber\\
&+\left(\frac{(x+y) (x+y+1)}{x^2 (z-1) (x+y-z+1)}\right)f^{\textup{ST2}\prime}_{30}
+\left(-\frac{z}{x^2 (x+y-z+1)}\right)f^{\textup{ST2}\prime}_{32}\nonumber\\
&+\left(\frac{y+1}{x^2 (y-1)}\right)f^{\textup{ST2}\prime}_{33}
+\left(\frac{x+y+1}{x^2 (x+y-z+1)}\right)f^{\textup{ST2}\prime}_{34}.
\end{align}
\normalsize
The resulting canonical forms, which are omitted here for brevity, exhibit the alphabets

\begin{align}
\mathcal{A}^\textup{ST1}=\,\,&\big\{x,\,y,\,x+y,\,x-z,\,y-z,\,x+y-z,\,1+x+y-z,\,-1+z,\nonumber \\ 
\,\,&\,\,\,\,z,\,1+x-z,\,y(-1+z)+(1+x-z)z\big\},
\end{align}

\begin{align}
\mathcal{A}^\textup{ST2}=\,\,&\big\{x,\,-1+y,\,y,\,x+y,\,x-z,\,1+x-z,\,y-z,\,x+y-z,\nonumber \\ 
\,\,&\,\,\,\,1+x+y-z,\,-1+z,\,z,\,x+y(1-z),\,x(-1+y)+y(y-z),\nonumber \\ 
\,\,&\,\,\,\,y(-1+z)+(1+x-z)z\big\}.
\end{align}
       \chapter{Conclusions}                                                                                                                                                                                                                                                                                                                                                                                                                                                                                                                                                                                                                                                                                                                                                                           
\label{chap:conclusion}

The calculation of higher-order corrections for the observables measured at the LHC is necessary to match the ever-increasing experimental precision. One of the main challenges in these calculations is posed by the evaluation of multi-loop Feynman integrals. Many phenomenologically interesting processes involve Feynman integrals depending on several kinematic invariants and mass scales such as the top-quark, the Higgs boson or the electroweak gauge bosons masses. Therefore, it is important to develop techniques for the evaluation of multi-scale Feynman integrals. In recent years the strategy to transform Feynman integrals to a canonical basis has been very successful. However, publicly available programs to compute such transformations have been limited to Feynman integrals depending on a single dimensionless scale. This thesis presents both an algorithm to transform multi-loop Feynman integrals depending on an arbitrary number of scales to a canonical basis and its implementation in the \texttt{Mathematica} package \textit{CANO\-NI\-CA}.

Analyzing the transformation law that governs transformations to a canonical form has been the main guiding principle in the construction of the algorithm. It has been shown that the expansion of a variant of the transformation law in the dimensional regulator leads to a finite number of differential equations for the coefficients in the expansion of the transformation. Using the decomposition of multivariate rational functions in terms of Leinartas functions, a method to generate an ansatz for the solution of these differential equations has been developed. The resulting equations in the parameters of the ansatz were argued to be linear upon fixing the freedom of a subsequent constant transformation. As a consequence, the transformation can be computed by solving only linear equations. It was suggested in previous approaches to compute the transformation in a recursion over sectors by exploiting the block-triangular structure of the differential equations. This strategy has been incorporated as well because it allows to make more specific ansatzes for the individual steps of the recursion and therefore leads to a better performance of the proposed algorithm.


It has been instrumental to pursue the construction of the algorithm from a rather broad perspective by first studying general properties of transformations to a canonical form. This strategy not only led to an algorithm with a wide range of applicability but also allowed to derive interesting results beyond the construction of the algorithm. Most notably, it has been proven that canonical forms of differential equations are \textit{unique} up to constant transformations. This property may prove useful as guiding principle for generalizing the notion of a canonical form beyond the realm of Chen iterated integrals.

The algorithm has been implemented in the publicly available \texttt{\justify Mathematica} package \textit{CANONICA}, which allows to compute rational transformations of differential equations of Feynman integrals into a canonical form. In addition to the main functionality, the package also provides numerous supplemental functions to perform frequently occurring computations with differential equations of Feynman integrals. The package has an extensive documentation including an interactive manual notebook, short usage messages of all functions and options as well as numerous examples of its usage. 
\textit{CANONICA} has been successfully tested on a variety of non-trivial multi-loop problems of which a selection has been discussed in this thesis. The presented examples are two- and three-loop integral topologies depending on up to three dimensionless scales including previously unknown topologies contributing to the NNLO QCD corrections to single top-quark production at the LHC.

In its current form, the algorithm is limited to differential equations that admit a \textit{rational} transformation to a canonical form. The presented applications illustrate that a wide range of phenomenologically relevant integral topologies can be transformed into a canonical form with a rational transformation. Moreover, as most of the main aspects of the algorithm carry over to the case of transformations with an algebraic dependence on the invariants, it can be expected that the generalization of the algorithm to the algebraic case is achievable in the future. Another possible direction of future work is to improve the choice of the ansatz in terms of rational functions. In particular, it would be desirable to constrain the set of rational functions that occur in the decomposition of the transformation to a finite set by deriving bounds on their numerator and denominator polynomials. If such bounds were known, the failure of the algorithm to compute a rational transformation to a canonical form would imply that no such transformation exists.

In conclusion, this thesis presented an algorithm that allows to drastically simplify the computation of multi-loop Feynman integrals by transforming them to a canonical basis. The implementation of this algorithm in the \textit{CANONICA} package is the first publicly available program to calculate such transformations for Feynman integrals depending on multiple scales. Therefore, it can be used for a much wider range of processes than alternative programs based on Lee's algorithm, which are restricted to single-scale problems. It has been demonstrated through complex multi-loop examples that \textit{CANONICA} is capable of handling state-of-the-art integral topologies. On this account, \textit{CANONICA} is expected to be a valuable contribution to the ongoing efforts to automatize the calculation of higher-order corrections beyond the next-to-leading order.

       \chapter*{Acknowledgments}
\label{sec:Acknowledgments}

First of all, I would like to thank my supervisor Peter Uwer for the opportunity to work on a project that suited my interests so well. More than that, I thank him for the freedom to pursue the topic in my own way and the trust expressed through that. I have learned a lot from him and am very grateful for his guidance and for always being available for advice. 

The work on this thesis has been financially supported by the German Research Foundation (DFG) via the Research Training Group `Masse Spektrum Symmetrie' which I gratefully acknowledge. 

The present version of this thesis incorporates suggestions on the manuscript made by Peter Uwer, Rasmus Bentmann and Markus Schulze for which I am very thankful. Moreover, I wish to thank Frank Brand for proofreading the final manuscript and for countless advice. I would also like to thank Bas Tausk for answering various questions mostly regarding references. The publications on which large portions of this thesis are based have benefited from the detailed remarks and suggestions of the referees, which I am grateful for. The Feynman diagrams in this thesis have been drawn with \texttt{JAXODRAW} \cite{Binosi:2008ig, Vermaseren:1994je}.

Furthermore, my gratitude goes to all current and former members of the particle physics phenomenology group at the Humboldt-Universit\"at zu Berlin for creating a friendly and pleasant working environment that has made my time much more enjoyable. I thank Till Martini and Felix Stollenwerk for many coffee breaks and for sharing the experience of pursuing a Ph.D. with me.

I would like to thank my friends for being understanding all the time and for taking my mind off physics when I needed it. In particular, I thank Dominik Sittel for always keeping my spirits high with his humor. 

Above all, I wish to thank my parents, grandparents, sister and uncle for their continuous and unconditional love and support. 
       
     \appendix
       \chapter{Massive tadpole integral} 
\label{App:Tadpole}
In this appendix, the calculation of the one-loop tadpole integral 
\begin{equation}
g_1=\int\frac{\textup{d}^dl}{\textup{i}\pi^{d/2}}\frac{1}{l^2-m^2},
\end{equation}
is presented, because this integral is needed as a boundary condition in \sec{sec:solitopoly}. Similar treatments may be found in most standard textbooks on quantum field theory \cite{Bjorken:1965zz, Peskin:1995ev, Weinberg:1995mt, Schwartz:2013pla}. The first step is to perform a Wick rotation~\cite{PhysRev.96.1124} to obtain a Euclidean loop momentum\footnote{The \emph{mostly minus} convention for the Minkowski metric is adopted here.}
\begin{equation}
l_0\rightarrow \textup{i}l_0\quad \Rightarrow\quad l^2\rightarrow -l_E^2,
\end{equation}
which allows to express the integral as
\begin{equation}
g_1=\frac{-1}{\pi^{d/2}}\int\textup{d}^dl_E\frac{1}{l_E^2+m^2}.
\end{equation}
Using spherical coordinates in $d$ dimensions, the integration is split into an integration over the $(d-1)$\nobreakdash-dimensional unit sphere and a radial integration
\begin{equation}
\int\textup{d}^dl_E=\int\textup{d}\Omega_{d-1}\int_0^\infty l_E^{d-1}.
\end{equation}
With the surface area of the $(d-1)$\nobreakdash-dimensional unit sphere given by
\begin{equation}
\Omega_{d-1}=\int\textup{d}\Omega_{d-1}=\frac{2\pi^{d/2}}{\Gamma\left(\frac{d}{2}\right)},
\end{equation}
this leads to
\begin{equation}
g_1=\frac{-2}{\Gamma\left(\frac{d}{2}\right)}\int_0^\infty\textup{d}l_E\frac{l_E^{d-1}}{l_E^2+m^2}.
\end{equation}
By virtue of the substitution
\begin{equation}
x=\frac{m^2}{l_E^2+m^2},
\end{equation}
the radial integration is cast in the form
\begin{equation}
g_1=-\frac{m^{d-2}}{\Gamma\left(\frac{d}{2}\right)}\int_0^1\textup{d}x(1-x)^{\frac{d}{2}-1}x^{-\frac{d}{2}},
\end{equation}
which resembles the following integral representation of the $\beta$\nobreakdash-function:
\begin{equation}
\beta(a,b)=\frac{\Gamma(a)\Gamma(b)}{\Gamma(a+b)}=\int_0^1\textup{d}x(1-x)^{a-1}x^{b-1}.
\end{equation}
Inserting this representation allows to express the tadpole integral in terms of a $\Gamma$\nobreakdash-function:
\begin{equation}
g_1=-m^{d-2}\Gamma\left(1-\frac{d}{2}\right).
\end{equation}
       \chapter{Polynomial rings} 
\label{App:PolyAlgebra}
For the reader's convenience, this appendix reproduces some standard definitions and results about polynomial algebra, which are used in \sec{sec:Leinartas}. For a more detailed exposition, the reader is referred to~\cite{Cox:2007:IVA:1204670}. The ring of polynomials in $d$ variables and coefficients in the field $K$ is denoted by $K[X]$ with $X=\{x_1,\dots,x_d\}$. Also, recall that in this thesis the natural numbers $\mathbb{N}$ include zero.
\begin{defn}[Ideal]
A subset $I\subseteq K[X]$ is called an ideal if the following conditions are satisfied:
\begin{enumerate}
\item $0\in I$.
\item If $f,g\in I$, then $f+g\in I$.
\item If $f\in I$ and $h\in K[X]$, then $hf\in I$.
\end{enumerate}
\end{defn}
\begin{defn}[Ideal generated by a set of polynomials]
Let $\{f_1,\dots,f_m\}\subset K[X]$ be a set of polynomials. Then 
\begin{equation*}
\langle f_1,\dots,f_m\rangle=\left\{\sum_{i=1}^mh_if_i \quad\big|\quad h_1,\dots h_m\in K[X]\right\}
\end{equation*}
is an ideal, which is called the ideal generated by $\{f_1,\dots,f_m\}$.
\end{defn}
\begin{defn}[Irreducible polynomial]
A polynomial $f\in K[X]$ is called irreducible over $K$ if f is non-constant and is not the product of two non-constant polynomials in $K[X]$.
\end{defn}
\begin{thm}[Factorization]
Every non-constant $f\in K[X]$ can be written as a product $f=f_1^{e_1}\dots f_m^{e_m}$ of irreducible polynomials over $K$. This factorization is unique up to multiplication with constant factors and reordering of the irreducible factors $f_i$. 
\end{thm}
\begin{thm}[Weak Nullstellensatz]
Let $I\subseteq K[X]$ be an ideal that satisfies 
\begin{equation*}
V(I)=\bigcap_{f\in I}V(f)=\emptyset,
\end{equation*}
then $I=K[X]$.
\end{thm}

\begin{defn}[Monomial order]
\label{def:monOrd}
A monomial order on $K[X]$ is a relation $>$ on the set of monomials $x^\alpha$, $\alpha\in\mathbb{N}^d$ which satisfies:
\begin{enumerate}
\item $>$ is a total order on $\mathbb{N}^d$.
\item \label{def:monOrd:2} If $\alpha>\beta$ and $\gamma\in\mathbb{N}^d$, then $\alpha+\gamma>\beta+\gamma$.
\item \label{def:monOrd:3} For all $A\subseteq\mathbb{N}^d$ there exists an $\alpha\in A$ such that $\beta>\alpha$ for all $\beta\neq\alpha$ in A.
\end{enumerate}
\end{defn}

While the considerations in \sec{sec:Leinartas} are agnostic about the monomial order, in practice the \emph{lexicographic order} has proven to be a good choice.
\begin{defn}[Lexicographic order]
For $\alpha=(\alpha_1,\dots,\alpha_d)$ and $\beta=(\beta_1,\dots,\beta_d)$ in $\mathbb{N}^d$, it is said that $\alpha>_\textup{lex}\beta$ if the leftmost non-zero entry of $\alpha-\beta\in\mathbb{N}^d$ is positive.
\end{defn}
Note that different orders of the variables give rise to different lexicographic orders.
\begin{defn}[]
Let $f=\sum_\alpha a_\alpha x^\alpha$ be a non-zero polynomial in $K[X]$ and $\alpha\in\mathbb{N}^d$ and let $>$ be a monomial order. 
\begin{enumerate}
\item The multidegree of $f$ is
\begin{equation*}
\textup{multideg}(f)=\textup{max}\left\{\alpha\in\mathbb{N}^d\,|\,a_\alpha\neq 0\right\}
\end{equation*}
the maximum is taken with respect to the monomial order $>$.
\item The leading coefficient of $f$ is
\begin{equation*}
\textup{LC}(f)=a_{\textup{multideg}(f)}\in K.
\end{equation*}
\item The leading monomial of $f$ is
\begin{equation*}
\textup{LM}(f)=x^{\textup{multideg}(f)}.
\end{equation*}
\item The leading term of $f$ is
\begin{equation*}
\textup{LT}(f)=\textup{LC}(f)\cdot\textup{LM}(f).
\end{equation*}
\end{enumerate}
\end{defn}
\begin{lemma}
\label{lem:multidegProperties}
Let $f,g\in K[X]$ be non-zero polynomials. Then 
\begin{equation*}
\textup{multideg}(fg)=\textup{multideg}(f)+\textup{multideg}(g).
\end{equation*}
\end{lemma}

\begin{lemma}
\label{lem:lowerboundmonoordering}
Let $>$ be a relation on $\mathbb{N}^d$ satisfying: 
\begin{enumerate}
\item $>$ is a total order on $\mathbb{N}^d$.
\item If $\alpha>\beta$ and $\gamma\in\mathbb{N}^d$, then $\alpha+\gamma>\beta+\gamma$.
\end{enumerate}
Then $\alpha\geq 0$ for all $\alpha\in\mathbb{N}^d$ if and only if for all $A\subseteq\mathbb{N}^d$ there exists an $\alpha\in A$ such that $\beta>\alpha$ for all $\beta\neq\alpha$ in A.
\end{lemma}
This lemma implies that $\alpha\geq 0$ holds for any monomial order and for all $\alpha\in\mathbb{N}^d$.

\begin{defn}[Set of leading terms]
Fix a monomial order on $K[X]$ and let $I\subseteq K[X]$ be an ideal other than $\{0\}$, then $\textup{LT}(I)$ denotes the set of leading terms of non-zero elements of $I$:
\begin{equation*}
\textup{LT}(I)=\left\{cx^\alpha\,\,|\,\,\exists f \in I\setminus \{0\}\,\, \textup{with}\,\, \textup{LT}(f)=cx^\alpha\right\}.
\end{equation*}
\end{defn}

\begin{defn}[Gr\"obner basis]
Fix a monomial order on $K[X]$. A finite subset $G=\{g_1,\dots,g_t\}$ of an ideal $I\subseteq K[X]$ other than $\{0\}$ is said to be a \emph{Gr\"obner basis} if
\begin{equation*}
\langle\textup{LT}(g_1),\dots,\textup{LT}(g_t)\rangle=\langle\textup{LT}(I)\rangle.
\end{equation*}
\end{defn}

       \chapter{\textit{CANONICA} quick reference guide}
\label{App:Lists}

\section{Installation}
\label{subsec:Installation}
\textit{CANONICA} is a \texttt{\justify Mathematica} package and requires an installation of version 10 or higher of \texttt{\justify Mathematica}. The \textit{CANONICA} repository can be copied to the local directory with
\begin{verse}
\begin{verbatim}
git clone https://github.com/christophmeyer/CANONICA.git
\end{verbatim}
\end{verse}
Alternatively, an archive file can be downloaded at
\begin{verse}
\begin{verbatim}
https://github.com/christophmeyer/CANONICA/archive/v1.0.tar.gz
\end{verbatim}
\end{verse}
which may be extracted with
\begin{verse}
\begin{verbatim}
tar -xvzf CANONICA-1.0.tar.gz
\end{verbatim}
\end{verse}
There is no further installation necessary, in particular, there are no dependencies other than \texttt{\justify Mathematica}. In a \texttt{\justify Mathematica} session, the package can be loaded by 
\begin{verse}
\begin{verbatim}
Get["CANONICA.m"];
\end{verbatim}
\end{verse}
provided the file \texttt{\justify CANONICA.m} is placed either in the current working directory or in one of the search paths. If this is not the case, \texttt{\justify Get} either has to be called with the full path of the file \texttt{\justify CANONICA.m}, or its location has to be added to the list of \texttt{\justify Mathematica's} search paths, which is stored in the global variable \texttt{\justify \$Path}, by
\begin{verse}
\begin{verbatim}
AppendTo[$Path,"/path/to/CANONICA/src/"]
\end{verbatim}
\end{verse}
Changes to \texttt{\justify \$Path} can be made permanent by adding them to the initialization file \texttt{\justify init.m}.

\section{Files of the package}
\label{subsec:Files}
The root directory of the \textit{CANONICA} package contains the following files and directories.
\begin{description}
\item[\texttt{./src/CANONICA.m}]~\\Contains the source code of the program, in particular, all function definitions as well as short usage messages for the public functions and options.
\item[\texttt{./manual.nb}]~\\An interactive manual in the \texttt{\justify Mathematica} notebook format explaining the usage of all functions and options with short examples.
\item[\texttt{./examples}]~\\Several examples are provided in this directory. The directory of each example contains a .m file with the corresponding differential equation and a .nb notebook file illustrating the application of \textit{CANONICA} to this example. The calculation of the full transformation can also be run in terminal mode with the script \texttt{\justify RunExample.m}. The script is started by calling
\begin{verse}
\begin{verbatim}
math -run "<<RunExample.m"
\end{verbatim}
\end{verse}
or
\begin{verse}
\begin{verbatim}
math -script RunExample.m
\end{verbatim}
\end{verse}
Some basic information about the examples, such as the master integrals and the definition of the kinematic invariants, is provided in the \texttt{\justify ./examples/ex\-amples.pdf} file.
\item[\texttt{./LICENSE}]~\\A copy of the third version of the GNU General Public License.
\item[\texttt{./README}]~\\A README file providing basic information on the package.
\end{description}

\section{List of functions provided by \textit{CANONICA}} 
\begin{description}
\item[CalculateDlogForm:]~\\\texttt{\justify CalculateDlogForm[a, invariants, alphabet]} returns a list of matrices of the same dimensions as \texttt{\justify a}, where each matrix is the matrix-residue of one of the letters. The ordering is the same as the one in \texttt{\justify alphabet}. Returns \texttt{\justify False} if \texttt{\justify a} cannot be cast in a dlog-form with the given \texttt{\justify alphabet}.
\item[CalculateNexta:]~\\\texttt{\justify CalculateNexta[aFull, invariants, sectorBoundaries, trafoPrevious, aPrevious]} applies \texttt{\justify trafoPrevious} to \texttt{\justify aFull} and returns the differential equation of the next sector. \texttt{\justify aPrevious} is used to recycle the transformation of lower sectors.
\item[CalculateNextSubsectorD:]~\\\texttt{\justify CalculateNextSubsectorD[a, invariants, sectorBoundaries, previousD]} computes the $D_k$ of the next sector, prepends it to \texttt{\justify previousD} and returns the result. The ansatz to be used can be specified with the optional argument \texttt{\justify userProvidedAnsatz}. If no ansatz is provided, an ansatz is generated automatically. The size of the automatically generated ansatz can be controlled with the option \texttt{\justify DDeltaNumeratorDegree}.
\item[CheckDlogForm:]~\\\texttt{\justify CheckDlogForm[a, invariants, alphabet]} tests whether the differential equation \texttt{\justify a} is in dlog-form for the given \texttt{\justify alphabet}. Returns either \texttt{\justify True} or \texttt{\justify False}.
\item[CheckEpsForm:]~\\\texttt{\justify CheckEpsForm[a, invariants, alphabet]} tests whether the differential equation \texttt{\justify a} is in canonical form with the given \texttt{\justify alphabet}. Returns either \texttt{\justify True} or \texttt{\justify False}.
\item[CheckIntegrability:]~\\\texttt{\justify CheckIntegrability[a, invariants]} tests whether \texttt{\justify a} satisfies the integrability condition $\textup{d}a-a\wedge a=0$ and returns either \texttt{\justify True} or \texttt{\justify False}.
\item[CheckSectorBoundaries:]~\\\texttt{\justify CheckSectorBoundaries[a, sectorBoundaries]} tests whether the \texttt{\justify sectorBoundaries} are compatible with \texttt{\justify a} and returns either \texttt{\justify True} or \texttt{\justify False}.
\item[ExtractDiagonalBlock:]~\\\texttt{\justify ExtractDiagonalBlock[a, boundaries]} returns the diagonal block of the differential equation \texttt{\justify a} specified by the \texttt{\justify boundaries} argument. \texttt{\justify boundaries} is expected to be of the format \texttt{\justify \{nLowest, nHighest\}}, where \texttt{\justify nLowest} and \texttt{\justify nHighest} are positive integers indicating the lowest and highest integrals of the diagonal block, respectively.
\item[ExtractIrreducibles:]~\\\texttt{\justify ExtractIrreducibles[a]} returns the irreducible denominator factors of \texttt{\justify a} that do not depend on the regulator. The option \texttt{\justify AllowEpsDependence->True} allows the irreducible factors to depend on both the invariants and the regulator.
\item[FindAnsatzSubsectorD:]~\\\texttt{\justify FindAnsatzSubsectorD[a, invariants, sectorBoundaries, previousD]} takes a differential equation \texttt{\justify a}, which is required to be in canonical form except for the off-diagonal block of the highest sector. Needs to be provided with all previous $D_k$ in the argument \texttt{\justify previousD} and computes the ansatz $\mathcal{R}_D$ for the computation of the next $D_k$. Takes the option \texttt{\justify DDeltaNumeratorDegree} to enlarge the ansatz. For more details see \sec{subsec:AnsatzOD}.
\item[FindAnsatzT:]~\\\texttt{\justify FindAnsatzT[a, invariants]} takes a differential equation \texttt{\justify a} in the \texttt{\justify invariants} and computes an ansatz $\mathcal{R}_T$ as described in \sec{subsec:AnsatzDB}. The ansatz can be enlarged with the options \texttt{\justify TDeltaNumeratorDegree} and \texttt{\justify TDeltaDenominatorDegree}.
\item[FindConstantNormalization:]~\\\texttt{\justify FindConstantNormalization[invariants, trafoPrevious, aPrevious]} calculates a constant diagonal transformation to minimize the number of prime factors present in the matrix-residues. The transformation is composed with \texttt{\justify trafoPrevious} and returned together with the resulting differential equation.
\item[FindEpsDependentNormalization:]~\\\texttt{\justify FindEpsDependentNormalization[a, invariants]} calculates a diagonal trans\-form\-ation depending only on the dimensional regulator in order to attempt to minimize the number of orders that need to be calculated in a subsequent determination of the transformation to a canonical form. Returns the transformation together with the resulting differential equation.
\item[RecursivelyTransformSectors:]~\\\texttt{\justify RecursivelyTransformSectors[aFull, invariants, sectorBoundaries, \{n\-SecStart, nSecStop\}]} calculates a rational transformation of \texttt{\justify aFull} to a canonical form in a recursion over the sectors of the differential equation, which have to be specified by \texttt{\justify sectorBoundaries}. The arguments \texttt{\justify nSecStart} and \texttt{\justify nSecStop} set the first and the last sector to be computed, respectively. If \texttt{\justify nSecStart} is greater than one, the result of the calculation for the sectors lower than \texttt{\justify nSecStart} needs to be provided in the additional arguments \texttt{\justify trafoPrevious} and \texttt{\justify aPrevious}. \texttt{\justify RecursivelyTransformSectors} returns the transformation of \texttt{\justify aFull} to a canonical form for the sectors up to \texttt{\justify nSecStop} and the resulting differential equation. The ansatzes for the individual blocks are generated automatically. The sizes of the ansatzes for the diagonal blocks can be controlled with the options \texttt{\justify TDeltaNumeratorDegree} and \texttt{\justify TDeltaDenominatorDegree}. Similarly, the sizes of the ansatzes for the off-diagonal blocks are controlled by the option \texttt{\justify DDeltaNumeratorDegree}.
\item[SectorBoundariesFromDE:]~\\\texttt{\justify SectorBoundariesFromDE[a]} returns the most fine-grained sector boundaries compatible with \texttt{\justify a}.
\item[SectorBoundariesFromID:]~\\\texttt{\justify SectorBoundariesFromID[masterIntegrals]} takes a list of \texttt{\justify masterIntegrals}, which need to be ordered by their sector-ids and returns the sector boundaries computed from the sector-ids.
\item[TransformDE:]~\\\texttt{\justify TransformDE[a, invariants, t]} applies the transformation \texttt{\justify t} to the differential equation \texttt{\justify a}. Returns $a^\prime=t^{-1}at-t^{-1}\textup{d}t$. The option \texttt{\justify SimplifyResult->False} deactivates the simplification of the result.
\item[TransformDiagonalBlock:]~\\\texttt{\justify TransformDiagonalBlock[a, invariants]} calculates a rational transformation to transform \texttt{\justify a} into canonical form and returns the transformation together with the resulting differential equation. With the optional argument \texttt{\justify userProvidedAnsatz}, the user can specify the ansatz to be used. If no ansatz is provided, an ansatz is generated automatically. The size of the automatically generated ansatz can be controlled with the options \texttt{\justify TDeltaNumeratorDegree} and \texttt{\justify TDeltaDenominatorDegree}.
\item[TransformDlogToEpsForm:]~\\\texttt{\justify TransformDlogToEpsForm[invariants, sectorBoundaries, trafoPrevious, aPrevious]} computes a transformation depending only on the regulator in order to transform \texttt{\justify aPrevious} from dlog-form into canonical form (cf.~\cite{Lee:2014ioa}). The transformation is composed with \texttt{\justify trafoPrevious} and returned together with the resulting differential equation. Per default, the transformation is demanded to be in a block-triangular form induced by \texttt{\justify sectorBoundaries}. This condition can be dropped with the option \texttt{\justify Enforce\-BlockTriangular->False}.
\item[TransformNextDiagonalBlock:]~\\\texttt{\justify TransformNextDiagonalBlock[aFull, invariants, sectorBoundaries, tr\-afoPrevious, aPrevious]} calls \texttt{\justify TransformDiagonalBlock} to compute the transformation of the next diagonal block into canonical form and composes it with \texttt{\justify trafoPrevious}. Returns the composed transformation together with the resulting differential equation. With the optional argument \texttt{\justify userProvidedAnsatz}, the user can specify the ansatz to be used. If no ansatz is provided, an ansatz is generated automatically. The size of the automatically generated ansatz can be controlled with the options \texttt{\justify TDeltaNumeratorDegree} and \texttt{\justify TDeltaDenominatorDegree}.
\item[TransformNextSector:]~\\\texttt{\justify TransformNextSector[aFull, invariants, sectorBoundaries, trafoPrevious, aPrevious]} transforms the next sector into canonical form, composes the calculated transformation with \texttt{\justify trafoPrevious} and returns it together with the resulting differential equation. With the optional argument \texttt{\justify userProvidedAnsatz}, the user can specify the ansatz to be used for the diagonal block. If no ansatz is provided, an ansatz is generated automatically. The size of the automatically generated ansatz for the diagonal block can be controlled with the options \texttt{\justify TDeltaNumeratorDegree} and \texttt{\justify TDeltaDenominatorDegree}. Similarly, the sizes of the ansatzes for the off-diagonal blocks are controlled by the option \texttt{\justify DDeltaNumeratorDegree}.
\item[TransformOffDiagonalBlock:]~\\\texttt{\justify TransformOffDiagonalBlock[invariants, sectorBoundaries, trafoPrevious, aPrevious]} assumes \texttt{\justify aPrevious} to be in canonical form except for the highest sector of which only the diagonal block is assumed to be in canonical form. Computes a transformation to transform the off-diagonal block of the highest sector into dlog-form. This transformation is composed with \texttt{\justify trafoPrevious} and returned together with the resulting differential equation. Proceeds in a recursion over sectors, which can be resumed at an intermediate step by providing all previous $D_k$ in the optional argument \texttt{\justify userProvidedD}. The sizes of the automatically generated ansatzes for the off-diagonal blocks are controlled by the option \texttt{\justify DDeltaNumeratorDegree}.
\end{description}

\section{List of options} 
\label{App:ListOptions}
\begin{description}
\item[AllowEpsDependence:]~\\\texttt{\justify AllowEpsDependence} is an option of \texttt{\justify ExtractIrreducibles} controlling whether irreducible factors depending on both the invariants and the regulator are returned as well. The default value is \texttt{\justify False}.
\item[DDeltaNumeratorDegree:]~\\\texttt{\justify DDeltaNumeratorDegree} is an option controlling the numerator powers of the rational functions in the ansatz used for the computation of $D$ for the transformation of off-diagonal blocks. The default value is \texttt{0}. For more details see \sec{subsec:AnsatzOD}. \texttt{\justify DDeltaNumeratorDegree} is an option of the following functions: \texttt{\justify CalculateNextSubsectorD}, \texttt{\justify FindAnsatzSubsectorD}, \texttt{\justify RecursivelyTransformSectors}, \texttt{\justify TransformNextSector}, \texttt{\justify TransformOffDiagonalBlock}.
\item[EnforceBlockTriangular:]~\\\texttt{\justify EnforceBlockTriangular} is an option of \texttt{\justify TransformDlogToEpsForm} controlling whether the resulting transformation is demanded to be in the block-triangular form induced by the \texttt{\justify sectorBoundaries} argument. The default value is \texttt{\justify True}.
\item[FinalConstantNormalization:]~\\\texttt{\justify FinalConstantNormalization} is an option of \texttt{\justify RecursivelyTransformSectors} controlling whether \texttt{\justify FindConstantNormalization} is invoked after all sectors have been transformed into canonical form in order to simplify the resulting canonical form. The default value is \texttt{\justify False}.
\item[PreRescale:]~\\\texttt{\justify PreRescale} is an option of \texttt{\justify TransformDiagonalBlock} controlling whether \texttt{\justify FindEpsDependentNormalization} is called prior to the main computation in order to attempt to minimize the number of orders that need to be calculated in a subsequent determination of the transformation to a canonical form. The default value is \texttt{\justify True}.
\item[SimplifyResult:]~\\\texttt{\justify SimplifyResult} is an option of \texttt{\justify TransformDE} controlling whether the resulting differential equation is simplified. The default value is \texttt{\justify True}.
\item[TDeltaDenominatorDegree:]~\\\texttt{\justify TDeltaDenominatorDegree} is an option controlling the denominator powers of the rational functions in the ansatz used for the computation of the transformation of diagonal blocks. The default value is \texttt{0}. For more details see \sec{subsec:AnsatzDB}. \texttt{\justify TDeltaDenominatorDegree} is an option of the following functions: \texttt{\justify FindAnsatzT}, \texttt{\justify RecursivelyTransformSectors}, \texttt{\justify TransformNextDiagonalBlock}, \texttt{\justify TransformNextSector}.
\item[TDeltaNumeratorDegree:]~\\\texttt{\justify TDeltaNumeratorDegree} is an option controlling the numerator powers of the rational functions in the ansatz used for the computation of the transformation of diagonal blocks. The default value is \texttt{0}. For more details see \sec{subsec:AnsatzDB}. \texttt{\justify TDeltaNumeratorDegree} is an option of the following functions: \texttt{\justify FindAnsatzT}, \texttt{\justify RecursivelyTransformSectors}, \texttt{\justify TransformNextDiagonalBlock}, \texttt{\justify TransformNextSector}.
\item[VerbosityLevel:]~\\\texttt{\justify VerbosityLevel} is an option controlling the verbosity of several main functions. Takes integer values from \texttt{0} to \texttt{12} with a value of \texttt{12} resulting in the most detailed output about the current state of the computation and a value of \texttt{0} suppressing all output but warnings about inconsistent inputs. The default value is \texttt{10}. The following functions accept the \texttt{\justify VerbosityLevel} option: \texttt{\justify CalculateNextSubsectorD}, \texttt{\justify FindConstantNormalization}, \texttt{\justify RecursivelyTransformSectors}, \texttt{\justify TransformDiagonalBlock}, \texttt{\justify TransformDlogToEpsForm}, \texttt{\justify TransformNextDiagonalBlock}, \texttt{\justify TransformNextSector}, \texttt{\justify TransformOffDiagonalBlock}.
\end{description}
\section{List of global variables and protected symbols} 
\label{App:GlobalVars}
\begin{description}
\item[\$ComputeParallel:]~\\\texttt{\justify \$ComputeParallel} is a global variable that needs to be set to \texttt{\justify True} to enable parallel computations. The number of kernels to be used is controlled by \texttt{\justify \$NParallelKernels}.
\item[\$NParallelKernels:]~\\\texttt{\justify \$NParallelKernels} is a global variable setting the number of parallel kernels to be used. \texttt{\justify \$NParallelKernels} has no effect if \texttt{\justify \$ComputeParallel} is not set to \texttt{\justify True}. If \texttt{\justify \$ComputeParallel} is \texttt{\justify True} and \texttt{\justify \$NParallelKernels} is not assigned a value, then all available kernels are used for the computation.
\item[eps:]~\\\texttt{\justify eps} is a protected symbol representing the dimensional regulator.
\end{description}

    \backmatter
      \bibliographystyle{JHEP}
      \bibliography{../../LoopIntegrals/DifferentialEqns/myPapers/literature.bib}
      \begingroup
\selectlanguage{ngerman}
\chapter*{Selbst\"andigkeitserkl\"arung}

Ich erkläre, dass ich die Dissertation selbständig und nur unter Verwendung der von mir gemäß § 7 Abs. 3 der Promotionsordnung der Mathematisch-Naturwissenschaftlichen Fakultät, veröffentlicht im Amtlichen Mitteilungsblatt der Humboldt-Universität zu Berlin Nr. 126/2014 am 18.11.2014 angegebenen Hilfsmittel angefertigt habe.

\vspace{5\baselineskip}
\noindent Berlin, den \today\hfill Christoph Meyer
\endgroup

\end{document}